\documentclass[conference]{IEEEtran}
\IEEEoverridecommandlockouts
\usepackage{mathptmx} % This is Times font

\usepackage{flushend}
\usepackage{balance}
\usepackage{booktabs}
\usepackage{todonotes}
\usepackage{epsfig}
\usepackage{graphicx}
\usepackage{amsmath}
\usepackage{amssymb}
\usepackage{lipsum}
\usepackage{subfig}
\usepackage{color}

\usepackage[normalem]{ulem}
\usepackage{bigstrut}
\usepackage{multirow}
\usepackage{siunitx}
\usepackage{ulem}
\usepackage{listings}
\usepackage{eso-pic}
\usepackage{tikz}
\usepackage{xspace}
\usepackage{fancyhdr}
\usepackage{txfonts}
\usepackage{mathtools}
\usepackage{soul}
\usepackage[leftcaption]{sidecap}

\lstdefinelanguage
   [x64]{Assembler}     % add a "x64" dialect of Assembler
   [x86masm]{Assembler} % based on the "x86masm" dialect
   {morekeywords={findrf, LOAD, SORT, MERGE, ACUM, SUBI, JMP, ASSIGN, findneuron%
                  }} % etc.

\lstdefinestyle{PythonStyle}{
    commentstyle=\color{mGreen},
    keywordstyle=\color{magenta},
    numberstyle=\tiny\color{mGray},
    stringstyle=\color{mPurple},
    basicstyle=\linespread{2}\ttfamily,
    breakatwhitespace=false,         
    breaklines=true,                 
    captionpos=b,                    
    keepspaces=true,                 
    numbers=none,                    
    numbersep=5pt,                  
    showspaces=false,                
    showstringspaces=false,
    showtabs=false,                  
    tabsize=2,
    otherkeywords={},
    language=Python
}

\usepackage[bookmarks=true,breaklinks=true,letterpaper=true,colorlinks,linkcolor=blue,citecolor=magenta,urlcolor=blue]{hyperref}

\title{\huge{Ptolemy: Architecture Support for Robust Deep Learning}
}

\graphicspath{{figs/}}

\begin{document}

\makeatletter
\newcommand{\linebreakand}{%
  \end{@IEEEauthorhalign}
  \hfill\mbox{}\par
  \mbox{}\hfill\begin{@IEEEauthorhalign}
}
\makeatother

\author{\IEEEauthorblockN{Yiming Gan\textsuperscript{*}}
\IEEEauthorblockA{University of Rochester\\ygan10@ur.rochester.edu}
\and
\IEEEauthorblockN{Yuxian Qiu\textsuperscript{*}}
\IEEEauthorblockA{Shanghai Jiao Tong University\\qiuyuxian@sjtu.edu.cn}
\linebreakand
\IEEEauthorblockN{Jingwen Leng}
\IEEEauthorblockA{Shanghai Jiao Tong University \\ Shanghai Qi Zhi Institute\\leng-jw@cs.sjtu.edu.cn}
\and
\IEEEauthorblockN{Minyi Guo}
\IEEEauthorblockA{Shanghai Jiao Tong University\\ Shanghai Qi Zhi Institute\\guo-my@cs.sjtu.edu.cn}
\and
\IEEEauthorblockN{Yuhao Zhu}
\IEEEauthorblockA{University of Rochester\\yzhu@rochester.edu}
}
\maketitle
\begingroup\renewcommand\thefootnote{*}
\footnotetext{Equal contribution}
\endgroup
%\thispagestyle{firstpage}
%\pagestyle{plain}

%%%%%% -- PAPER CONTENT STARTS-- %%%%%%%%

%!TEX root=paper.tex

\newcommand{\website}[1]{{\tt #1}}
\newcommand{\program}[1]{{\tt #1}}
\newcommand{\benchmark}[1]{{\it #1}}
\newcommand{\fixme}[1]{{\textcolor{red}{\textit{#1}}}}

\newcommand{\algrule}[1][.2pt]{\par\vskip.5\baselineskip\hrule height #1\par\vskip.5\baselineskip}
\newcommand{\funcname}[1]{\textsc{\textbf{\textcolor{RoyalBlue}{#1}}}}
\newcommand{\inst}[1]{\textbf{\texttt{#1}}}
\newcommand{\sys}[1]{\underline{\textsc{#1}}}

\newcommand*\circled[2]{\tikz[baseline=(char.base)]{
            \node[shape=circle,fill=black,inner sep=1pt] (char) {\textcolor{#1}{{\footnotesize #2}}};}}

%https://tex.stackexchange.com/questions/58702/creating-gears-in-tikz
\newcommand{\drawgear}[5]{
\foreach \i in {1,...,#1} {
  [rotate=(\i-1)*360/#1]  (0:#2)  arc (0:#4:#2) {[rounded corners=1.5pt]
             -- (#4+#5:#3)  arc (#4+#5:360/#1-#5:#3)} --  (360/#1:#2)
}}

\newcommand{\gear}[5]{
\begin{tikzpicture}
  \draw[thick] \drawgear{#1}{#2}{#3}{#4}{#5};
\end{tikzpicture}
}

\newcommand{\knob}[1]{
  \noindent\gear{8}{0.08}{0.1}{10}{2}~\textbf{#1}
}

\ifx\figurename\undefined \def\figurename{Figure}\fi
\renewcommand{\figurename}{Fig.}
\renewcommand{\paragraph}[1]{\textbf{#1}~~}
\newcommand{\figline}{{\vspace*{.05in}\hline}}

\newcommand{\Sect}[1]{Sec.~\ref{#1}}
\newcommand{\Fig}[1]{Fig.~\ref{#1}}
\newcommand{\Tbl}[1]{Table~\ref{#1}}
\newcommand{\Equ}[1]{Equ.~\ref{#1}}
\newcommand{\Apdx}[1]{Apdx.~\ref{#1}}
\newcommand{\Algo}[1]{Alg.~\ref{#1}}
\newcommand{\Ques}[1]{Question~\ref{#1}}
\newcommand{\Comm}[1]{Comment~\ref{#1}}
\newcommand{\Lst}[1]{Lst.~\ref{#1}}

\newcommand{\specialcell}[2][c]{\begin{tabular}[#1]{@{}l@{}}#2\end{tabular}}
\newcommand{\note}[1]{\textcolor{red}{#1}}

\newcommand{\greenweb}{{\fontfamily{cmtt}\selectfont GreenWeb}\xspace}
\newcommand{\autogreen}{\textsc{AutoGreen}\xspace}
\newcommand{\proj}{\textsc{Ptolemy}\xspace}
\newcommand{\prednet}{\textsc{Prednet}\xspace}
\newcommand{\scalesim}{\textsc{ScaleSim}\xspace}

\newcommand{\no}[1]{#1}
\renewcommand{\no}[1]{}
\newcommand{\RNum}[1]{\uppercase\expandafter{\romannumeral #1\relax}}

% checkmark and xmark in the pifont package
%\newcommand{\cmark}{\ding{51}}
%\newcommand{\xmark}{\ding{55}}

\begin{abstract}

Deep learning is vulnerable to adversarial attacks, where carefully-crafted input perturbations could mislead a well-trained Deep Neural Network (DNN) to produce incorrect results. Adversarial attacks jeopardize the safety, security, and privacy of DNN-enabled systems. Today's countermeasures to adversarial attacks either do not have the capability to \textit{detect} adversarial samples at inference-time, or introduce prohibitively high overhead to be practical at inference-time.

We propose \proj, an algorithm-architecture co-designed system that \textit{detects adversarial attacks at inference time with low overhead and high accuracy}. We exploit the synergies between DNN inference and imperative program execution: an input to a DNN uniquely activates a set of neurons that contribute significantly to the inference output, analogous to the sequence of basic blocks exercised by an input in a conventional program. Critically, we observe that adversarial samples tend to activate distinctive paths from those of benign inputs. Leveraging this insight, we propose an adversarial sample detection framework, which uses canary paths generated from offline profiling to detect adversarial samples at runtime. \no{We provide a high-level programming interface that allows programmers to explore the critical trade-off between detection accuracy and efficiency under the general algorithm framework.} The \proj compiler along with the co-designed hardware enable efficient execution by exploiting the unique algorithmic characteristics. Extensive evaluations show that \proj achieves higher or similar adversarial sample detection accuracy than today's mechanisms with a much lower (as low as 2\%) runtime overhead.

\end{abstract}

% keywords

\begin{Artifact}
\url{https://github.com/Ptolemy-DL/Ptolemy}
\end{Artifact}

%\IEEEraisesectionheading{
\section{Introduction}
\label{sec:intro}

%While Deep Neural Networks (DNN) have demonstrated remarkable capability in a wide range of tasks such as computer vision~\cite{alexnet} and natural language processing \cite{word2vec},

%Fundamentally, DNNs are results of non-linear, non-convex optimization problems that do not exactly simulate human brains. Thus, 
%that are indistinguishable to human 
Deep Neural Networks (DNN) are not robust. Small perturbations to inputs could easily ``fool'' DNNs to produce incorrect results. By manipulating the perturbation, a range of so-called \textit{adversarial attacks} have been demonstrated to lead DNNs to mis-predict~\cite{papernot2016distillation, carlini2017towards, kurakin2016adversarial, intriguring, nguyen2015deep, hu2020deepsniffer}, which could result in potentially severe consequences. For instance, physically putting a sticker on a stop sign could lead a well-trained object recognition DNN to misclassify the stop sign as a yield sign~\cite{kurakin2016adversarial}. Beyond mission-critical scenarios such as autonomous driving, the robustness issue also obstructs the deployment of DNN in privacy/security-sensitive domains such as biometric authentication~\cite{deepface,deeppid}.
% and surveillance~\cite{rajpoot2014security}.

%Perhaps the most infamous example is the fatal collision of a Tesla Model S with a tractor trailer, whose white side is mistaken for the brightly-lit sky by Tesla's Autopilot system~\cite{tesladeath}. 
%, or more controversially, predicting how likely someone is going to commit a crime~\cite{??}.

%In all these scenarios, not only do incorrect inferences lead to security/privacy vulnerabilities~\cite{??} and life-threatening consequences~\cite{??}, but also they can easily occur.

%cause severe effects. However, the robustness and reliability of DNN have been proved to be weak by many works \cite{polluting,2014towards,2016measure} which may result in serious security vulnerability.

%The single most important attacking approach is adversarial attack which indicates that one can engineer small perturbations to the input data and lead to the malfunction of the high-performing DNNs found by Szegedy \textit{et al.} \cite{intriguring}. Different policies of perturbations \cite{adversarialphysical,jsma,onepixel} have been proposed to mislead deep neural networks proving that there is still significant gap between machine and human perception.

We take a first step toward architectural support for robust deep learning. For a robustness scheme to be effective in practice, it not only has to accurately \textit{detect} adversarial inputs, but must also do so efficiently \textit{at inference time} so that proper measure could be taken. This paper proposes \proj, an algorithm-architecture co-design system that \textit{detects adversarial attacks at inference time with low overhead and high accuracy}. This enables applications to reject incorrect results produced by adversarial attacks during inference.~\Fig{fig:overview} provides an overview of the system.
%, significantly improving their robustness in practical deployment.

%, and thus have limited applicability in practice.

%To counter the adversarial attacks, people have been proposing techniques to improve the robustness of DNNs including injecting adversarial examples into training data to increase the robustness \cite{adversarialtraining,improvingstability}, applying random modifications to model weights \cite{randomdefense} and leveraging modular redundancy (MR) with majority vote. However, none of the approaches described above is efficient and cheap to be deployed during inference resulting it less useful in most real world scenarios. Another direction of defense is to identify specialized online pattern during inference and compare with offline database to detect adversarial examples \cite{effectpath2019} which can be utilized during online inference procedure but with high latency and memory overhead.

%We capture the dynamic behaviors of DNN inference using techniques inspired by classic profiling-driven optimizations. 
Existing countermeasures to adversarial attacks are unable to detect adversarial samples at inference time~\cite{carlini2017adversarial,he2017adversarial}. Fundamentally, they treat DNN inferences as black boxes, ignoring their runtime behaviors. To enable efficient online adversarial detection, this paper takes a different, white-box approach. We exploit the fact that each input to a DNN uniquely exercises an \textit{activation path}---a collection of neurons that contribute significantly to the inference output, analogous to the sequence of basic blocks exercised by an input in a conventional program. Analyzing ``hot'' activation paths in DNNs, our \textbf{key observation} is that inputs that lead to the same inference class tend to exercise a group of paths that are distinctive from other inference classes.

%similar to how an input to a conventional program would exercise a unique control-flow path consisting of basic blocks, 

\begin{figure*}[t]
%\vspace{-5pt}
\centering
\includegraphics[trim=0 0 0 0, clip, width=2\columnwidth]{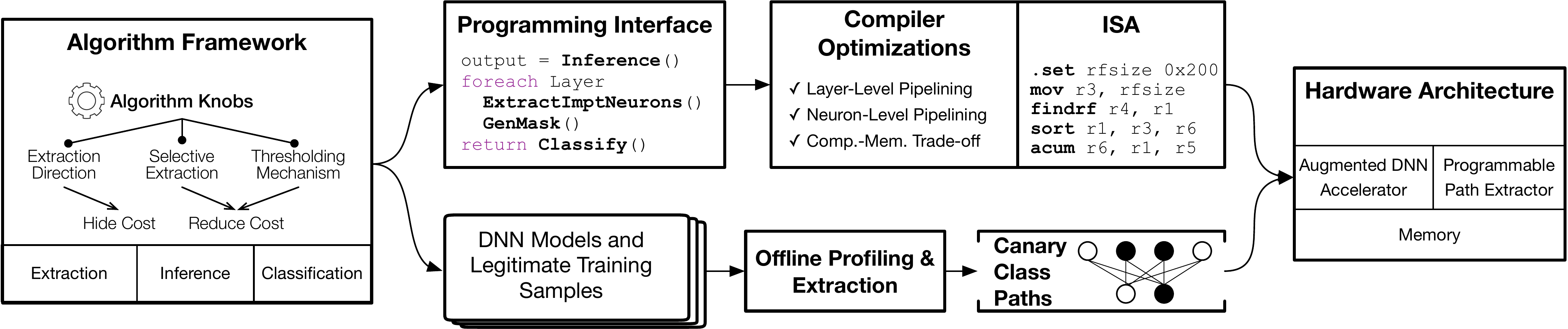}
%\vspace{-5pt}
\caption{\proj system overview.}
\label{fig:overview}
\vspace{-10pt}
\end{figure*}

We propose a general algorithmic framework that exploits the runtime path behaviors for efficient online adversarial sample detection. The detection framework constructs a canary \textit{class path} offline for each inference class by profiling the training data. At runtime, it builds the activation path for an input, and detects the input as an adversary if the activation path is different from the canary path associated with the predicted class. The general algorithm framework exposes a myriad of design knobs affecting the critical trade-off between detection accuracy and compute cost, such as how a path is formulated and when the path is constructed. To widen the applicability of our detection framework, \proj provides a high-level programming interface, which allows programmers to calibrate the algorithmic knobs to explore the accuracy-cost trade-off that best suits an application's needs.

\proj provides an efficient execution substrate. The key to the execution efficiency is the \proj compiler, which hides and reduces the detection overhead by exploiting the unique parallelisms and redundancies exposed by the detection algorithms. We show that with the aggressive compile-time optimizations and a well-defined ISA, detection algorithms can be implemented on top of existing DNN accelerators with a set of basic, yet principled, hardware extensions, further widening the applicability of \proj.

\proj enables highly accurate adversarial detection with low performance overhead. Compared to today's defense mechanisms that introduce over 10 $\times$ performance overhead, we demonstrate a system that achieves higher accuracy with only 2\% performance overhead. \proj defends not only existing attacks, but also \textit{adaptive} attacks that are specifically designed to defeat our defense~\cite{carlini2019evaluating}. We also demonstrate the \proj framework's flexibility by presenting a range of algorithm variants that offer different accuracy-efficiency trade-offs. For instance, \proj could trade 10\% performance overhead for 0.03 higher detection accuracy.

%We evaluate the \proj over a general-purpose system consisting of an Intel Xeon Silver 4110 CPU and an Nvidia RTX 2080 Ti GPU. We show that \proj achieves xx speedup and xx power reduction. We also show that user could build a new detection algorithm easily with the knobs we provide, and we propose an example with xx speedup compared to the original algorithm with only xx accuracy loss.

%To the best of knowledge, this is the first paper that enables low-overhead, high-accu on systems and architecture support for robust deep learning. By co-designing the algorithmic framework with the systems stack, 
The \proj artifact, including the pre-trained models, offline-generated class paths, code to generate adaptive and non-adaptive attacks, and the detection implementation is available at \mbox{\url{https://github.com/Ptolemy-DL/Ptolemy}}. In summary, \proj provides a generic framework for low-overhead, high-accuracy online defense against adversarial attacks with the following contributions:

\begin{itemize}
    %\setlength\itemsep{-2pt}
	%\item We throughly understand the adversarial sample detection algorithm and propose three different optimization knobs to form a framework.
	\item We propose a novel static-dynamic collaborative approach for adversarial detection by exploiting the unique program execution characteristics of DNN inferences that are largely ignored before.
	\item We present a general algorithmic framework, along with a high-level programming interface, that allows programmers to explore key algorithm design knobs to navigate the accuracy-efficiency trade-off space.
	%\item We propose a novel and lightweight ISA having strong descriptive capability of our detection algorithm.
	\item We demonstrate that with a carefully-designed ISA, compiler optimizations could enable efficient detection by exploiting the unique parallelisms and redundancies exposed by our detection algorithm framework.
	\item We present a programmable hardware to achieve low-latency online adversarial defense with principled extensions to existing DNN accelerators.
\end{itemize}

%The rest of this paper is organized as follows. \Sect{sec:motiv} introduces multiple defense methods against adversarial attacks and motivates this work. \Sect{sec:sw} gives an overview of the software framework and its variants, we also introduce two algorithm level optimization.\Sect{sec:inter} introduces the interface between software and hardware in which we introduce both high level API, low level ISA and compilation. \Sect{sec:arch} describes the architecture support and architectural optimization we applied. \Sect{sec:setup} describes the evaluation methodology and setup. \Sect{sec:eval} evaluates this work. \Sect{sec:related} puts this work in the context of related work and \Sect{sec:diss} concludes this paper.

\section{Background}
\label{sec:bg}

%This section provides a brief background on adversarial attacks of deep learning models (\Sect{sec:bg:attack}), followed by an overview of existing defense methods, which are not suitable for inference-time detection (\Sect{sec:bg:defense}).

%\subsection{DNN Robustness}
%\label{sec:bg:attack}

%\begin{figure*}[t]
%%\vspace{-2pt}
%  \begin{minipage}[t]{0.3\columnwidth}
%    \centering
%    \includegraphics[trim=0 0 0 0, clip, height=1.2in]{stopexample}
%    \caption{Adversarial sample generated with the FGSM \cite{fgsm} attack.}
%    \label{fig:example}
%  \end{minipage}
%  %\hfill
%  \begin{minipage}[t]{1.6\columnwidth}
%    \centering
%    \includegraphics[trim=0 0 0 0, clip, height=1.2in]{overview}
%    \caption{Overview of the \proj system.}
%    \label{fig:overview}
%  \end{minipage}
%%\vspace{-3pt}
%\end{figure*}

 \paragraph{Adversarial Attacks} DNNs are not robust to adversarial attacks, where DNNs mis-predict under slightly perturbed inputs~\cite{carlini2017towards, kurakin2016adversarial, papernot2016distillation, deepfool}.~\Fig{fig:example} shows one such example, where the two slightly different images are both perceived as stop signs to human eyes, but the second image is mis-predicted by a DNN model as a yield sign. The perturbations could be the result of carefully engineered attacks, but could also be an artifact of normal data acquisition such as noisy sensor capturing and image compression/resizing~\cite{thang2019image}.

\begin{figure}[h]
%\vspace{-5pt}
\centering
\includegraphics[trim=0 0 0 0, clip, width=0.9\columnwidth]{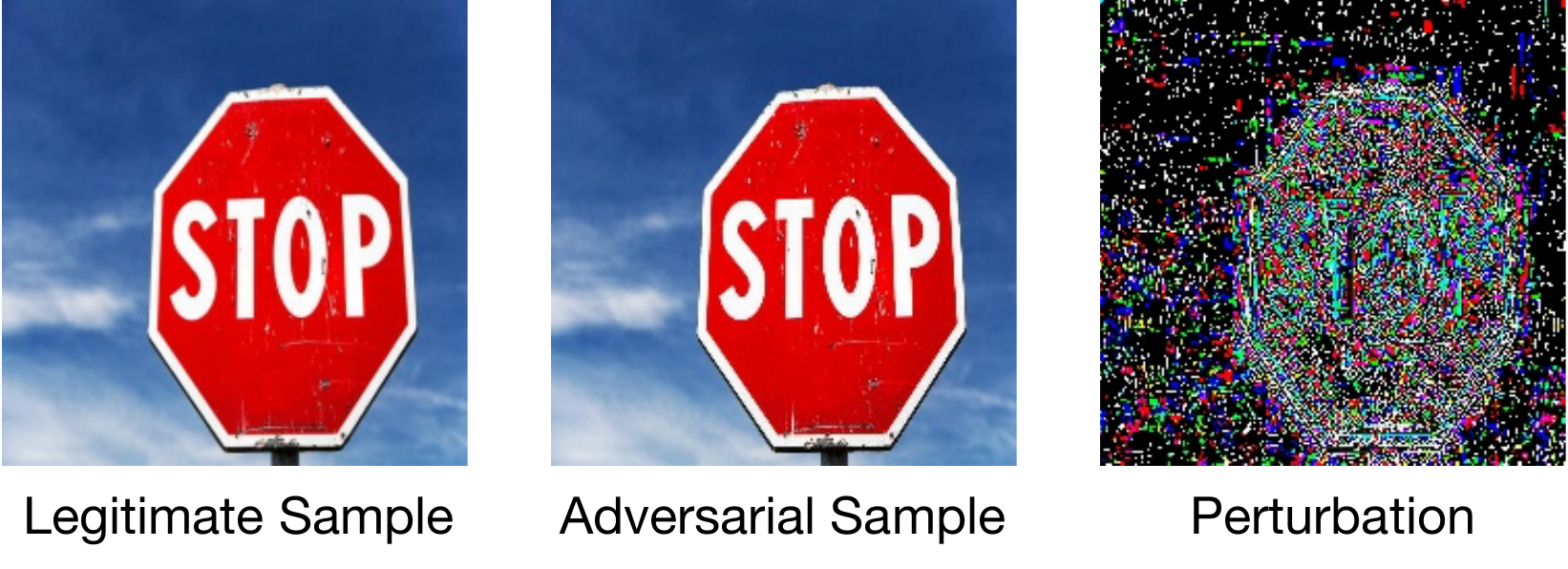}
%\vspace{-5pt}
\caption{Adversarial example using the FGSM \cite{fgsm} attack.}
\label{fig:example}
%\vspace{-15pt}
\end{figure}

Formally, given a DNN $\mathcal{C}$, an input $x'$ is defined as an adversarial sample if it is close to $x$ yet makes $\mathcal{C}^*(x) = \mathcal{C}(x) \neq \mathcal{C}(x')$, where $\mathcal{C}^*(x)$ is the correct class of $x$. Different adversarial samples differ in their measures of the distance between $x$ and $x'$. The distance could be small, where the input perturbations are imperceptible to humans (as in the example above), but could also be large, where the perturbations are visible to humans but still ``fool'' a DNN. For instance, physically putting a sticker on a stop sign could mislead a DNN to misclassify the stop sign as a yield sign~\cite{kurakin2016adversarial,li2019adversarial}. \proj targets the general robustness issue that introduces mis-predictions through input perturbations---small or large, inadvertent or malicious. For simplicity, we refer to all of them as adversarial attacks throughout this paper.

An adversarial attack is a black-box attack if it does not assume knowledge of the attacked model; white-box attacks in contrast assume full knowledge of the model. Orthogonally, adaptive attacks have complete knowledge of the defense's inner workings, i.e., are specifically designed to attempt to defeat a defense, while non-adaptive attacks do not~\cite{tramer2020adaptive, carlini2019evaluating, carlini2017adversarial}. We show that our detection scheme can defend against a range of different attacks, including the strongest form of attack: white-box adaptive attacks.

\begin{figure*}[t]
  %\vspace{-5pt}
  \centering
  \includegraphics[trim=0 0 0 0, clip, width=2\columnwidth]{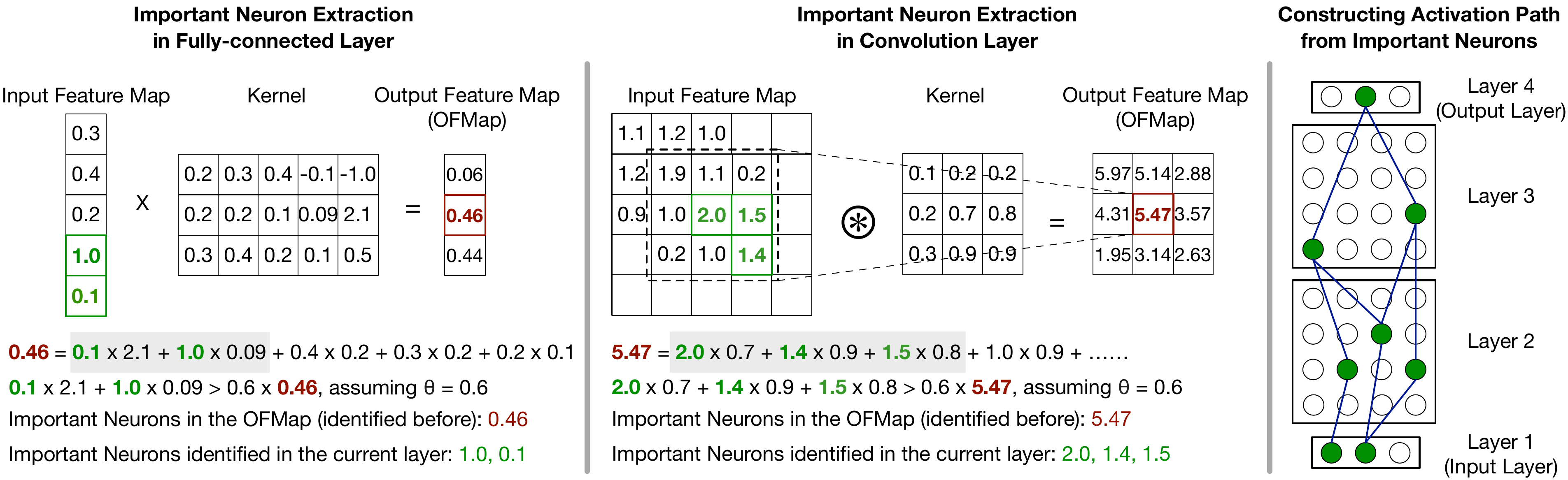}
  \caption{Extracting important neurons from a fully connected layer (left) and a convolution layer (middle), and constructing the activation path from important neurons across layers (right). Activation paths are input-specific. This figure illustrates backward extraction using a cumulative threshold. Forward extraction would start from the first layer rather than from the last year. Absolute thresholding would select important neurons based on absolute partial sums rather than cumulative partial sums.}
  \label{fig:illustration}
  \vspace{-10pt}
\end{figure*}

\paragraph{Countermeasures} We aim to enable fast and accurate systems that can \textit{detect} adversarial examples at \textit{inference-time} such that proper measures could be taken. Today's defense mechanisms largely fall under two categories, neither of which meets this goal. The first class of defenses improves the robustness of DNN models at \textit{training time} (e.g., adversarial retraining)~\cite{adversarialtraining, improvingstability} by incorporating adversarial examples into the training data. However, re-training is not suitable at inference-time and requires accesses to the training data. Another class of defenses uses redundancies to defend against adversarial attacks~\cite{thang2019image, rouhani2018deepfense}, similar to the multi-module redundancy used in classic fault-tolerant systems~\cite{sorin2009fault}. This scheme, however, introduces high overhead, limiting its applicability at inference time.

\section{Algorithmic Framework}
\label{sec:sw}

%To enable deep neural network based systems with the ability of countering adversarial examples during on-line inference, a general, flexible and fast method needed to be proposed. The method should be able to counter all kinds of attacks without adding anything into training receipts and it should add minor timing and energy overhead to the original system. This section first gives an overview on the algorithms we use to detect adversarial examples in \Sect{sec:overview} and further proposes the basic algorithm alternative knobs supported by \proj in \Sect{sec:basic}.

This section introduces the \proj algorithm framework, which enables adversarial attack detection at inference-time with high accuracy and low latency. \proj provides a set of principled design knobs to allow programmers to customize the accuracy vs. efficiency trade-off.

We first describe the intuition and key concepts behind our algorithm framework~(\Sect{sec:sw:ov}). We then introduce the algorithm framework, and show that a basic algorithm under the framework introduces excessive compute and memory cost (\Sect{sec:sw:cost}). We further introduce key algorithmic knobs that enable different algorithm variants to offer different accuracy-efficiency trade-offs (\Sect{sec:sw:algo}). Finally, we introduce a high-level programming interface to flexibly express detection algorithms within our framework~(\Sect{sec:sw:lang}).

\subsection{Intuition and Key Concepts}
\label{sec:sw:ov}

%Understanding DNN inference behaviors at runtime rather than treating DNN as a blackbox helps design efficient adversarial detection methods. We introduce the intuition behind our algorithm framework.

%Our detection algorithm is inspired by the classic profile-guided optimizations~\cite{??}. 
\paragraph{Intuition} Each input to a DNN activates a sequence of neurons. We find that inputs that are correctly predicted as the same class tend to activate a unique set of neurons distinctive from that of other inputs. This is a manifestation of recent work on \textit{class-level} model sparsity~\cite{effectpath2019, cdrp}, which shows that a small, but distinctive, portion of the network contributes to each predicted class. Taking this perspective, the way adversarial samples alter the inference result can be thought of as activating a sequence of neurons different from the canonical sequence associated with its predicted output. Analyzing dynamic paths in DNN inferences thus allows us to detect adversaries.

A sequence of activated neurons is analogous to a sequence of basic blocks exercised by an input to a conventional program. The frequently exercised basic block sequences, i.e., ``hot paths''~\cite{ball1996efficient, fisher1981trace, chang1988trace}, can be used to improve performance in classic profile-guided optimizations and dynamic compilers~\cite{smith2000overcoming, smith2005virtual,donovan2000profile}. Our approach shares a similar idea, where we treat a DNN as an imperative program, and leverage its runtime paths (sequence of neurons) to guide adversarial sample detection. Conventional countermeasures largely ignore the program execution behaviors of DNN inferences.
%Critically, a path in a DNN inference consists of neurons as opposed to basic blocks.

%We find that when an adversarial input successfully alters the prediction result, the network usually activates a significantly different path compared to the benign samples, allow us to detect the adversary.

%Our adversarial sample detection algorithm is inspired from a line of recent work~\cite{effectpath2019, cdrp} that leverages the insight of \emph{image/class-level sparsity}.

%It is well understood that DNN models are highly sparse when inspecting the model output from all training samples.
%As such, it is natural the model is even more sparse when inspecting an individual image.
%However, prior work has made the fundamental observation that images in the same class activate the similar portion of the network.
%We choose to use the existing method with the highest detection accuracy and reasonable performance overhead.

\begin{figure*}[t]
  %\vspace{-5pt}
  \centering
  \includegraphics[trim=0 0 0 0, clip, width=1.9\columnwidth]{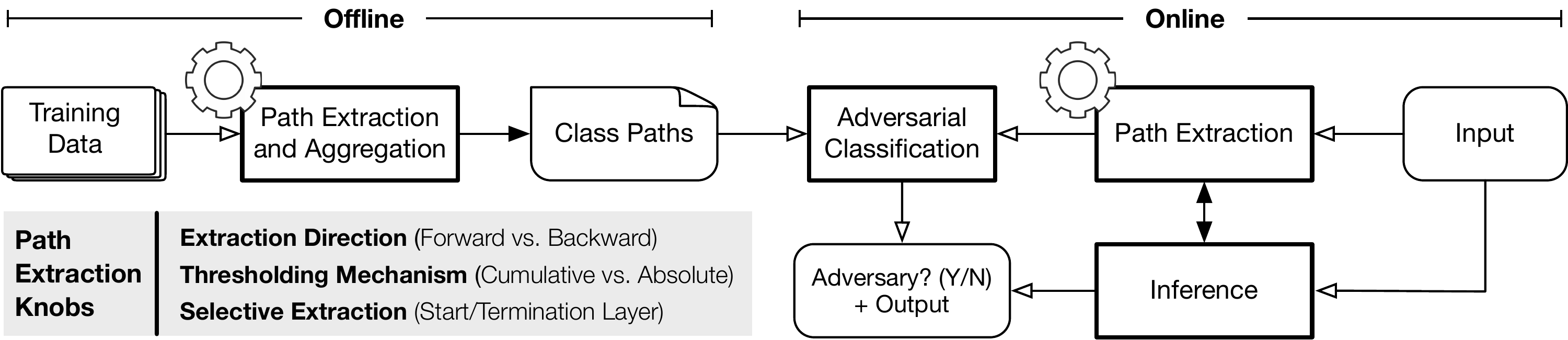}
  \caption{Adversarial detection algorithm framework. It provides a range of knobs for path extraction, which dominates the runtime overhead. Note that the path extraction methods in both the offline and online phases must match.}
  \label{fig:algo}
  \vspace{-10pt}
\end{figure*}

\paragraph{Important Neurons} The premise of our detection algorithm framework is the notion of \textit{important neurons}, which denote a set of neurons that contribute significantly to the inference output. Important neurons are extracted in a backward fashion. The last layer $L_n$ has only one important neuron, which is the neuron $\mathbf{n}$ that corresponds to the predicted class. At the second last layer $L_{n-1}$, the important neurons are the minimal set of neurons in the input feature map that contribute to at least $\theta$ ($0 \leq \theta \leq 1$) of $\mathbf{n}$. Here, $\theta$ controls the coverage of important neurons. To extract the important neurons of layer $L_{n-1}$, we simply rank the partial sums used to calculate $\mathbf{n}$, and choose the minimal number of neurons whose partial sums collectively contribute to at least $\theta \times \mathbf{n}$.
%, and is a key parameter that decides the detection accuracy vs. efficiency trade-off as we will discuss later

The left panel in~\Fig{fig:illustration} shows an example using a fully-connected layer. Assuming $\theta = 0.6$ and the second neuron in the output feature map (0.46) is the important neuron identified in the next layer. The fourth (1.0) and the fifth (0.1) neurons in the input feature map are identified as the important neurons in the current layer, because they contribute the two large partial sums and their cumulative partial sum (0.3) contribute to more than 60\% of the important neuron in the output feature map. The same process can be extended to convolution layers. The middle panel in~\Fig{fig:illustration} shows an example. For the important neuron in the output feature map, we first find its receptive field in the input feature map, and then identify the minimal set of neurons in the receptive field whose cumulative partial sums contribute to at least $\theta \times \mathbf{n}$.

%An element in the input feature map could be in multiple receptive fields, and thus could be an important neuron of multiple output elements.

%Essentially, for every important neuron in the output feature map, we would identify its receptive field and apply the same method 

This process is repeated \textit{backwards} from the last layer to the first layer, as shown in the right panel in~\Fig{fig:illustration}. The important neurons identified at layer $L_i$ are used to determine the important neurons at layer $L_{i-1}$.
%~\Algo{alg:backward_ex_code} shows the pseudo-code of the general algorithm.

%https://tex.stackexchange.com/questions/381104/algorithm2e-why-are-some-of-my-texts-are-italicized-and-some-are-not
%\SetNlSty{}{}{}
%\newcommand{\mykwsty}[1]{\textcolor{OliveGreen}{\textbf{#1}}}
%\SetKwSty{mykwsty}
%
%\begin{algorithm2e}[t]
%\SetAlgoLined
%\SetKwProg{Fn}{function}{}{}
%\KwIn{$ImptN_{l+1}$: Important neurons at layer $l+1$;\newline$\theta$: cumulative threshold.}
%\KwResult{$ImptN_l$: Important neurons at layer $l$}
%
%%\Fn{\funcname{BwdExtImptNeurons}($ImptN_{l+1}$, $\theta$)}{
%$ImptN_l \leftarrow \emptyset$;\\
%\ForEach {$O \in ImptN_{l+1}$} {
%    $RankedPSum \leftarrow$ all the partial sums used to generate $O$ ranked in descending order;\\
%    $ImptN_l \leftarrow ImptN_l \bigcup$ \funcname{FindImptNeurons}($RankedPSum$, $\theta$, $O$);\\
%}
%\Return $ImptN_l$;
%%}
%
%\Fn{\funcname{FindImptNeurons}($RankedPSum$, $\theta$, $O$)}{
%$AccumSum \leftarrow 0$;\\
%$ImptN \leftarrow \emptyset$;\\
%\ForEach {$p \in RankedPSum$} {
%    $AccumSum$ += $p$;\\
%    $ImptN$.add(neuron corresponding to $p$);\\
%    \If{$AccumSum > \theta \times O$}{
%        break;
%    }
%}
%\Return $ImptN$;
%}
%
%\caption{Backward important neuron extraction.}
%\label{alg:backward_ex_code}
%%\vspace{-4pt}
%\end{algorithm2e}

\begin{figure}[t]
%\vspace{-5pt}
\centering
\subfloat[\small{AlexNet @ ImageNet.}]
{
  \includegraphics[trim=0 0 0 0, clip, width=0.48\columnwidth]{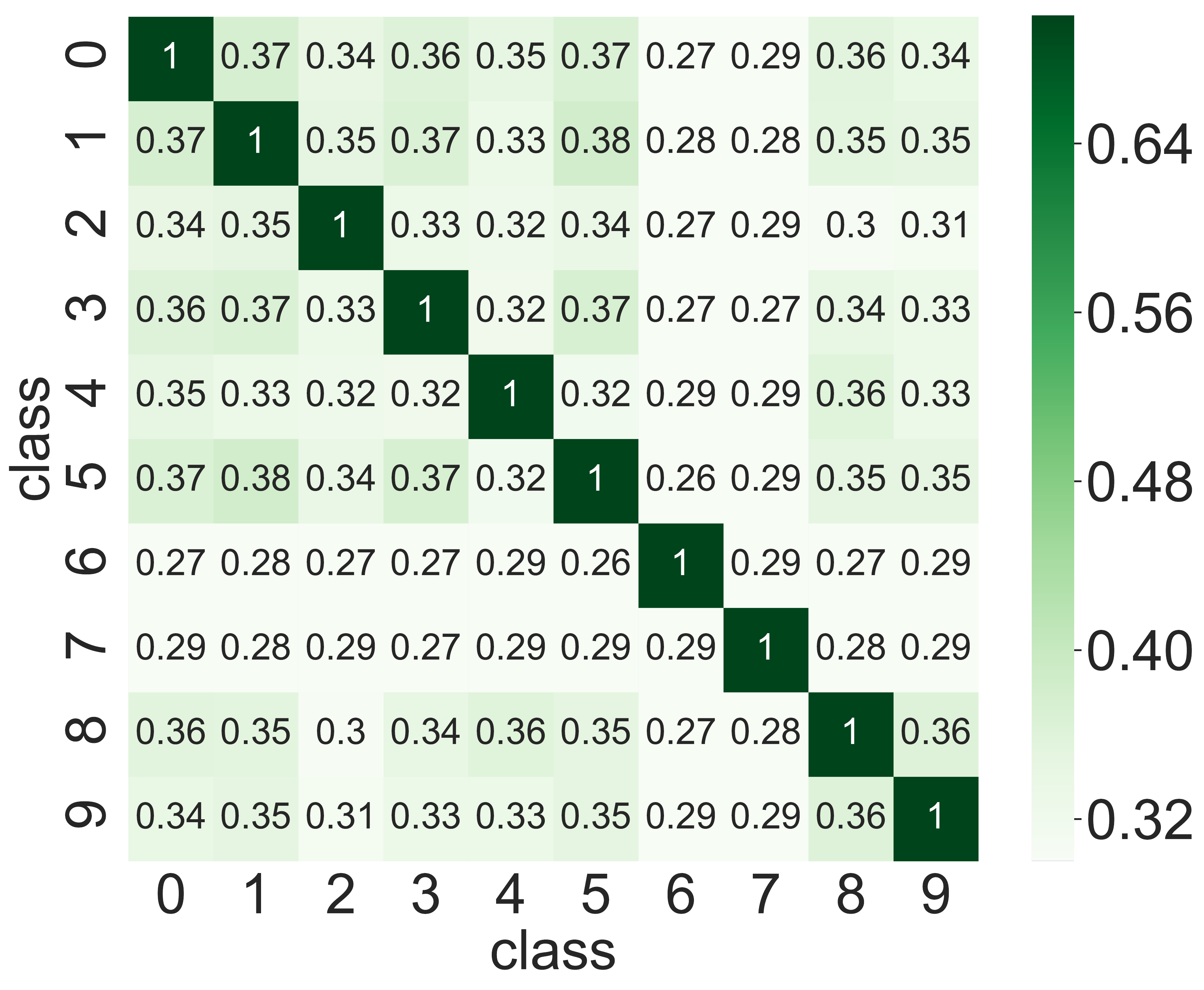}
  \label{fig:simi_alex}
}
%\hspace{3pt}
\subfloat[\small{ResNet18 @ CIFAR-10.}]
{
  \includegraphics[trim=0 0 0 0, clip, width=0.48\columnwidth]{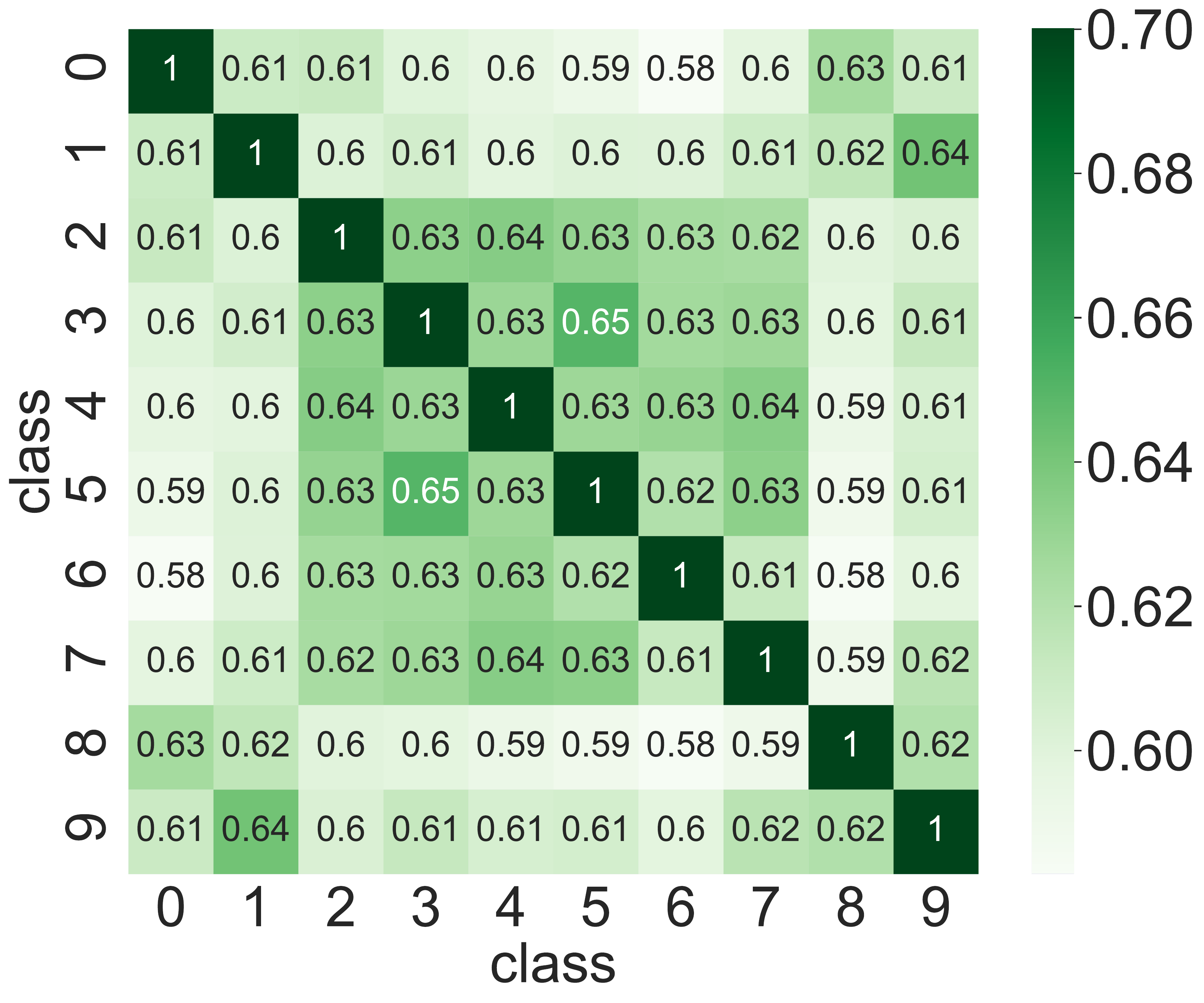}
  \label{fig:simi_res}
}
%\vspace{-5pt}
\caption{Class path similarity ($\theta$ = 0.5).}
\label{fig:memover}
%\vspace{-5pt}
\end{figure}
%that every input to a DNN uniquely activates a set of neurons from the input layer to the output layer. Those neurons that contribute significantly to the inference output are what we call \textit{important} neurons. Intuitively, not all the neurons are ``important''; some are either unactivated or only weakly activated after the non-linear activation function, and therefore contribute little to the inference output~\cite{??}. We will precisely define important neurons later, but a simple example of capturing the ``important'' neurons would be based on their magnitudes.~\Fig{fig:xx} shows a fully connected layer, where we define important neurons to be those have the highest three partial sums (absolutely values).

\paragraph{From Neurons to Paths} The collection of important neurons across all the layers under a given input constitutes an \textit{activation path} of that input, similar to how a sequence of basic blocks constitutes a path/trace in a program. We represent a path using a bitmask, where each bit $m_{i, j}$ indicates whether the neuron (input feature map element) at layer $i$ position $j$ is an important neuron.

From individual activation paths, we introduce the concept of a \textit{class path} for a class $c$, which aggregates (bitwise OR) the activation paths of different inputs that are correctly predicted as class $c$. That is: $\mathrm{P}_c = \bigcup_{x\in \bar{x}_c}\mathrm{P}(x)$, where $\mathrm{P}(x)$ denotes the activation path of input $x$, $\bar{x}_c$ denotes the set of all the correctly predicted inputs of class $c$, $\bigcup$ denotes bitwise OR, and $\mathrm{P}_c$ denotes the class path of class $c$. 
We observe that $\mathrm{P}_c$ starts to saturate around 100 images and including more images from the training dataset does not result all bits being 1. We do not manually stop filling the bits.

Critically, class paths are significantly different from each other.~\Fig{fig:simi_alex} shows the path similarity in AlexNet~\cite{alexnet} across 10 randomly-sampled classes from ImageNet~\cite{imagenet}.~\Fig{fig:simi_res} shows the path similarity in ResNet18~\cite{resnet} across the 10 classes in CIFAR-10~\cite{cifar}. All the results are obtained on the training set. The average inter-class path similarity is only 36.2\% (max 38.2\%, 90-percentile 36.6\%) for AlexNet on ImageNet and 61.2\% (max 65.1\%, 90-percentile 63.4\%) for ResNet18 on CIFAR-10, suggesting that class paths are distinctive. In an attempt to normalize the dataset, we also perform the same experiment on ResNet50 on ImageNet. The average inter-class path similarity is 37.6\% (max 40.9\%, 90-percentile 39.1\%), similar to those of AlexNet on ImageNet.

The class path similarity is much higher in CIFAR-10 than in ImageNet. This is because ImageNet has 1,000 classes that cover a wide range of objects and CIFAR-10 has only 10 classes, which are similar to each other (e.g., cat vs. dog). The randomly picked 10 classes in ImageNet are more likely to be different from each other than the 10 classes in CIFAR-10. Across all the 1,000 classes in ImageNet, the maximum inter-class path similarity is still only 0.44, suggesting that our random sampling of ImageNet is representative.

\subsection{Detection Framework and Cost Analysis}
\label{sec:sw:cost}

We leverage the clear distinction across different class paths to detect adversarial inputs. If an input $x$ is predicted as class $c$ while its activation path $\mathrm{P}(x)$ does not resemble the class path $\mathrm{P}_c$, we hypothesize that the input is an adversary.

\paragraph{Framework} \Fig{fig:algo} shows an overview of the algorithm framework, which requires static-dynamic collaboration. The static component profiles the training data to extract activation paths $\mathrm{P}(x)$ for each correctly predicted sample $x$, and generates the class path $\mathrm{P}_c$ for each class $c$ as described before. The class paths are stored offline and reused over time. Critically, our profiling method can easily integrate new training samples, whose activation paths would simply be aggregated (\texttt{OR}-ed) with the existing class paths without having to re-generate the entire class paths from scratch.
%, by taking the union of the activation paths of all correctly predicted training samples from class $c$.  That is: $\mathrm{P}_c = \bigcup_{x\in \bar{x}_c}\mathrm{P}(x)$, where $\bar{x}_c$ denotes the set of all the correctly predicted inputs of class $c$.

At inference-time, the dynamic component extracts the path for a given input. Note that activation paths are extracted only after the entire DNN inference finishes, because the identification of important neurons starts from the predicted class in the last layer and propagates backward. We will show other variants in~\Sect{sec:sw:algo} that relax this restriction.

%With the dynamically extracted activation path for an input $x$, a classification module then classifies the input as either adversarial or benign based on $\mathrm{P}(x)$ and the class paths constructed offline. Generally, the classifier compares the similarity between $\mathrm{P}(x)$ and $\mathrm{P}_c$, where $c$ is the predicted class of $x$. We discuss implementation details in \Sect{sec:arch:ctrl}.

Given the activation path $\mathrm{P}(x)$ of an input $x$ and the canary class path $\mathrm{P}_c$, where $c$ is the predicted class of $x$, a classification module then decides whether $x$ is an adversary or not based on the similarity between $\mathrm{P}(x)$ and $\mathrm{P}_c$. While a range of similarity metrics and algorithms could be used, we propose a lightweight algorithm that is extremely efficient to compute while providing high accuracy. Specifically, we first estimate the similarity $\mathrm{S}$ between $\mathrm{P}(x)$ and $\mathrm{P}_c$: $\mathrm{S} = \|\mathrm{P}(x)~\&~\mathrm{P}_c\|_1 / \|\mathrm{P}(x)\|_1$, where $\|P\|_1$ denotes the number of 1s in the vector $P$, and $\&$ denotes bitwise \texttt{AND}. $\mathrm{S}$ is fed into a learned classifier, for which we use the lightweight random forest method~\cite{liaw2002classification}, for the final classification. The classification module is lightweight, contributing to less than 0.1\% of the total detection cost.
%\footnote{We evaluated other definitions of $\mathrm{P}_c$, such as tracking the activation counts for each neuron. The accuracy generally became lower.}
%predominately (over 99.9\%) introduced by the path extraction process; the final classification step involves very lightweight computation.
%We call this similarity metric "Intersection over Canary", inspired by the classic intersection over union (IOU) metric~\cite{Everingham15}.

%This algorithm is extremely light; it consumes about 2,000 operations on AlexNet, five orders of magnitude lower than the inference, and could execute on an MCU in $\mu s$.

\no{\begin{figure}[b]
\vspace{-10pt}
\centering
\subfloat[\small{Memory overhead on different networks (constant w.r.t. $\theta$).}]
{
  \includegraphics[trim=0 0 0 0, clip, width=0.46\columnwidth]{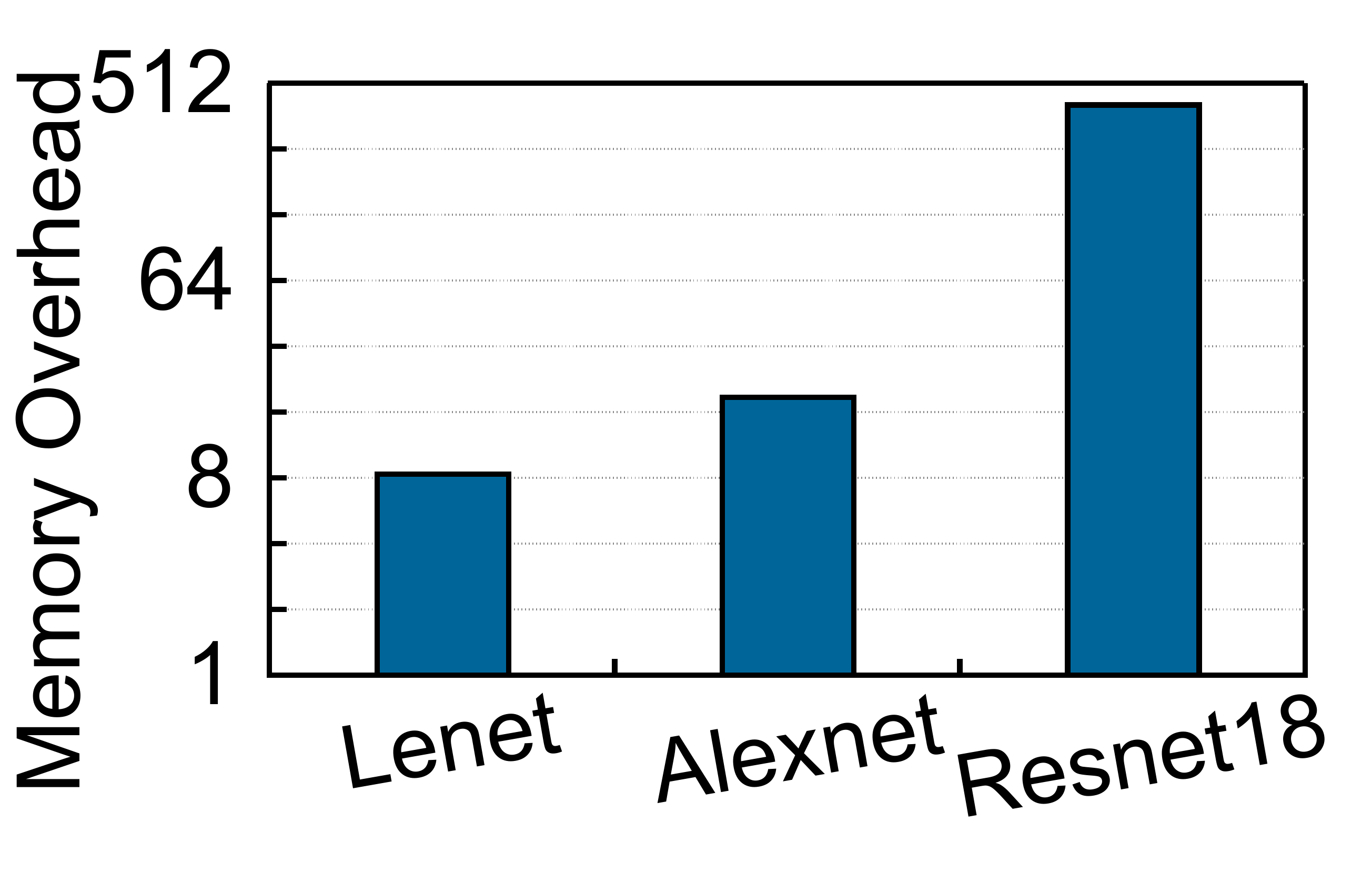}
  \label{fig:memover}
}
\hspace{3pt}
\subfloat[\small{Operation overhead on AlexNet under different $\theta$s.}]
{
  \includegraphics[trim=0 0 0 0, clip, width=0.46\columnwidth]{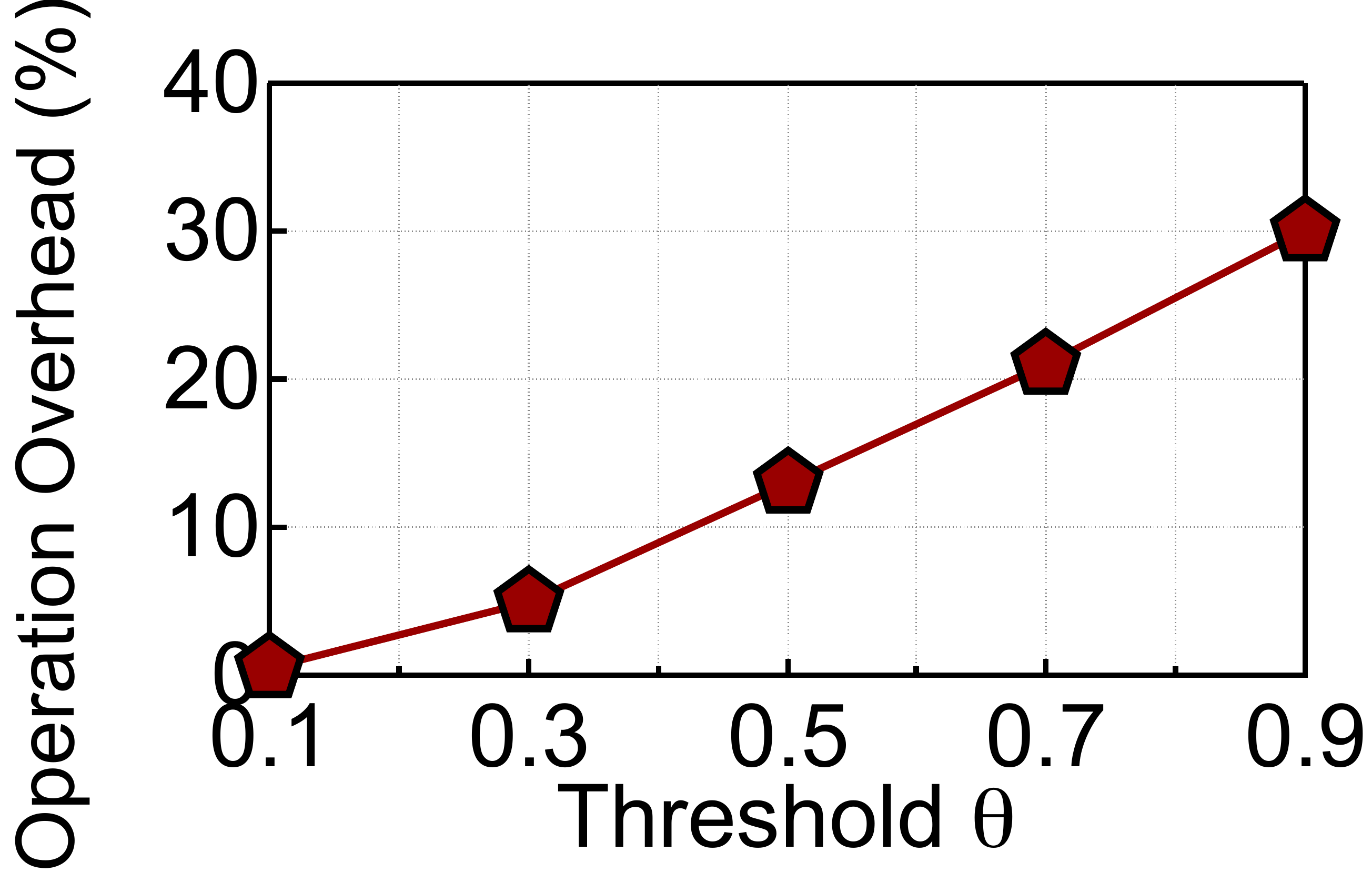}
  \label{fig:macover}
}
\vspace{-5pt}
\caption{Memory and operation overhead of adversarial detection over usual DNN inference without detection.}
\label{fig:over}
%\vspace{-10pt}
\end{figure}}

\paragraph{Cost Analysis} The algorithm described above is able to achieve accuracy higher than state-of-the-art methods~(see \Sect{sec:eval}). However, runtime extraction of activation paths also introduces significant memory and compute costs.
% The cost is predominately (over 99.9\%) introduced by the path extraction process; the final classification step involves very lightweight computation.

%The partial sums are used to identify important neurons, which are only known once the inference finishes. 
The memory cost is significant because every single partial sum generated during inference must be stored in the memory before the path extraction process begins.\no{~\Fig{fig:memover} shows the amount of data that is generated by the detection algorithm normalized to the amount of data required by usual DNN inference (i.e., model weights, input, and output feature maps) across different networks.} The detection algorithm introduces 9 $\times$ to 420 $\times$ memory overhead, which is a lower bound of the actual memory traffic overhead in real systems because the massive partial sums will not be buffered completely on-chip. Storing partial sums will also stall the computing units and increase latency.
%We will introduce algorithmic~(\Sect{sec:sw:algo}) and compiler optimizations~(\Sect{sec:interface:compiler}) to mitigate the memory overhead.

%; multiple accesses to one partial sum are likely needed in reality. Nevertheless, 
% as the network becomes deeper (e.g., 5-layer LeNet vs, 18-layer ResNet18), the memory overhead grows with it because the important neurons will increase as backward extraction process goes deeper.

Path extraction also introduces compute overhead due to sorting and accumulating partial sums. Using AlexNet as an example,\no{~\Fig{fig:macover} shows the number of operations introduced by path extraction ($y$-axis) normalized to inference while varying the $\theta$ value from 0.1 to 0.9.} at $\theta=0.9$, the compute overhead could be as high as 30\%. At first glance, it might be surprising that the compute overhead is ``only'' 30\%. Further investigations show that percentage of important neurons in a network is generally below 5\% even with $\theta=0.9$. Thus, the expensive sorting and accumulation operations are applied to only a small portion of partial sums. Note that the compute cost shown here leads to much higher latency overhead in reality because, while inference is massively parallel, sorting and accumulating are much less so. A pure software implementation of the detection algorithm introduces 15.4$\times$ and 50.7$\times$ overhead over inference on AlexNet and ResNet50, respectively.

\no{\begin{table*}[t]
\vspace{-3pt}
\Huge
\centering
\caption{Programming APIs \proj provides to allow conveniently expressing a wide range of detection algorithms.}
\renewcommand*{\arraystretch}{1.1}
\renewcommand*{\tabcolsep}{20pt}
\resizebox{2\columnwidth}{!}
{
  \begin{tabular}{ll}
  \toprule[0.15em]
  \textbf{API} & \textbf{Semantics} \\
  \midrule[0.05em]
  \textsc{Inference}($model$, $input$)    &  \llap{\textbullet} DNN inference on $model$ using $input$ as the input.        \\
  \textsc{ExtractImptNeurons}($dir$, $thd$, $para$, $l$)    &  \specialcell{\llap{\textbullet} Extract important neurons at layer $l$. Binary $dir$ specifies the extraction direction;\\~Binary $thd$ specifies the threshold type; $para$ specifies the exactly threshold value.}         \\
  \textsc{LoadClassPath}($class$)    & \llap{\textbullet} Load the class path from storage for $class$.      \\
  \textsc{Classify}($classPath$, $inputPath$) & \llap{\textbullet} Classify $inputPath$ as either adversarial or benign against the $classPath$. \\
  \textsc{GenMask}($neurons$) & \specialcell{\llap{\textbullet} Generate the bitmask from important neurons $neurons$.}\\
  %\midrule[0.05em]
  %\multirow{2}{*}{Custom Data Type} & \textsc{enum Dir \{FWD = 0, BWD = 1\}};    &  Extraction direction: forward (\textsc{FWD}) vs. backward (\textsc{BWD}).        \\
  %~ & \textsc{enum ThdType \{ABS = 0, CUM = 1\}};    &  Threshold type: absolute threshold (\textsc{ABS}) vs. cumulative threshold (\textsc{CUM}).       \\
  \bottomrule[0.15em]
  \end{tabular}
}
\label{tab:lang}
%\vspace{4pt}
\end{table*}

%The mask has the same dimension\\~as the input feature map, where 1 indicates the corresponding neuron is an important\\~neuron, and vice versa.
}

\subsection{Algorithmic Knobs and Variants}
\label{sec:sw:algo}

To trade little accuracy loss for significant efficiency gains, we introduce three algorithmic knobs that control how activation paths are extracted, which dominates the runtime performance/energy overhead. The result is a set of algorithm variants that follow the same algorithm framework described in~\Fig{fig:algo}, but that differ in how the paths are extracted.

%The general framework leaves many design decisions to be made, including 1) how to define important neuron, 2) how is an activation path constructed from important neurons, 3) how are individual paths aggregated as a class path, and 4) how is an input classified as adversarial or benign based on its dynamically extract path? This section describes these design decisions.

%Critically, these design decisions could affect the trade-off between detection accuracy and efficiency. The goal of \proj is to provide a general framework that allows programmers to flexibly tune these knobs. Building on top of the general framework, we will develop a few concrete detection algorithms.

\knob{Hiding Detection Cost: Extraction Direction}

The cost introduced by the basic detection algorithm directly increases the inference latency because path extraction and inference must be serialized. We identify a key algorithmic knob that provides the opportunity to hide the compute cost of detection by overlapping detection with inference.

%\begin{algorithm2e}[t]
%\SetAlgoLined
%\SetKwProg{Fn}{function}{}{}
%\KwIn{$\theta$: cumulative threshold.}
%\KwResult{$ImptN_l$: Important neurons at layer $l$}
%
%%\Fn{\textsc{\textbf{\textcolor{RoyalBlue}{FwdExtImptNeurons}}}($\theta$)}{
%$ImptN \leftarrow$ top-K ranked neurons in the output feature map of Layer $l$;\\
%\ForEach {$O \in ImptN$} {
%    $RankedPSum \leftarrow$ all the partial sums used to generate $O$ ranked in descending order;\\
%    $ImptN_l \leftarrow ImptN_l \bigcup$ \funcname{FindImptNeurons}($RankedPSum$, $\theta$, $O$);\\
%}
%\Return $ImptN_l$;
%%}
%\caption{Forward extraction of important neurons.}
%\label{alg:forward_ex_code}
%%\vspace{-4pt}
%\end{algorithm2e}

%~\Algo{alg:forward_ex_code} shows the pseudo-code of the forward extraction algorithm. 
The key to the new algorithm is to extract important neurons in a \textit{forward} rather than a backward manner. Recall that in the original backward extraction process, we use the important neurons in layer $L_i$'s output (which is equivalent to layer $L_{i+1}$'s input) to identify the important neurons in layer $L_i$'s input. In our new forward extraction process, as soon as layer $L_{i}$ finishes inference we first determine the important neurons in its output by simply ranking output neurons according to their numerical values and selecting the largest neurons, instead of waiting until after the extraction of layer $L_{i+1}$. In this way, the extraction of important neurons at layer $L_{i}$ and the inference of layer $L_{i+1}$ can be overlapped.

%to identify important neurons at layer $L_{i}$ we must know the important neurons in $L_{i}$'s output (equivalent to $L_{i+1}$'s input), which are only obtained after extracting the important neurons in $L_{i+1}$.

\no{Forward extraction does not guarantee the exact same important neurons as those produced by backward extraction. Instead of deriving important neurons by identifying those neurons that contribute most significantly to the inference output, forward extraction infers the important neurons from those that respond the most actively to the input. The forward method generally leads to slightly lower accuracy than the backward method, but allows for significant performance improvement that enables real-time adversary detection~(\Sect{sec:eval:acc}).}
%because important neurons generated by the former does not directly correlate with the inference output

\knob{Reducing Detection Cost: Thresholding Mechanism}

The forward extraction process hides the extraction behind inference, but does not reduce the detection cost, which could significantly increase the energy overhead.

%The cost in extraction important neurons is dominated by ranking neurons according to their partial sums and accumulating the partial sums until the relatively threshold $\theta$ is reached, as shown in~\Fig{fig:backward_ex}.

To reduce the detection cost, we propose to extract important neurons using absolute thresholds rather than cumulative thresholds. Whenever a partial sum is generated during inference it is compared against an absolute threshold $\phi$. A single-bit mask is stored to the memory based on the comparison result. Later during path extraction, the masks (as opposed to partial sums) are loaded to determine important neurons. Thresholding can be specified at each layer, and can be applied to both extraction directions.
%is orthogonal to the extraction direction, and 

Using absolute thresholds significantly reduces both the compute and memory costs~(\Sect{sec:eval:lat}), because comparing partial sums against a threshold is much cheaper than sorting and accumulating them, and writing single-bit masks rather than partial sums significantly reduces the memory accesses.
%This optimization reduces the compute overhead by 2.2$\times$ and memory overhead by 11.4$\times$.

%each input neuron whose absolute partial sum is greater than an absolute threshold $\phi$ will be regarded as an important neuron without having to sort all the partial sums. 

%~\Algo{alg:forward_ex_abs_code} shows the pseudo-code of this process on top of the forward extraction algorithm, but the same idea can be integrated with backward extraction as well.

%\begin{algorithm2e}[t]
%\SetAlgoLined
%\SetKwProg{Fn}{function}{}{}
%\KwIn{$\phi$: absolute threshold.}
%\KwResult{$ImptN_l$: Important neurons at layer $l$}
%\Fn{\textsc{\textbf{\textcolor{RoyalBlue}{FwdExtImptNeuronsAbs}}}($\phi$)}{
%$ImptN \leftarrow$ top-K ranked neurons in the output feature map of Layer $l$;\\
%\ForEach {$O \in ImptN$} {
%    $PSum \leftarrow$ partial sums used to generate $O$;\\
%    \ForEach {$p \in PSum$} {
%        \If{$p > \phi$}{
%            $ImptN_l$.add(neuron corresponding to $p$);\\
%        }
%    }
%}
%\Return $ImptN_l$;
%}
%\caption{Forward extraction algorithm using absolute thresholds.}
%\label{alg:forward_ex_abs_code}
%%\vspace{-4pt}
%\end{algorithm2e}

\no{Theoretically, the absolute thresholds ($\phi$) can be carefully chosen in a way such that the extraction result matches that generated from using cumulative thresholds ($\theta$). The trade-off is that cumulative thresholds could be more intuitive from a programmer's perspective as $\theta$ is indicative of important neuron coverage. In contrast, $\phi$s' values are less intuitive, and often have to be tuned for each DNN model. For instance, $\phi = 0.1$ could be a very low threshold for layers where partial sums are generally above 1, but is a high threshold for layers where partial sums are mostly below 0.05.}

\knob{Reducing Detection Cost: Selective Extraction}

%\paragraph{Algorithmic Variants} In summary, our algorithmic framework provides two design knobs, \textit{extraction direction} (backward vs. forward) and the \textit{thresholding mechanism} (cumulative vs. absolute). Generally, backward- and cumulative threshold-based methods have higher accuracy and higher cost; forward- and absolute threshold-based methods have the opposite trade-off.

An orthogonal way to reduce the cost is to skip important neurons from certain layers altogether. In many networks, later layers have a more significant impact on the inference output than earlier layers~\cite{raghu2017svcca}. Thus, one could extract important neurons from just the last a few layers to further reduce the cost~(\Sect{sec:eval:et}). When combined with forward extraction, this is equivalent to starting extraction later (``late-start''); when combined with backward extraction, this is equivalent to terminating extraction earlier (``early-termination''). This knob specifies the start/termination layer.
% during extraction.

\paragraph{Summary} The \proj framework provides three different knobs to explore the accuracy-efficiency trade-off. While the \textit{extraction direction} applies to the entire network and hides the detection cost behind the inference cost, the \textit{thresholding mechanism} and the \textit{extracted layer} are specified at the layer level to reduce the detection cost.

\no{They impact the workload that the hardware underneath would see:}

\no{\begin{itemize}
    \setlength\itemsep{-2pt}
	\item Extraction direction: forward extraction decouples inference and extraction, which our software pipelining optimization exploits.
	\item Thresholding mechanism: absolute thresholding saves compute operations compared to cumulative thresholding. Assuming a 3x3x3 receptive field, using absolute threshold saves 78\% compute operations compared to using cumulative threshold on average.
	\item Selective extraction: on AlexNet, extracting only the second half of the network (layer 5-8) saves 92\% compute operations compared to extracting all layers.
\end{itemize}}

%The two design knobs described above are orthogonal. Combined, they provide four different extraction methods. Critically, the four methods could be individually applied to \textit{each DNN layer}. In fact, it is even possible to skip extracting important neurons from certain layers altogether. The final knob governs how extraction methods are used in each layer.

%Thus, one could use more accurate but heavier extraction methods on later layers while using lightweight methods on early layers, or omitting important neurons in early layers altogether.
%Overall, the \proj algorithm framework (\Fig{fig:algo}) provides a large design space.
%In total, our framework provides a total of $N^5$ different ways of implementing the detection algorithm, where $N$ is the number of DNN layers.

%The fine-grained is motivated by the observation that different layers contribute differently to the inference output. For instance, 

\begin{figure}[t]
  %\vspace{-5pt}
  \centering
  \includegraphics[trim=0 0 0 0, clip, width=\columnwidth]{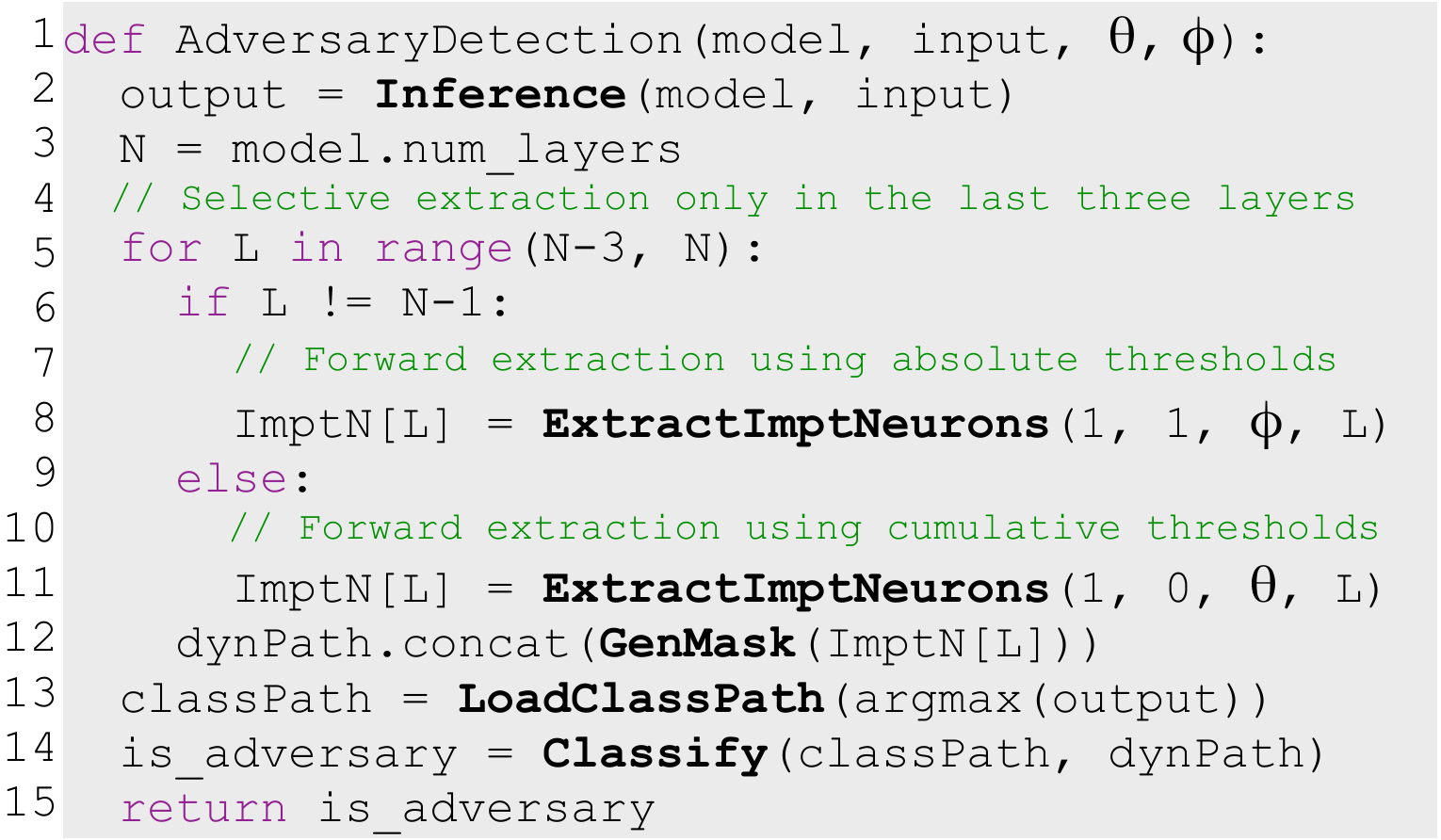}
  \caption{An adversarial detection algorithm expressed using the programming interface.}
  \label{fig:ex}
  %\vspace{-10pt}
\end{figure}

\begin{table*}[]
\caption{Summary of \proj instructions. Operands in the first three instruction classes are registers to simplify encoding.}
\renewcommand*{\arraystretch}{1.1}
\renewcommand*{\tabcolsep}{4pt}
\resizebox{2\columnwidth}{!}
{
\label{tab:isa}
\begin{tabular}{llllllll}
\toprule[0.15em]
\textbf{Class}                                                    & \textbf{Name}       & \textbf{23-20} & \textbf{19-16}                           & \textbf{15-12}                   & \textbf{11-8}                          & \textbf{7-4}                  & \textbf{3-0}    \\
\midrule[0.05em]
\multirow{3}{*}{Inference}                               & \inst{inf}        & 0000  & Input addr.                   & Weight addr.          & Output addr.                & \multicolumn{2}{l}{Unused}    \\
~ & \inst{infsp}      & 0001  & Input addr.                   & Weight addr.          & Output addr.                & First partial sum addr. & Unused \\
~ & \inst{csps}       & 0010    & Output neuron ID                & Layer ID                & First partial sum addr.          & \multicolumn{2}{l}{Unused}    \\
\midrule[0.05em]

\multirow{5}{*}{Path Construction}                       & \inst{sort}       &  0011     & Unsorted seq. start addr. & Seq. length         & Sorted seq. start addr. & \multicolumn{2}{l}{Unused}    \\
~ & \inst{acum}       & 0100      & Input addr.                   & Output addr.          & Cumulative threshold          & \multicolumn{2}{l}{Unused}    \\
%~ & \inst{compare}    & 0101      & Input addr.                   & Output addr.          & Absolute threshold            & \multicolumn{2}{l}{Unused}    \\
~ & \inst{genmasks}    & 0101      & Input addr.                   & Output addr.          & \multicolumn{3}{l}{Unused}    \\
%\midrule[0.05em]
%\multicolumn{1}{c}{\multirow{2}{*}{Address Calculation}} & \inst{findneuron} &       & Layer ID                        & Neuron position         & Target neuron addr.         & \multicolumn{2}{l}{Unused}    \\
~ & \inst{findneuron} &  0110     & Layer ID                        & Neuron position         & Target neuron addr.         & \multicolumn{2}{l}{Unused}    \\
\multicolumn{1}{c}{}                                     & \inst{findrf}     &  0111     & Neuron addr.                  & Receptive field addr. & \multicolumn{2}{l}{Unused}                           &        \\
\midrule[0.05em]

Classification                                                 & \inst{cls}        &  1000     & Class path addr.              & Activation path addr. & Result                        & Unused               &        \\
\midrule[0.05em]
Others                                                   & \multicolumn{6}{c}{Omitted for simplicity (\inst{mov}, \inst{dec}, \inst{jne}, etc.)}                                                                                                           &\\
\bottomrule[0.15em]
\end{tabular}
}
\end{table*}

\subsection{Programming Interface}
\label{sec:sw:lang}

\proj provides a (Python-based) programming interface that allows programmers to express a range of different algorithmic design knobs described above.\no{~\Tbl{tab:lang} shows the set of APIs that \proj provides.} Our programming interface is designed with two principles in mind, which we will explain using an actual detection algorithm expressed using the programming interface shown in~\Fig{fig:ex}.

%\SetNlSty{}{}{}
%\newcommand\mycommfont[1]{\footnotesize\ttfamily\textcolor{black}{#1}}
%\SetCommentSty{mycommfont}
%\SetKw{kwRange}{range}
%
%\begin{algorithm2e}[h!]
%\SetAlgoLined
%\SetKwProg{Fn}{function}{}{}
%\KwIn{$model$: the DNN model; $in$: model input; $\theta$: cumulative threshold; $\phi$: absolute threshold.}
%\KwResult{$is\_adversary$}
%
%\Fn{\funcname{AdversaryDetection}($model$, $in$, $\theta$, $\phi$)}{
%
%%\ForEach{layer $l \leftarrow 1$ \KwTo $N$}{
%%	output[$l$] = \funcname{Inference}($l$, $in$);
%%}
%output = \funcname{Inference}($model$, $in$);\\
%$N = model.num\_layers$;\\
%
%\For {layer $l$ in \kwRange{[N-2, N]}} { \tcp{selective important neuron extraction}
%    \If{$l$ is not $N$} {
%        ImptN[$l$] = \funcname{ExtractImptNeurons}(\texttt{FWD}, \texttt{ABS}, $\phi$, $l$);\\
%    }
%    \Else {
%        ImptN[$l$] = \funcname{ExtractImptNeurons}(\texttt{FWD}, \texttt{CUM}, $\theta$, $l$);\\
%    }
%    dynPath.concat(\funcname{GenMask}(ImptN[$l$]));\\
%}
%%PredictedClass $c$ = output[$N$];\\
%classPath = \funcname{LoadClassPath}(output[$N$]);\\
%is\_adversary = \funcname{Classify}(classPath, dynPath);\\
%\Return is\_adversary;
%}
%\caption{A adversarial detection algorithm expressed using the programming interface.}
%\label{alg:detalgo}
%%\vspace{-4pt}
%\end{algorithm2e}

\paragraph{Decoupled Inference/Detection} The \proj programming interface decouples inference with detection, which allows programmers to focus on expressing the functionalities of the detection algorithm while leaving optimizations to the compiler and runtime. For instance, while the inference code (Line 2) and the path extraction code (Line 3--15) are expressed sequentially in the program, our compiler will understand that the program uses the forward extraction algorithm (Line 8 and 11), and thus will automatically pipeline inference with important neuron extraction across layers (see~\Sect{sec:interface:compiler}).

\paragraph{Per-Layer Extraction Granularity} Our programming interface provides the flexibility to specify the important neuron extraction method \textit{for each layer} to leverage the three knobs described above to explore the efficiency-accuracy trade-off space. We will demonstrate its effectiveness in~\Sect{sec:eval:et}.

For instance in~\Fig{fig:ex}, the programmer selectively extracts important neurons only for the last three layers (Line 5). In addition, only the last layer uses the cumulative threshold to extract important neurons (Line 11), which is more accurate but requires more computations than using absolute thresholds, which is the method used by the other two layers (Line 8). Note that we do not allow backward extraction and forward extraction to be combined in one network to avoid discrepancies in the layer where they join.

\section{ISA and Compiler Optimizations}
\label{sec:interface}

This section describes how \proj efficiently maps detection algorithms expressed in the high-level programming interface to the hardware architecture. To that end, we first introduce the software-hardware interface, i.e., the Instruction Set Architecture (ISA)~(\Sect{sec:interface:isa}), followed by the compiler optimizations~(\Sect{sec:interface:compiler}).

%If first describes the high level programming model that is programmable to the users (\Sect{sec:prog}). We further introduces the compilation process in \Sect{sec:comp} that transfers the high level API into low level instruction set architectures (ISA) shown in \Sect{sec:isa} that is designed to communicate with the hardware.

\subsection{Instruction Set Architecture}
\label{sec:interface:isa}

\proj provides a custom CISC-like ISA to allow efficient mapping from high-level detection algorithms to the hardware architecture. The design principles of the ISA are two-fold. First, it abstracts away hardware implementation details; the semantics are closer to high-level DNN programmers, and the instructions will be decomposed by micro-instructions controlled by an FSM. Second, it exposes opportunities for compiler and hardware to exploit parallelisms.

%The design principle is to be accessible to programmers while relying on compiler and hardware for optimizations. Both instructions you mentioned (findneuron and findrf) carry high-level detection semantics, and will be broken down into microinstructions (i.e., CISC-like).

%The ISA defines two key components: a set of instructions and the runtime memory layout of various data structures.

%We design an flexible and efficient instruction set architecture (ISA) enabling the programmable hardware support for adversarial detection algorithm. The principles we apply designing the ISA is to construct a simple and uniform customized ISA that could separate the hardware with complex algorithms. We show an overview of the ISA in \Tbl{tab:isa}.

The \proj ISA contains four types of instructions: \textit{Inference}, \textit{Path Construction}, \textit{Classification}, and \textit{Others}. They are high-level instructions in the CISC style that perform complex operations. We use a 24-bit fixed length encoding, and provide 16 general-purpose registers.~\Tbl{tab:isa} summarizes the instructions. We highlight key design decisions.
%One can think of each instruction as a command sent to the hardware. Each command operates a state machine that 

%For the control and data transfer logic, we follow the design fashion of corresponding MIPS instructions. These high-level instructions 

\begin{itemize}
	\item \underline{Inference} These instructions dictate the inference process. In addition to support usual inferences (\inst{inf}), \proj also provides an instruction that stores the partial sums to memory (\inst{infsp}) during inference for backward extraction. Each inference instruction operates on one layer to match the per-layer extraction semantics in the high-level programming interface. Finally, the ISA also provides a special instruction that calculates and stores all the partial sums given an output feature map element (\inst{csps}), which will be used by the compiler for memory optimizations.
	\item \underline{Path Construction} This class of instructions is used to construct activation path dynamically at runtime for any given input. To construct path, the ISA provides instructions to identify important neurons (sorting \inst{sort}, accumulate \inst{acum}) and to generate the masks from the identified important neurons to form an activation path (\inst{genmasks}). There are also instructions to calculate neuron addresses, which are convenient in finding the start address of a receptive field for a given neuron (\inst{findrf}) and finding a given neuron given its position in the network (\inst{findneuron}).%While these are simple arithmetics that could be implemented using a sequence of primitive arithmetic operations, these high-level instructions ease the compiler code generation.
	\item \underline{Classification} The classification instruction (\inst{cls}) is used to classify an input as either adversarial or benign.
    \item \underline{Others} The ISA provides a set of control-flow instructions (e.g., and \inst{jne}), arithmetic instructions (e.g., \inst{dec}), and scalar data movement instructions (e.g., \inst{mov}).
	%\item \underline{Control} The ISA provides a set of control-flow instructions (\inst{jmp}, \inst{beq}, \inst{bne}, etc.), mainly used to implement loops whose iterations have to be serialized.
	%\item \underline{Numeral Calculation} The ISA also provides a set of numeral calculation (\inst{add}, \inst{sub}, etc.), mainly used in bounds calculation when implementing bounds.
\end{itemize}

\paragraph{Example}~\Lst{lst:example} shows a sample code that uses cumulative thresholds to extract important neurons. Through a loop, it iteratively finds a receptive field (\inst{findrf}), sorts partial sums in the receptive field (\inst{sort}), and uses the sorted partial sums to identify important neurons whose cumulative partial sums exceed the threshold (\inst{acum}).

\begin{lstlisting}[label={lst:example}, caption={Generating important neurons using a cumulative threshold. \texttt{.set} is a directive setting compiler-calculated constants. \texttt{[code]} indicates code omitted for simplicity.}, captionpos=b, frame=single]
.set rfsize 0x200
.set thrd 0x08
mov r3, rfsize
mov r5, thrd
<start>
[update r7&r2 for next output neuron]
findneuron r2, r7, r4
mul r5, (r4)
findrf r4, r1
sort r1, r3, r6
acum r6, r1, r5
dec r11 
jne <start>
\end{lstlisting}

It highlights an important design decision of the \proj ISA: all the detection related instructions use register operands. This design simplifies instruction encoding with little performance impact. For instance, the \inst{findrf} instruction requires the receptive field size as an operand, which can be statically calculated by the compiler given the DNN model configurations. Since the receptive field size could be arbitrarily large and thus does not always fit in a reasonable, fixed-length encoding, a \inst{mov} instruction is used to move the statically calculated immediate value to a register (\texttt{r3}), which is later used in the \inst{sort} instruction. While a more complex instruction encoding that limits the range of immediate operands could eliminate this \inst{mov} instruction, the performance overhead introduced by this \inst{mov} instruction is negligible compared to the heavy-duty \inst{sort} and \inst{acum} instructions.

\subsection{Code Generation and Optimization}
\label{sec:interface:compiler}

%The compiler has two phases: 1) code generation, and 2) code optimization. The code generation phase maps each high-level language API to a set of corresponding instructions, and calculates effective addresses for memory-access instructions. The effective addresses can be statically computed because the size of each data structure is statically known given a model's parameters. We now focus on the optimizations that the compiler performs to maximize performance.

%\begin{figure}[t]
%\vspace{-5pt}
%\centering
%\subfloat[\small{Pipeline execution among neurons. N1 means neuron one; S1 means stage one.}]
%{
%  \includegraphics[trim=0 0 0 0, clip, width=0.38\columnwidth]{figs/pipe_neuron}
%  \label{fig:pipe_neu}
%}
%\hfill
%\subfloat[\small{Pipeline execution among layers. L1 means layer one; Inf means inference; Ext means extraction.}]
%{
%  \includegraphics[trim=0 0 0 0, clip, width=0.49\columnwidth]{figs/pipe_layer}
%  \label{fig:pipe_lay}
%}
%\vspace{-5pt}
%\caption{Execution model among neurons and layers.}
%\label{fig:exe_neu}
%\vspace{-5pt}
%\end{figure}

%\textbf{Effective Address Calculation} The compiler will calculate the effective address for each instruction in the process of compilation. The effective address including the starting address of the input, weight, output, intermediate data and the masks. All the memory address of each part is statically calculated during compilation by compiler and will not be changed in the runtime.

The compiler maximizes performance by exploiting unique parallelisms and redundancies inherent to the detection algorithms. This is achieved through statically scheduling instructions, which minimizes runtime overhead and hardware complexity. Static scheduling is possible because the compute and memory access behaviors of both DNN inference and detection are known at the compile time.

\paragraph{Layer-Level Pipelining} A key characteristic of algorithms that use the forward extraction method is that inference and extraction of different layers can be overlapped. While the high-level programming interface decouples inference (\textsc{Inference}) and extraction (\textsc{ExtractImptNeurons}), and expresses them sequentially, our compiler will reorder instructions to enable automatic pipelining at runtime, in a way similar to classic software-pipelining technique~\cite{allan1995software}.

\Fig{fig:compiler_opt_layer} shows an example. We use pseudo-code to remove unnecessary details. \texttt{<\inst{extraction} for j>} indicates the code block for extracting important neurons at layer \texttt{j}, and \inst{inf}\texttt(j) indicates inference at layer \texttt{j}. By simply reordering instructions, inference of layer \texttt{j+1} and extraction of layer \texttt{j}, which are independent, could be pipelined. At the hardware level, once \mbox{\inst{inf}}\texttt(j) is issued to execute on the DNN accelerator, \texttt{<\inst{extraction} for j>} could be issued and executed immediately on our hardware extension (\mbox{\Sect{sec:arch:acc}}).

Note that our software pipelining technique does not fully hide the instruction latency to guarantee that a new instruction can be dispatched every cycle. Both inference and the extraction code block take tens of millions of cycles. Fully hiding latencies requires expensive optimizations in classic compiler literature~\cite{triantafyllis2003compiler,hoste2008cole}. We find that our simple static instruction reordering is able to largely overlap inference with extraction, leading to very low performance overhead. A side effect of not fully hiding the instruction latencies is that our hardware would still have the logic to check dependencies and stall the pipeline if necessary. But the hardware remains in-order without the expensive out-of-order instruction scheduling logic.

%each layer's important neurons rely on their previous layer. However, for the methods with a forward execution direction, each layer's important neurons are independent. Therefore, the extraction process and inference can happen at the same time since they will not face structural hazard. We apply layer level pipeline execution and show the execution model in \Fig{fig:pipe_lay}. Whenever the layer finishes inference, the pattern extraction process for that layer starts while the next layer's inference can happen in parallel.

%(e.g., the different iterations of the \texttt{for} loop at Line 13 in~\Algo{alg:backward_ex_code})
\paragraph{Neuron-Level Pipelining} Similar to layer-level pipelining, our compiler will also automatically pipeline the extraction of different important neurons within a layer.~\Fig{fig:compiler_opt_neuron} shows an example, where cumulative thresholds are used. The two steps needed to extract important neurons, sorting all the partial sums (\inst{sort}) and accumulating the partial sums until the threshold is reached (\inst{acum}), have data dependencies. The compiler overlaps the extraction across different important neurons (iterations), improving hardware utilization and performance.

%in pattern extraction process is that there is no neuron level dependency, each neuron's weight-input mask generation process is independent. This enable the possibility of execute the pattern extraction among different neurons in a software pipeline way. 

%We show the execution model among neurons of using sorting on weight-input in \Fig{fig:pipe_neu}. Each neuron's mask generation process will have three stages: sort, merge and accumulate. During the sort stage , \proj will sort all the weight-input of the neuron into multiple fixed-length sorted sequences. The fixed-length sorted sequence will be merged into a large sorted sequence in merge stage and be accumulated in accumulation stage. After the sorting stage finishes for the first neuron, the second neuron's sorting could start and meanwhile the second neuron's merge stage can be executed. 

\begin{figure}[t]
\centering
\subfloat[Overlapping inference with extraction across layers in forward extraction. \texttt{L} is the total number of DNN layers.]
{
  \includegraphics[trim=0 0 0 0, clip, height=2in]{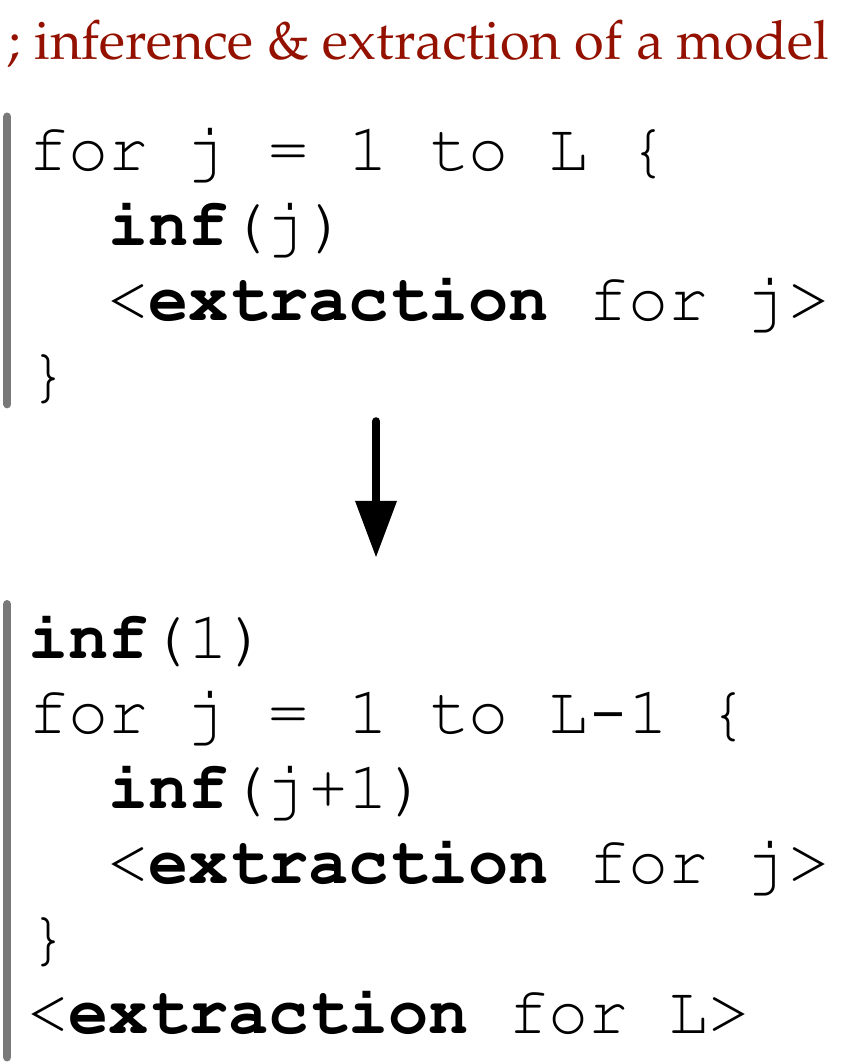}
  \label{fig:compiler_opt_layer}
}
\hfill
\subfloat[Neuron-level pipelining in important neuron extraction. \texttt{N} denotes the number of important neurons in the current layer's output.]
{
  \includegraphics[trim=0 0 0 0, clip, height=2in]{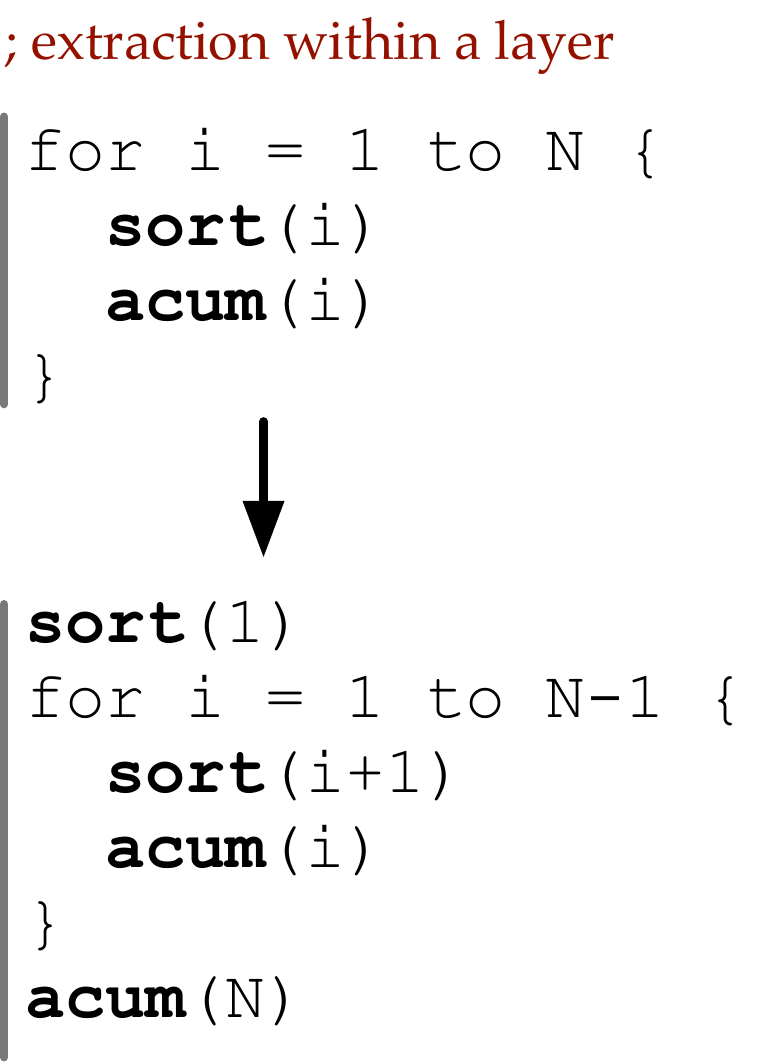}
  \label{fig:compiler_opt_neuron}
}
\vspace{-5pt}
\caption{Pseudo-code of instruction scheduling examples. The code in \protect\subref{fig:compiler_opt_neuron} is the extraction block simplified in \protect\subref{fig:compiler_opt_layer}.}
\label{fig:compiler_opt}
%\vspace{-15pt}
\end{figure}

\paragraph{Trading-off Compute for Memory} Algorithms that use cumulative thresholds have high memory cost because all the partial sums must be stored to memory (\Fig{fig:memover}). However, if a receptive field does not correspond to an important neuron in the output feature map, its partial sums will not be used later. We observe that fewer than 5\% of the partial sums stored are used later to extract important neurons.
%not all the partial sums will later be used. 

We propose to use redundant computation to reduce memory overhead. Instead of storing all the partial sums during inference, we re-compute the partial sums during the extraction process only for the receptive fields that are known to correspond to important neurons in the output feature map. The compiler implements this by generating \inst{csps} instructions to re-compute partial sums.
%Since this optimization increases compute overhead, we provide a compiler switch for programmers to control the trade-offs.

%The essence of the algorithms \proj support requires all the intermediate data during inference from DNN accelerators which violates the nature of state-of-the-art systolic array based DNN accelerator design. It will add multiple times of output data transfer under the situation that the SRAM bandwidth and DRAM bandwidth remains the same. The systolic array will have to stall to satisfy the bandwidth requirements, thus the systolic array utilization will be reduced and the inference latency will be increased. 

%To address the problem that the inference latency will be increased because of much higher write requirements, we utilize re-computation techniques during compilation process. We observe that the important output neuron has a low density among output feature map. Thus, instead of reading the intermediate data for all output neurons, we only read intermediate data for important output neurons. We achieve this by changing the order of instructions. After the regular inference finish, we sort or use threshold to identify the important output neuron and we re-compute these neurons to get the intermediate data and further generate the masks using these intermediate data.
\section{Architecture Support}
\label{sec:arch}

This section introduces the \proj hardware architecture. Following an overview~(\Sect{sec:arch:ov}), we describe the designs of major hardware components~(\Sect{sec:arch:acc} -- \Sect{sec:arch:ctrl}).

\subsection{Overview}
\label{sec:arch:ov}

Our architecture builds on top of a conventional DNN accelerator.~\Fig{fig:arch} provides an overview of the \proj architecture, which consists of an augmented DNN accelerator, a Path Constructor that builds the activation path for an input, and a Controller that dispatches instructions, runs state machines that control the hardware blocks, and executes the final classifier. An off-chip memory stores all the data structures that are needed for inference and detection. Both the DNN accelerator and the Path Constructor use double-buffered on-chip SRAMs to capture data reuse and to overlap DMA transfer with computation. The controller's SRAM stores the compiled detection program and activation/class paths for classification.

%All the components are connected to the DRAM through a common bus. The instructions generated by compiler will be running on microcontroller and microcontroller will issue command to other components. The DNN accelerator will be responsible for the inference and write all the weight-input out to the global buffer. The extractor will use the output and weight-input data from global buffer to generate mask for each layer. Finally the detector will be using the masks to generate a vector and this vector will be feeded into a random forest classifier for the final detection result.

\begin{figure}[t]
  %\vspace{-5pt}
  \centering
  \includegraphics[trim=0 0 0 0, clip, width=1\columnwidth]{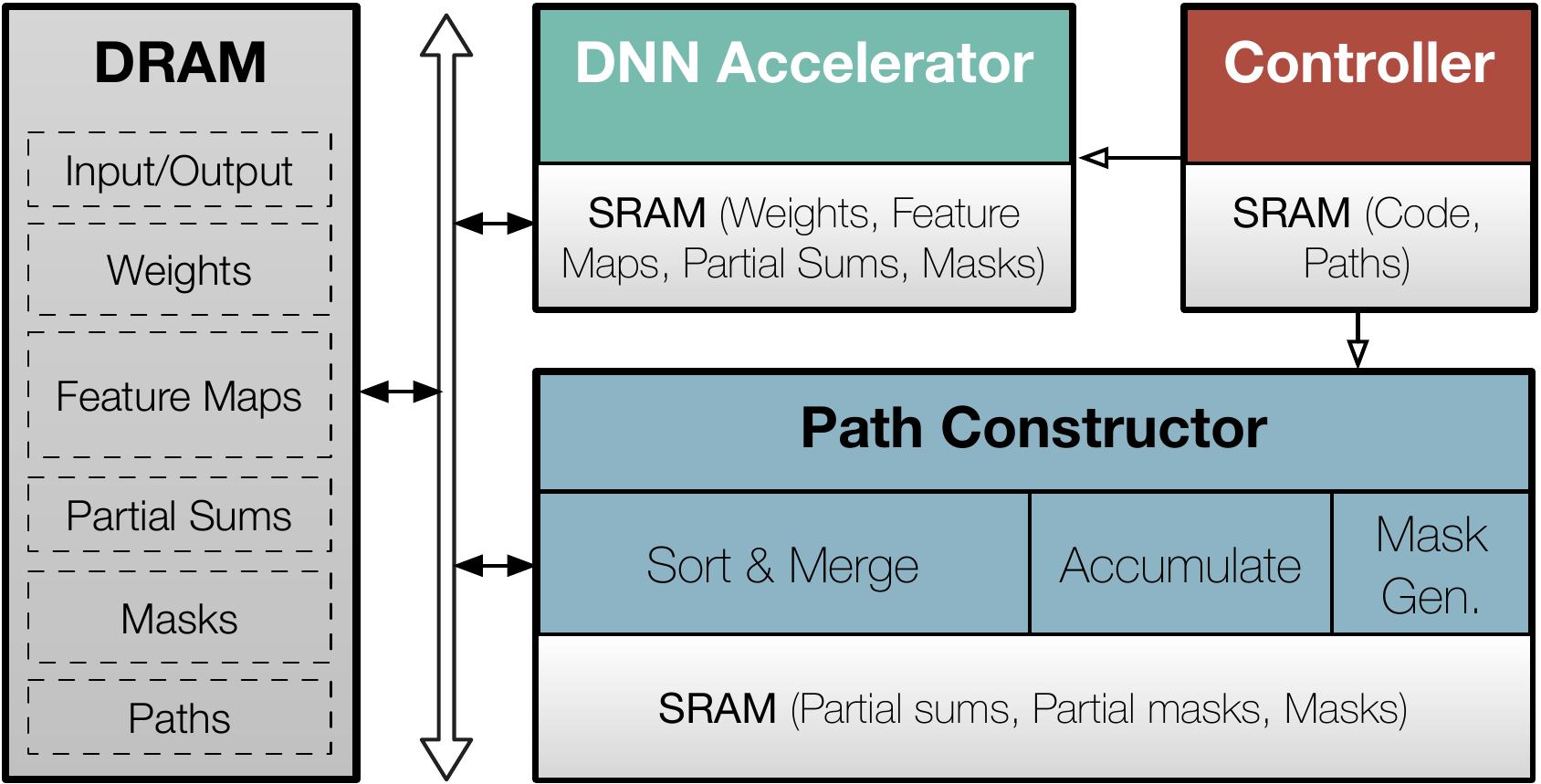}
  \caption{\proj architecture overview.}
  \label{fig:arch}
  %\vspace{-10pt}
\end{figure}

\subsection{Enhanced DNN Accelerator}
\label{sec:arch:acc}

\proj can be integrated into general DNN accelerator designs. Without losing generality we assume a TPU-like systolic array design~\cite{jouppi2017datacenter}. Each PE consists of two 16-bit input registers, a 16-bit fixed-point MAC unit with a 32-bit accumulator register, and simple trivial control logic.

\proj minimally extends each MAC unit.~\Fig{fig:pe} shows the simple MAC unit augmentations (shaded). Specifically, algorithms that use absolute thresholds compare each partial sum with the threshold and store the single-bit mask to the SRAM; algorithms that use cumulative thresholds require each partial sum to be stored to the SRAM. Note that with the re-computation optimization, partial sums are recomputed at extraction time only for important neurons instead of being stored during inference.

%We use a DNN accelerator which has a similar with the one designed in Eyeriss \cite{eyeriss}. However, to address the unique weight-input request in \proj, we enhanced the systolic array architecture in two folds. Firstly, we add an output wire for all the PEs at the first column and first row to write the weight-input to the on-chip weight-input buffer. Secondly, we add an comparison logic inside each PE to compare the multiplication result with a threshold as shown in \Fig{fig:arch_sys}. 

%For the methods that use sorting as the weight-input identify method, all the intermediate weight-input will be needed. During the inference procedure, the systolic array will read from input buffer and weight buffer to calculate the output of the neural networks, meanwhile the output will be stored into the output buffer.

%
%The systolic array would have experienced frequent stalls for algorithms that use cumulative thresholds, because partial sums must be stored to the memory every cycle. 

To avoid the SRAM becoming a scalability bottleneck, the partial sums and the masks are double-buffered in the SRAM and doubled-buffered to the DRAM through a DMA. Later, the partial sums and/or masks are double-buffered back to the SRAM, similar to how feature maps and kernels are accessed. The extra DRAM space required to store partial sums is small as we will show in~\Sect{sec:eval:overhead}. The additional DRAM traffic incurred by storing and reading partial sums is negligible ($<$0.1\%) compared to the original DRAM traffic since each partial sum is read and stored only once.

%The SRAM is carefully banked to avoid bank conflicts, both in usual inference and in partial sum store.
%Alternatively, one could use different algorithm variants (e.g., absolute threshold) to mitigate the high memory traffic.

%While increasing the SRAM size could to some extent accommodate the additional traffic, the most effective mitigations are introduced by the re-computation optimizations in the compiler, or using 

The PE array is used both for the usual inference and for re-computing partial sums as instructed by the \inst{clps} instruction~(\Sect{sec:interface:compiler}). During re-computation, only the first row in the PE array is active because only a selected few elements in the output feature maps are to be re-computed.

%During the re-computation procedure, because of the sparsity and irregularity of the output neuron, only the first row and first column of PEs will be used to re-produce the intermediate weight-input. The intermediate weight-input will be written out to the on chip weight-input buffer and further write to the global buffer in a double-buffered fashion. 

%For the methods that use threshold as the weight-input identify method, we move the comparison procedure onto the systolic array. With the comparison enhancement built up in each PE, the multiplication result will be compared with a threshold that is written in ahead of time before it is sent into the adder. If the multiplication result is larger than the threshold, the PE will write a one into the mask buffer in the on-chip SRAM. Otherwise the PE will write a zero into the mask buffer. Since each PE will only need to write one bit at a time, the systolic array do not add much overhead on the SRAM bandwidth and DRAM bandwidth. There will be no re-computation process for the algorithms using threshold as the identify method for weight-input.

\subsection{Path Constructor}
\label{sec:arch:cons}

The goal of the path constructor is to extract important neurons and to construct activation paths. Algorithms that use cumulative thresholds requires sorting partial sums in receptive fields. Since receptive fields in modern DNNs are usually large (tens of thousands of elements), sorting all the elements on one piece of hardware could become a latency bottleneck as the sequence length increases. Our design splits a long sequence into multiple subsequences, which are sorted in parallel and merged together.~\Fig{fig:sort} shows the sort unit organization. The sort unit uses the classic sorting network~\cite{knuth2014art}, and the merge unit uses a standard merge tree, both have efficient hardware implementations~\cite{mueller2012sorting, chen2015energy, koch2011fpgasort}.

The path constructor uses lightweight mask generation hardware, which generates the important neuron masks for each layer, from which the entire activation path (a bit vector) is constructed. The path constructor also integrates hardware that calculates similarities between an activation path and a canary class path, which is a highly bit-parallel operation. The SRAM in the path constructor is separate from the SRAM used by the DNN accelerator to avoid resource contention, and is also doubled-buffered.

\begin{figure}[t]
  %\vspace{-5pt}
\centering
\subfloat[Enhanced MAC unit.]
{
  \includegraphics[trim=0 0 0 0, clip, height=1.2in]{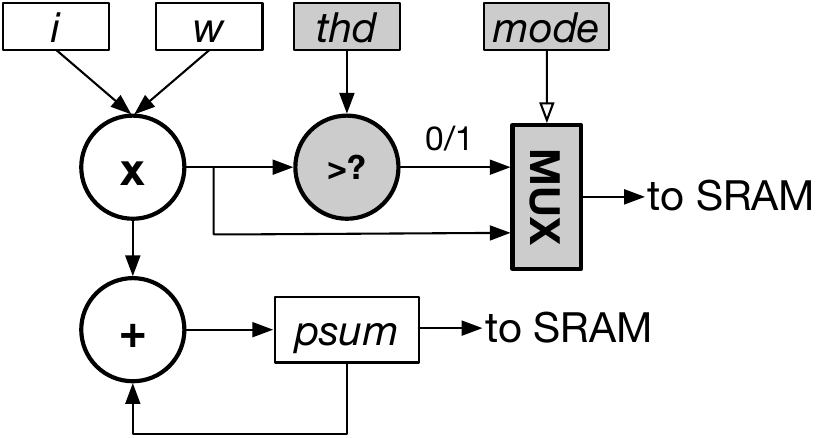}
  \label{fig:pe}
}
%\hfill
\subfloat[Sorting logic.]
{
  \includegraphics[trim=0 0 0 0, clip, height=1.2in]{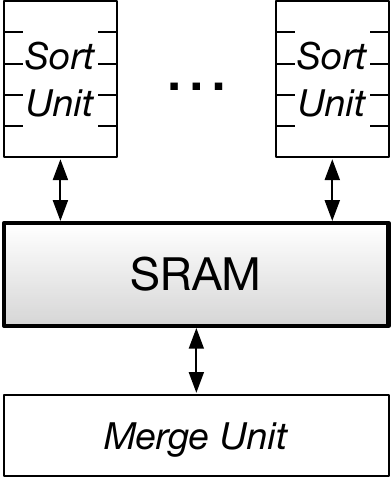}
  \label{fig:sort}
}
\vspace{-5pt}
\caption{Microarchitecture details. MAC and sorting constitutes 99.9\% of the operations in our detection algorithm.}
\label{fig:uarch}
%\vspace{-15pt}
\end{figure}

\subsection{Controller}
\label{sec:arch:ctrl}

We assume a micro-controller unit (MCU) in the baseline hardware, as is common in today's DNN-based Systems-on-a-chip (SoCs)~\cite{xaviersoc}. We piggyback two key tasks on the MCU: dispatching instructions and executing the final classifier to detect adversaries. Both are lightweight tasks that can be executed efficiently on an MCU without extra hardware.

\paragraph{Dispatching Instructions} Thanks to the simple ISA encoding~(\Tbl{tab:isa}), the compiled programs can be interpreted on the MCU (i.e., software decoding) efficiently while avoiding extra hardware cost. The overhead of interpreting the code is negligible compared to the total execution time. The programs are very small in size. The largest one, which uses cumulative thresholds and backward extraction, is about 30 static instructions (below 100 bytes).

\paragraph{Classification} The similarity between an activation path and the canary class path calculated from the path constructor is fed into a random forest (RF) for the final classification (\Sect{sec:sw:cost}). Our particular RF implementation uses 100 decision trees, each of which has an average depth of 12. In total, RF consumes about 2,000 operations on AlexNet (five orders of magnitude lower than inference), and could execute on an MCU in microseconds.

%Inspired by the classic intersection over union (IOU) metric~\cite{Everingham15}, the classifier calculates a similarity metric $\mathrm{S}$ between $\mathrm{P}(x)$ and $\mathrm{P}_c$: $\mathrm{S} = \|\mathrm{P}(x)~\&~\mathrm{P}_c\|_1 / \|\mathrm{P}_c\|_1$, where $\|P\|_1$ denotes the number of 1s in the vector $P$. $\mathrm{S}$ is fed into a random forest~\cite{liaw2002classification} for the final classification. This algorithm is extremely light; 

%We develop a method utilizing the concept of  in object detection \cite{Everingham15}. For each layer, the total number of ones in the AND (\&) result of two paths is devided by the total number of ones in the class path to get a value between zero and one. We feed the vector with the same length of layers into s simple random forest to get the final classification results. 

%While there are a range of potential classifier implementations, our design uses a lightweight random forest classifier.

%in case not to add more overhead, which has only two thousand operations, and can be executed with simple logic in a short time. 

\section{Evaluation Methodology}
\label{sec:exp}

This section explains the basic hardware and software setup (\Sect{sec:exp:setup}) and the evaluation plan~(\Sect{sec:exp:plan}).

\subsection{Experimental Setup}
\label{sec:exp:setup}

\textbf{Hardware Implementation} We develop RTL implementation using Synposys synthesis and Cadence layout tools with Silvaco’s Open-Cell 15nm technology~\cite{15nmcell}. The on-chip SRAM is generated using an ARM memory compiler and the off-chip DRAM is modeled after four Micron 16 Gb LPDDR3-1600 channels. 
We assume an ARM Cortex M4-like micro-controller (MCU) as the controller in the hardware (\Sect{sec:arch:ctrl}). The synthesis and memory estimation results are used to drive a cycle-level simulator for performance and energy analyses.
%Overall, the \proj has a total area of xx mm$^2$. 
%for interpreting the compiled program and classification 

\textbf{Networks and Datasets} We evaluate \proj using two networks: 1) ResNet18~\cite{resnet} on the CIFAR-100 dataset~\cite{cifar} with 100 different classes and 50,000 training images, and 2) AlexNet~\cite{alexnet} on the ImageNet dataset~\cite{imagenet} with 1000 different classes and 1 million training images. The networks and datasets we evaluate are at the high end of the benchmark scale evaluated by today's countermeasure mechanisms~\cite{carlini2017adversarial, he2017adversarial, metzen2017detecting}, which mostly use much smaller datasets and networks (e.g., MNIST, CIFAR-10)~\cite{mnist,svhn} that are less effective in exercising the capability of our system. The test sets are evenly split between adversarial and benign inputs, following the common setup of adversarial attack research.

The clean AlexNet without attacks has an accuracy of 55.13\% on ImageNet; ResNet18 has an accuracy of 94.49\% and 75.87\% on CIFAR-10 and CIFAR-100, respectively.

\textbf{Attacks} We evaluate \proj against a wide range of attacks. We first evaluate using five common non-adaptive attacks: BIM~\mbox{\cite{kurakin2016adversarial}}, CWL2~\mbox{\cite{cwl2}}, DeepFool~\mbox{\cite{deepfool}}, FGSM~\mbox{\cite{jsma}}, and JSMA~\mbox{\cite{jsma}}, which comprehensively cover all three types of input perturbation measures ($L_0$, $L_2$, and $L_\infty$)~\mbox{\cite{akhtar2018threat}}.

We also specifically construct attacks that attempt to defeat our detection mechanism (a.k.a., \textit{adaptive} attacks~\cite{carlini2017adversarial}). In particular, we assume an adversary that has a complete knowledge of \proj's detection algorithms and the attacked model, and thereby generates adversarial samples by incorporating path similarities into the loss function.

\textbf{Metrics} We use the standard ``area under curve'' (AUC) accuracy metric (between 0 and 1) for adversarial detection~\cite{huang2005using}, which captures the interaction between true positive rate and false positive rate. Unless otherwise noted, we report the average accuracy across all attacks. We confirm that the accuracy trend is similar across attacks.

%We evaluate on non-targeted attacks with three different norms: we use $l_{\infty}$ in FGSM and BIM, $l_{2}$ norm for DeepFool and CWL2 and $l_{0}$ norm for JSMA. \fixme{no idea what this is.}

%and Resnet \cite{resnet} as the classification networks. We evaluate on widely-adopted datasets. We use CIFAR-100 \cite{cifar} on ResNet18 which has 100 different classes and 60000 images, it was separated into 50000 images for training and 10000 for test. Another dataset we use is Imagenet \cite{imagenet} on Alexnet with 1000 different classes and 1.15 milllion images. The training set contains 1 million images and test set contains 150000 images.

\subsection{Evaluation Plan}
\label{sec:exp:plan}

Our evaluation is designed to demonstrate that 1) \proj achieves similar or higher accuracy than today's detection mechanisms with a much lower performance penalty, and 2) the general framework allows for a large accuracy-efficiency trade-off. To that end, we develop and evaluate four algorithm variants using our programming model. All the compiler optimizations (\Sect{sec:interface:compiler}) are enabled when applicable.

\begin{itemize}
	\item \sys{BwCu}: Backward extraction with cumulative thresholds.
	\item \sys{BwAb}: Backward extraction with absolute thresholds.
	\item \sys{FwAb}: Forward extraction with absolute thresholds.
	\item \sys{Hybrid}: Hybrid algorithm where \sys{BwAb} is used on the first half of a network and \sys{BwCu} is used on the rest.
\end{itemize}

\paragraph{Baselines} We compare against three state-of-the-art adversarial detection mechanisms: EP~\cite{effectpath2019}, CDRP~\cite{cdrp},  DeepFense~\cite{rouhani2018deepfense}. Both EP and CDRP leverage class-level sparsity. CDRP requires retraining and thus is not able to detect adversaries at inference-time. Note that we evaluate \mbox{\proj} using the exact same attacks used in the above papers.

%To our best knowledge, EP achieves the state-of-the-art detection accuracy at the time of writing, but introduces excessive overhead by serializing extraction and inference. 

%, and generally out-performs other redundancy-based mechanisms such as image transformation and randomization~\cite{xie2017mitigating, buckman2018thermometer}
DeepFense represents a class of detection mechanisms that use modular redundancy. DeepFense employs multiple latent models as redundancies. We directly use the accuracy results reported in their papers. Note that DeepFense is evaluated using ResNet18 on CIFAR-10, on which we perform additional experiments for a fair comparison.

\section{Evaluation}
\label{sec:eval}

We first show the area and DRAM space overhead introduced by \proj's hardware extensions (\Sect{sec:eval:overhead}) are small. We show that \proj provides more accurate detection (\Sect{sec:eval:acc}) with lower latency and energy overhead than prior work (\Sect{sec:eval:lat} -- \Sect{sec:eval:df}). We show that \proj is robust against adaptive attacks that are specifically designed to defeat it (\Sect{sec:eval:adaptive}). \proj provides a large accuracy-efficiency trade-off space (\Sect{sec:eval:et}). We further study the sensitivity and scalability of \proj (\Sect{sec:eval:sen}). Finally, we report additional results on several other models (\Sect{sec:eval:large}).
%the benefits of fine-grained layer control that \proj provides (\Sect{sec:eval:et}) as well as 

\subsection{Overhead Analysis}
\label{sec:eval:overhead}

\paragraph{Area Overhead} The baseline DNN accelerator incorporates a 20$\times$20 MAC array operating at 250MHz. The accelerator has an SRAM size of 1.5~MB, which is banked at a 64 KB granularity. \proj augments the baseline hardware with a 32~KB SRAM banked at 2KB granularity for storing partial sums/masks, and a 64~KB SRAM used by the path constructor, which includes two 16-element sort units, one 16-way merge tree, and an accumulation unit. This accelerator is used in evaluating both \mbox{\proj} and all our baselines.
%, and lightweight control logic.
%All the hardware extensions are clocked at 250MHz.

On top of the baseline DNN accelerator, \proj introduces a total area overhead of 5.2\% (0.08 $mm^{2}$), of which 3.9\% is contributed by the additional SRAM. The rest of the area overhead is attributed to the MAC unit augmentation (0.4\%) and other logic (0.9\%).
%The hardware incurs about 66.4\% average power overhead due to the additional work that the detection algorithm introduces. The path constructor contributes to 58.5\% (593.8~mW) of the power overhead.

%introduce an overhead of 6.25\% (17.1 $\mu m^{2}$) per unit, which contributes to 0.4\% (0.007 $mm^{2}$) of the total overhead, and the extra logic introduces an overhead of 0.9\% (0.014 $mm^{2}$).
% The intermediate data, mask buffer we add on DNN accelerator, and the extrator on-chip SRAM caused 33.9\% area overhead (0.52 $mm^{2}$).

% The enhanced systolic array leads to a power overhead of 8\% (0.2 mW) per PE which results in an total power overhead of 7.9\% (80 $mw$). The extractor causes 58.5\% (593.8 $mw$). 

\paragraph{DRAM Space} Under \sys{BwAb} and \sys{FwAb}, AlexNet and ResNet18 require 1.6 MB and 2.2 MB extra DRAM space. To show scalability, we also evaluated VGG19, which is 13$\times$ larger than ResNet18 and requires only 18.5 MB extra DRAM space. With the recompute optimization, AlexNet, ResNet18, and VGG19 require only an extra 12.8 MB, 17.6 MB, and 148.0 MB in DRAM, respectively under \sys{BwCu}. The additional DRAM traffic is less than 0.1\% (\Sect{sec:arch:acc}).

\begin{figure}[t]
%\vspace{-15pt}
\centering
\subfloat[\small{AlexNet on ImageNet.}]
{
  \includegraphics[trim=0 0 0 0, clip, width=0.47\columnwidth]{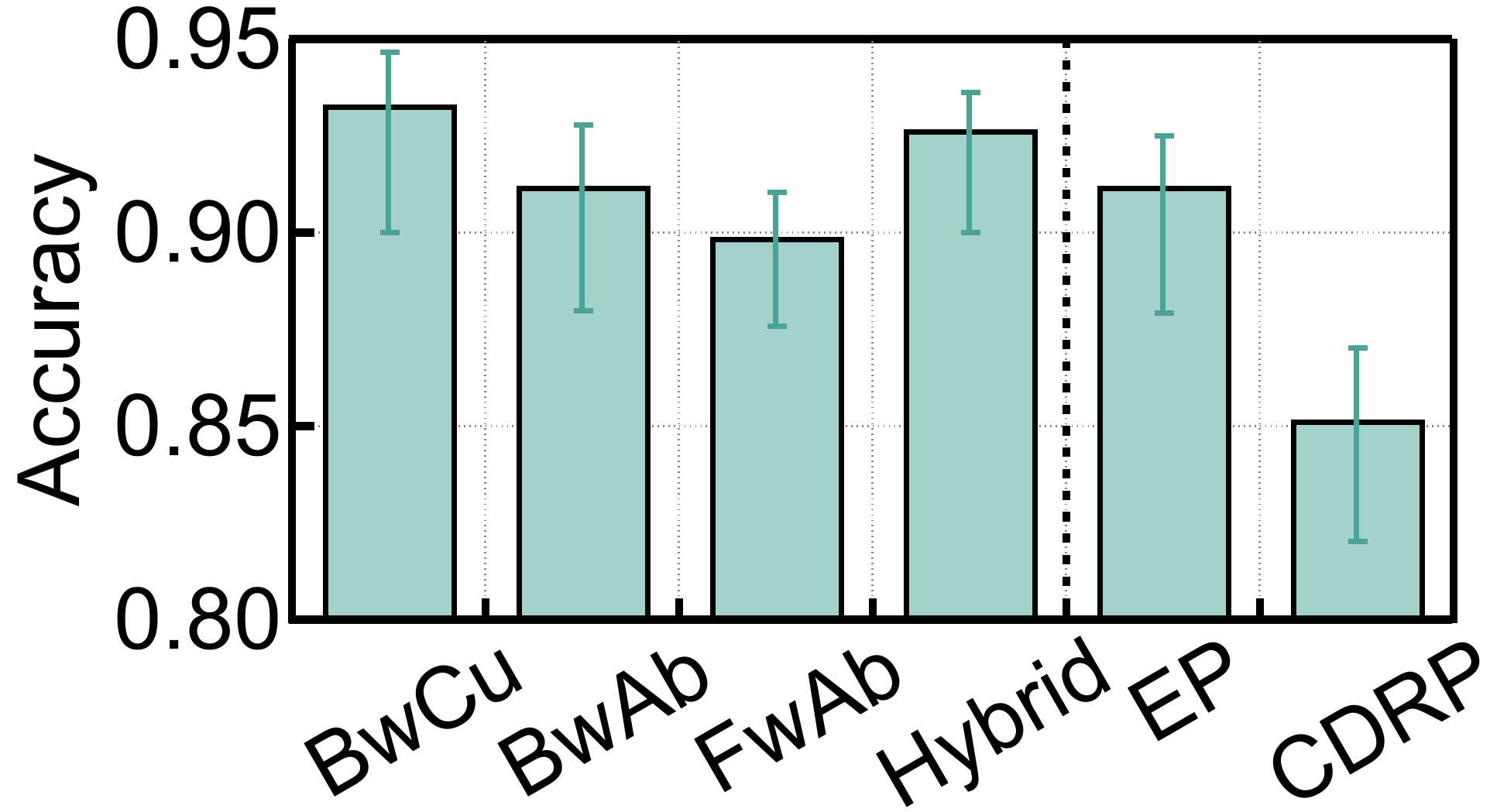}
  \label{fig:acc_imagenet}
}
%\hfill
\subfloat[\small{ResNet18 on CIFAR-100.}]
{
  \includegraphics[trim=0 0 0 0, clip, width=0.47\columnwidth]{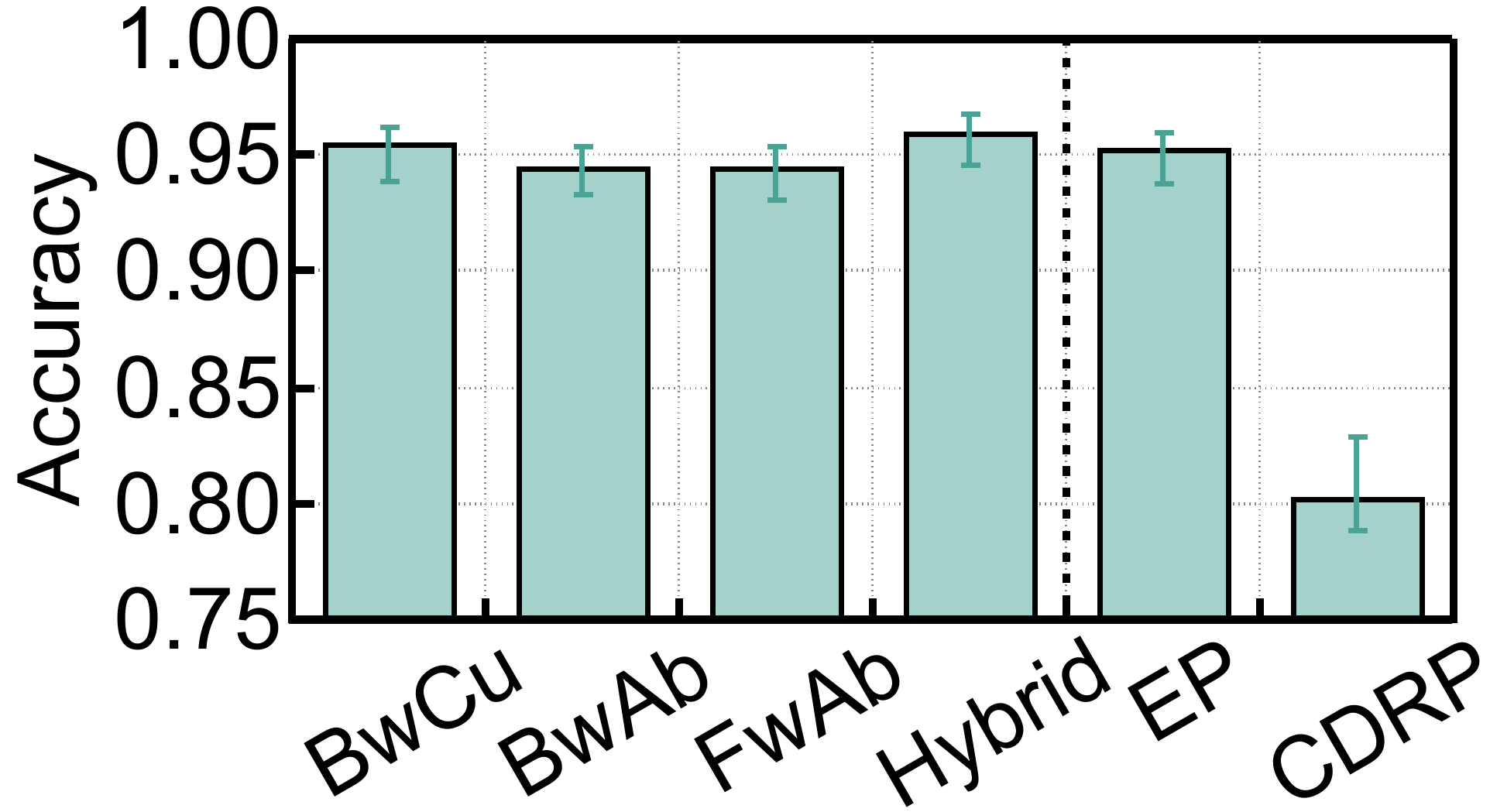}
  \label{fig:acc_cifar100}
}
%\vspace{-5pt}
\caption{Accuracy comparisons with EP and CDRP. Error bars indicate the max and min accuracies of all the attacks.}
\label{fig:acc_results}
%\vspace{-10pt}
\end{figure}

\subsection{Accuracy}
\label{sec:eval:acc}

\proj's accuracy varies with the choice of $\theta$ and $\phi$, which control the coverage of important neurons. Using \sys{BwCu} as an example,~\Tbl{tab:theta} shows how its accuracy changes as $\theta$ varies from 0.1 to 0.9. As $\theta$ initially increases from 0.1 to 0.5 the accuracy also increases, because a higher $
\theta$ captures more important neurons. However, as $\theta$ increases to 0.9, the accuracy slightly drops. This is because a high $\theta$ value causes different class paths to overlap and become less distinguishable. Meanwhile, the latency and energy consumption increase almost proportionally as $\theta$ increases. We thus use $\theta=0.5$ for the rest of our evaluation. The trend with respect to $\phi$ is similar, but is omitted due to limited space.

%https://tex.stackexchange.com/questions/394119/caption-at-left-side-of-the-table
%\begin{table}[h]
\begin{SCtable}[][h]
%\vspace{-5pt}
\centering
%\captionsetup{width=.5\columnwidth}
\caption{Sensitivity of accuracy, latency, and energy of \sys{BwCu} as $\theta$ varies. Latency and Energy are normalized to inference.}
\renewcommand*{\arraystretch}{1}
\renewcommand*{\tabcolsep}{2pt}
\resizebox{0.5\columnwidth}{!}
{
\label{tab:theta}
\begin{tabular}{cccc}
\toprule[0.15em]
\textbf{$\theta$} & \textbf{Accuracy} & \textbf{Latency} & \textbf{Energy} \\
\midrule[0.05em]
0.1   & 0.86     & 4.7$\times$    & 2.9$\times$    \\
0.5   & 0.94     & 12.3$\times$    & 7.7$\times$    \\
0.9   & 0.91     & 25.7$\times$   & 15.6$\times$  \\ \bottomrule[0.15em]
\end{tabular}
}
%\vspace{-5pt}
%\end{table}
\end{SCtable}

\proj variants achieve similar or better accuracy than existing defense mechanisms.~\Fig{fig:acc_results} shows the accuracy comparison. On AlexNet across all attacks (\Fig{fig:acc_imagenet}), the three backward extraction-based variants (\sys{BwCu}, \sys{BwAb}, and \sys{Hybrid}) outperform EP and CDRP by up to 0.02 and 0.1, respectively. \sys{FwAb} uses forward extraction and has 0.03 lower accuracy than EP (0.06 higher than CDRP), indicating the accuracy benefits of backward extraction. On ResNet18 (\Fig{fig:acc_cifar100}), \proj consistently achieves higher (0.14 -- 0.16) accuracy than CDRP, and has similar or higher accuracy than EP (at most 0.01 accuracy loss).

Note that adversarial attacks generated by CWL2 have low confidence of the rank1 class, and the confidence of rank1 class is similar to that of the rank2 class. Thus, evaluating CWL2 let us understand \mbox{\proj}'s robustness against adversarial attacks launched by ``low-confidence'' images. On Imagenet against CWL2, \mbox{\proj}'s accuracy is 0.95, while the baselines are 0.94 (EP) and 0.85 (CDRP); on CIFAR10, \mbox{\proj}'s accuracy is 0.96 while DeepFense is 0.93.

%\Fig{fig:acc_imagenet} shows the AUC on Y-axis for Alexnet and different attacks on X-axis and \Fig{fig:acc_cifar100} shows the results for ResNet18.

%where BWCU achieves 0.02 to 0.04 higher AUC compared with EP. BWAB achieves almost same AUC compared with EP and FORWARD achieves slightly less AUC compared with EP. The detection \proj all significantly outperforms CDRP among all attacks. For ResNet18, under all attacks, BWCU outperforms EP as well with . BWAB, FORWARD, and HYBRID achieve comparable AUC with EP, the largest AUC loss is less than 0.02. All setups significantly outperform CDRP on ResNet18 as well.

\begin{figure}[t]
\vspace{-5pt}
\centering
\subfloat[\small{AlexNet on ImageNet.}]
{
  \includegraphics[trim=0 0 0 0, clip, height=1.05in]{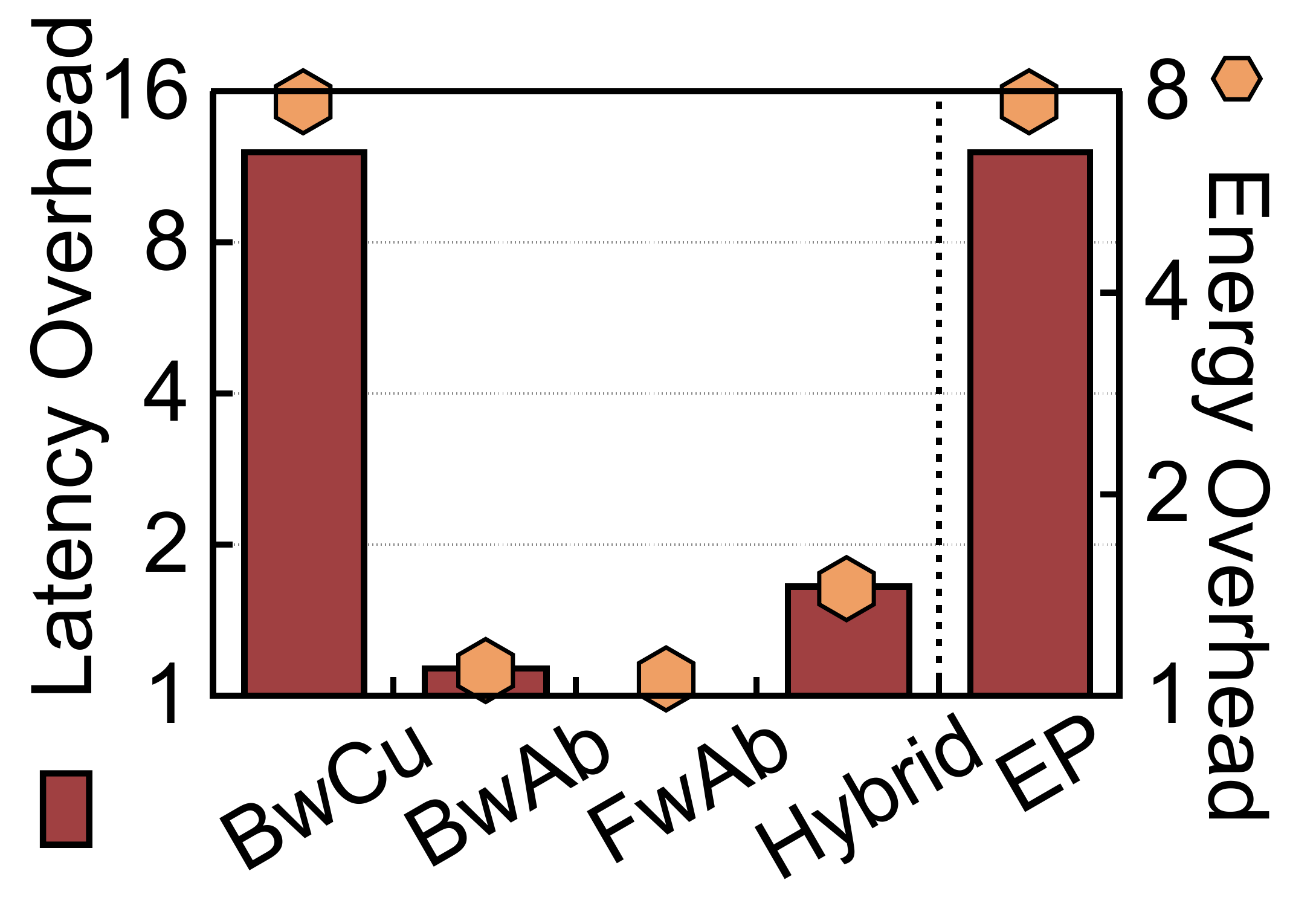}
  \label{fig:lat_imagenet}
}
%\hfill
\subfloat[\small{ResNet18 on CIFAR-100.}]
{
  \includegraphics[trim=0 0 0 0, clip, height=1.05in]{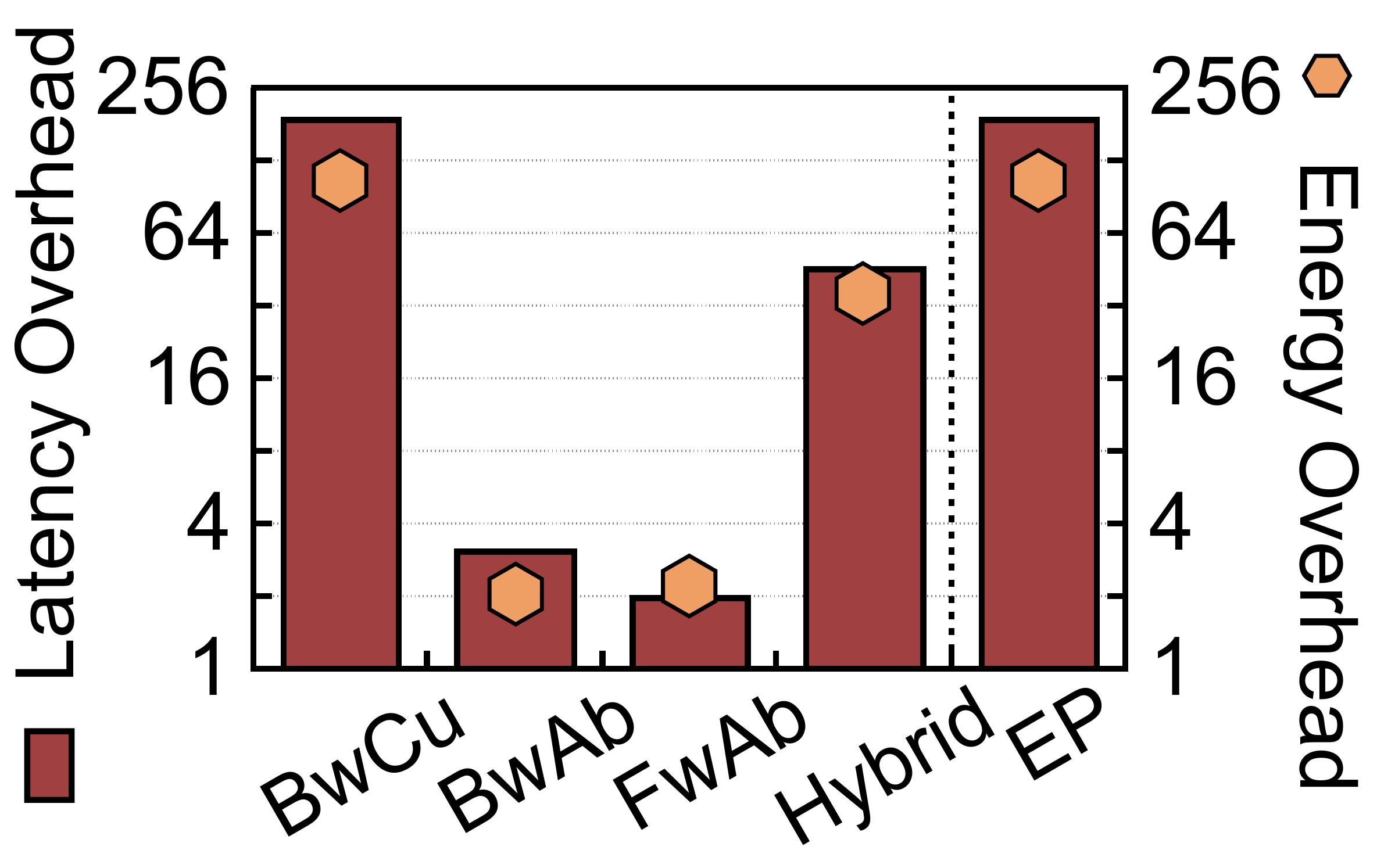}
  \label{fig:lat_cifar100}
}
%\vspace{-5pt}
\caption{Latency and energy comparisons with EP.}
\label{fig:lat_results}
%\vspace{-12pt}
\end{figure}

\subsection{Latency and Energy}
\label{sec:eval:lat}

\proj could achieve low performance and energy overhead over usual DNN inference.~\Fig{fig:lat_imagenet} and~\Fig{fig:lat_cifar100} show the latency and energy consumption of the four \proj variants normalized to DNN inference, respectively. For comparison purposes, we also show the latency and energy of EP. We do not show the results of CDRP because CDRP requires retraining and is not suitable for online detection.

%We show the latency results of \proj for Imagenet dataset on Alexnet in \Fig{fig:lat_imagenet} and CIFAR-100 dataset on ResNet18 in \Fig{fig:lat_cifar100}. All the latency results shown in this section is end-to-end latency meaning the latency of inference, extraction and classification are all included. We normalize the latency and energy to the inference latency and energy to show the relative overhead compared with inference procedure and showing log scale on Y-axis for better comparison.

Although having the highest accuracy, \sys{BwCu} also has the highest latency and energy overhead due to the expensive partial sum sorting and accumulation operations during extraction, which is serialized with inference. On AlexNet, \sys{BwCu} introduces 12.3$\times$ latency overhead and increases the energy by 7.7$\times$. The corresponding results on ResNet18 are 195.4$\times$ and 105.9$\times$, respectively. The overhead on ResNet18 (18 layers) is higher than on AlexNet (8 layers), because as the network becomes deeper the amount of important neurons increases, which in turn increases the extraction time.

The overhead of \sys{BwCu} is similar to EP, while \sys{BwAB}, \sys{FwAB} and \sys{Hybrid} all achieve much lower latency and energy overhead. \sys{BwAb} uses absolute thresholds to avoid sorting and storing partials sums. \sys{BwAb} reduces the latency and energy overhead on AlexNet to only 1.2$\times$ and 1.1$\times$, respectively, and 3.2$\times$ and 2.0$\times$ on ResNet18, respectively.

\sys{FwAb} further reduces the latency overhead to only 2.1\% and 2.1$\times$ on the two networks, respectively, by using forward extraction to overlap extraction with inference. The latency overhead on ResNet18 is higher because ResNet18 is deeper with a higher important neuron density (explained above), leading to longer extraction latency that is harder to hide behind the inference latency. \sys{FwAb} does not reduce energy overhead significantly comparing to \sys{BwAb}, because it hides, rather than reducing, the amount of compute.

% The reason is that FORWARD has to identify important output neuron each layer, while BWAB can trace its important output neuron from previous layer.
%The energy overhead on Alexnet and ResNet18 are 1.1$\times$ and 2.2$\times$. 

Finally, \sys{Hybrid} provides a design point that balances efficiency with accuracy by combining cumulative thresholds and absolute thresholds. It leads to 1.7$\times$ latency overhead and 1.4$\times$ energy overhead on AlexNet, and the overheads are 47.3$\times$ and 36.1$\times$ on ResNet18, respectively.

\subsection{DeepFense Comparison}
\label{sec:eval:df}

%We compare \proj with two state-of-the-art online defense frameworks.The first one is DeepFense, which utilizes modular redundancies to defend against adversarial examples.

We compare against the three default DeepFense variants, which differ in the number of redundant networks: 1 in \textit{DFL}, 8 in \textit{DFM}, and 16 in \textit{DFH}. DeepFense is originally implemented on FPGA/GPUs; we perform a best-effort reimplementation on our hardware substrate for a fair comparison.

%We evaluate DeepFense under our hardware substrate to make fair comparisons. Since DeepFense use up to 16 different defense network to achieve high detecting accuracy, we have three different settings for DeepFense: \textit{DFL} with only one defense network, \textit{DFM} with eight defense networks and \textit{DFH} with sixteen defense networks. 

%We show the accuracy between \proj and DeepFense. We use CIFAR-10 \cite{cifar} as the datasets since DeepFense has only been evaluated on MNIST\cite{mnist}, SVHN\cite{svhn} and CIFAR-10. We only use four attacks since JSMA was not evaluated on DeepFense.

\Fig{fig:DF_acc} shows the accuracy comparison between \proj and DeepFense using ResNet18 on CIFAR-10. All \proj variants achieve significantly higher detection accuracy over DeepFense. Specifically, \sys{FwAb}, which has the lowest accuracy among all \proj variants, outperforms \textit{DFH}, which is the most accurate setup of DeepFense, by 0.11 on average.
%over four out of five attacks except JSMA, which is not reported by DeepFense

%0.15, 0.11, 0.03, 0,14 on BIM, CWL2, DeepFool, and FGSM attacks, respectively.

%DeepFense shows comparable detection accuracy on Deepfool with \proj.  

\Fig{fig:DF_lat} shows the latency and energy of \proj and DeepFense variants normalized to usual inference. With higher accuracy, \sys{BwAb} and \sys{FwAb} are also faster and consume less energy compared to all three DeepFense variants. For instance, \sys{FwAb} reduces latency and energy overhead by 89.0\% and 59.0\%, respectively, compared with \textit{DFL}, the most light version of DeepFense. The better efficiency of \proj over DeepFense indicates the effectiveness of exploiting the runtime behaviors of DNN inferences.
% compared to treating inference as a blackbox.

%BWCU results on higher latency overhead (11.0$\times$ compared with 7.9$\times$) and similar energy overhead (7.0$\times$ compared with 8.0$\times$) with DFM, but both are lower than DFH (15.91$\times$ latency overhead and 15.91$\times$ energy overhead). Besides BWCU, BWAB, FORWARD, and HYBRID all cause less latency and energy overhead compared with DFL.

%\paragraph{DNNGuard} For a fair comparison, we normalize our hardware design to have the same number of MAC units as in DNNGuard. \Fig{fig:DNNGuard} compares the latency overhead and accuracy between \sys{BwCu} and DNNGuard. Accuracy uses the true positive rate (TPR) referenced in the DNNGuard paper, which does not report the AUC results. We also verify that under the same TPR, we have lower (better) false positive rate (FPR). FPR data is omitted due to space limitations.

%The different \sys{BwCu} variants in \Fig{fig:DNNGuard} represent different termination layers, where the left-most marker represents a variant that terminates at the last layer (i.e., extracting one layer only). Even when extracting only one layer \proj achieves higher accuracy with lower overhead compared to DNNGuard. As more layers are extracted, \proj's accuracy improves and the overhead also increases, but it still Pareto-dominates DNNGuard.
% If the termination layer is set to be less than 15 layers, \proj achieves lower overhead compared with DNNGuard. However, as termination layer goes up, the overhead of \proj increases and will have higher overhead compared with DNNGuard.

\begin{figure}[t]
%\vspace{-5pt}
\centering
\subfloat[\small{Accuracy.}]
{
  \includegraphics[trim=0 0 0 0, clip, height=.9in]{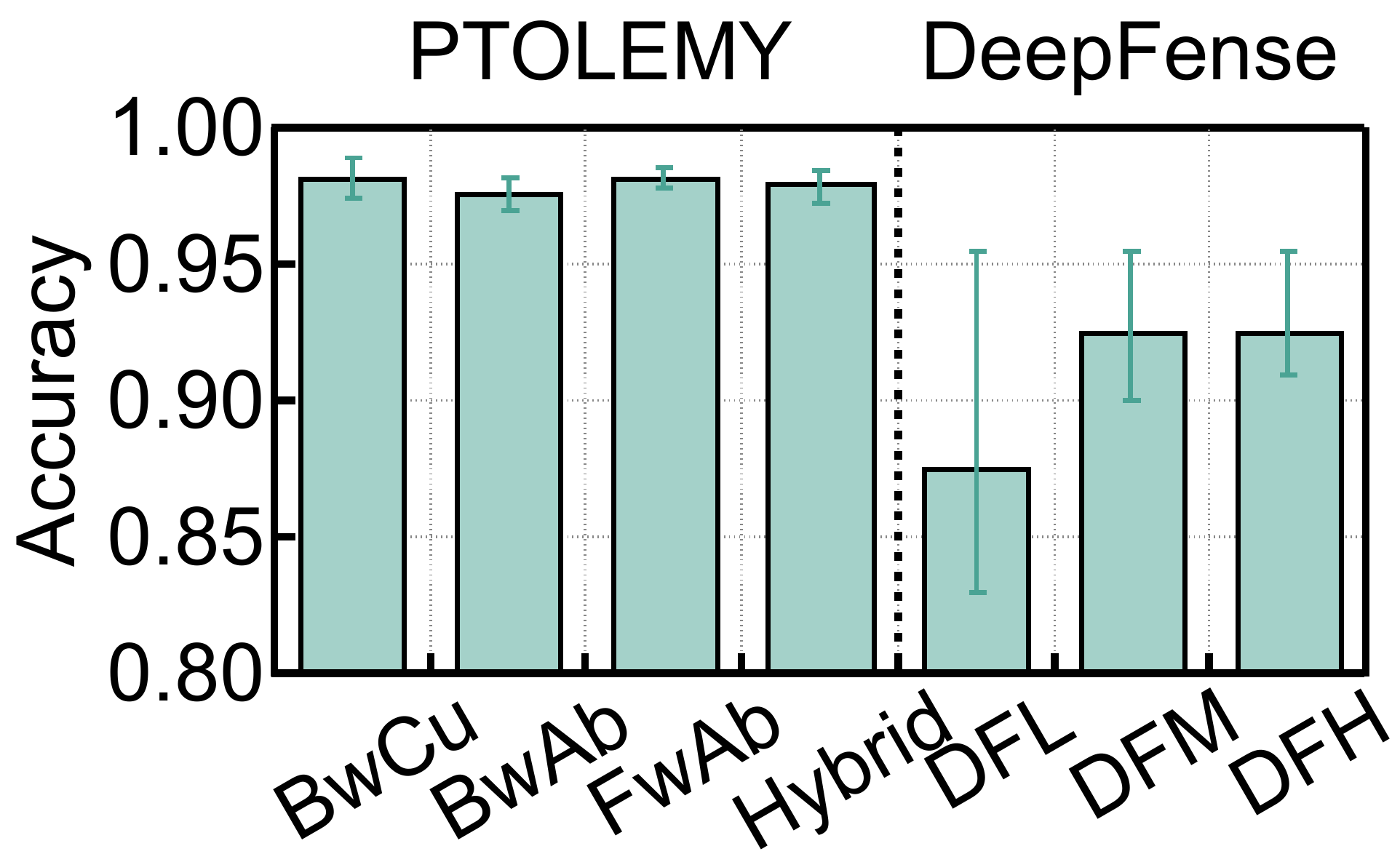}
  \label{fig:DF_acc}
}
\hfill
\subfloat[\small{Latency and energy.}]
{
  \includegraphics[trim=0 0 0 0, clip, height=.9in]{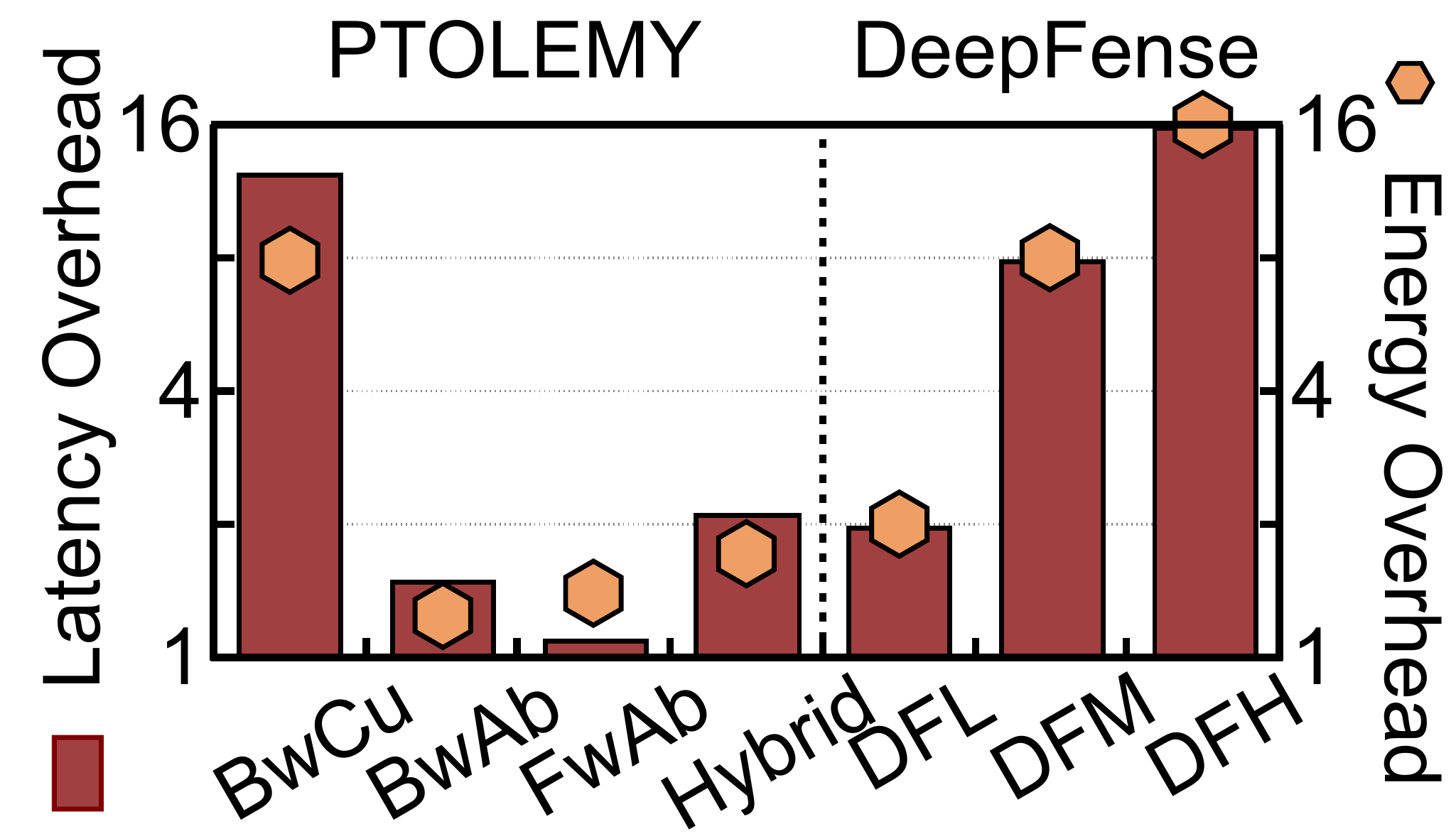}
  \label{fig:DF_lat}
}
%\subfloat[\small{DNNGuard.}]
%{
%  \includegraphics[trim=0 0 0 0, clip, width=0.3\columnwidth]{DNNguard}
%  \label{fig:DNNGuard}
%}
%\vspace{-5pt}
\caption{DeepFense comparison.}
\label{fig:lat_results}
%\vspace{-12pt}
\end{figure}

\subsection{Defending Against Adaptive Attacks}
\label{sec:eval:adaptive}

Adaptive attacks refer to attacks that have complete knowledge of how a defense mechanism works and attempt to defeat that specific defense~\cite{carlini2019evaluating, tramer2020adaptive}. We perform a best-effort construction of adaptive attacks against \proj, and show that \proj can effectively defend against adaptive attacks.
%, which attempt to defeat \proj's path-based detection method by incorporating the path similarity as part of the loss function when generating the adversarial samples.

\paragraph{Constructing the Attacks} To attempt to defeat \mbox{\proj}, we force an adversarial sample to have the same activation path as a benign input. However, since our path construction requires ranking/thresholding, which are non-differentiable, we opt for a differentiable approximation--a common practice in adversarial ML~\mbox{\cite{athalye2018robustness, tramer2020adaptive}}. We experiment with several heuristics, and find that the most effective one is to force all the activations of an adversary to be the same as a benign input, i.e., a sufficient but not necessary condition.

Specifically, to generate an adversarial sample from an input $x$ that has a true class $c$, we first randomly choose a benign input $x_t$ of target class $t$ from the training dataset, where $c \neq t$. We then add noise $\delta x$ to $x$ to generate $x_a$ such that $x_a$'s activations are as close to that of $x_t$ as possible. This is achieved by minimizing the L2 loss $\sum_{i} \left \| z_i(x + \delta x) - z_i(x_t) \right \|^2_2$ as the objective function, where $z_i(\star)$ denotes the activations of $\star$ at layer $i$. To strengthen the attack, we choose five different $x_t$ of different classes to generate five different $x_a$, and select the $x_a$ with the smallest loss. We use projected gradient descent (PGD)~\cite{madry2017towards} as the optimization method.

%, and verify that this method successfully generates adversarial samples for 100\% of the inputs in the datasets (i.e., 100\% success rate in adversarial ML parlance).
%We perform a binary search over epsilon and step size for PGD, with 100 iterations in each trial, $0.3$ as the initial epsilon and $0.01$ as the initial step size.

\begin{figure}[t]
  \centering
  \includegraphics[width=\columnwidth]{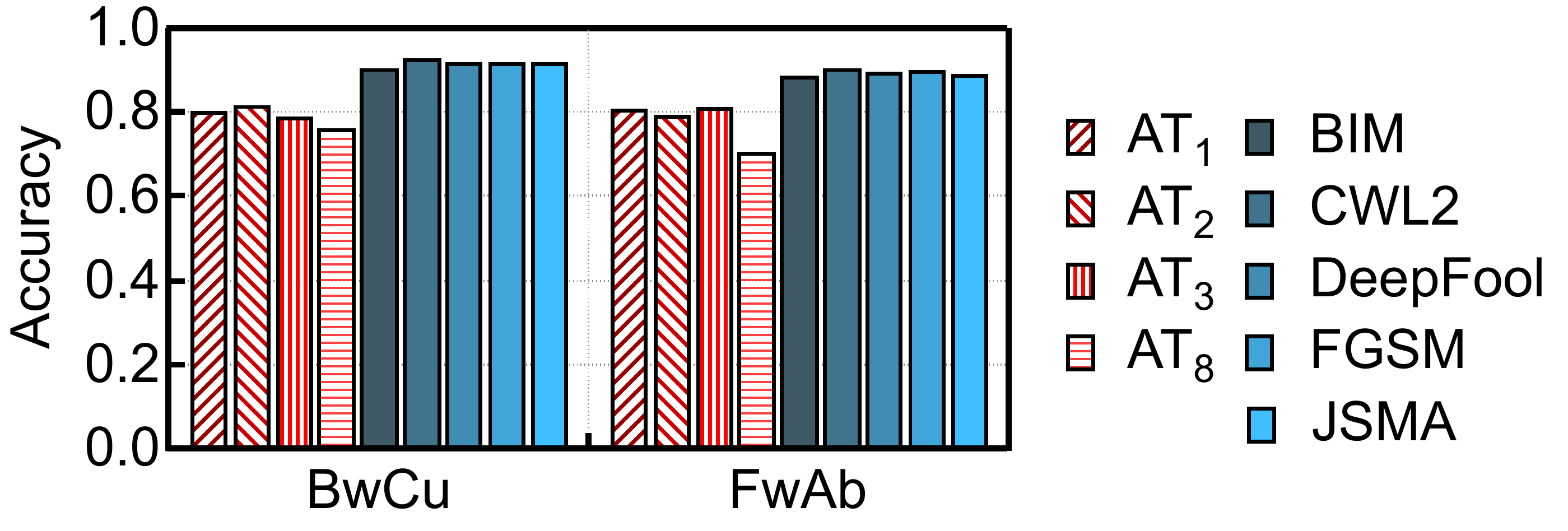}
  \caption{Detection accuracy of \mbox{\proj} on various adaptive attacks (AT) compared to the five existing attacks.}
  \label{fig:at}
\end{figure}

\paragraph{Results} \proj detects these adaptive adversarial samples, even though they are generated specifically to ``fool'' \proj by having activation paths that are similar to their benign counterparts. Using AlexNet on ImageNet as an example, \Fig{fig:at} shows the detection accuracy of \sys{BwCU} and \sys{FwAb} on the adaptive attacks ($AT$). $AT_n$ denotes that activations of the last $n$ layers are considered in the loss function when generating adversarial samples. Since AlexNet has 8 layers, $AT_8$ is the strongest adaptive attack. The detection accuracies on existing attacks are shown as for comparison.
%The trends on ResNet18 are similar.

Overall, the detection accuracy decreases as more layers are considered in generating the adaptive attacks, i.e., attacks become more effective. When only the first three layers are considered by the adaptive attack, the adversaries are more easily detected by \proj than existing attacks. The detection accuracies on adaptive attacks are lower than those on non-adaptive attacks, confirming that adaptive attacks are more effective, matching the intuition~\cite{carlini2019evaluating}.

%Even on the strongest attack $AT_8$, \proj has detection accuracies similar to or higher than on existing attacks.

%While \proj requires sorting/thresholding activations, which are non-differentiable operations, we construct a best-effort adversarial by approximating these non-differentiable operations using 

\paragraph{Validating and Analyzing the Attacks} Our adaptive attack does not bound perturbation, i.e., is an unbounded attack. Following the guideline in Carlini et al.~\mbox{\cite{carlini2019evaluating}} that ``\textit{The correct metric for evaluating unbounded attacks is the distortion required to generate an adversarial example, not the success rate (which should always be 100\%)}'', we verify the validity of our adaptive attack in two ways. First, we verify that the constructed attacks do reach 100\% success rate; the average distortion, measured in Mean Square Error (MSE), is 0.007, and the maximum MSE 0.035.

Second, we show how the detection accuracy of \mbox{\proj} is impacted by the distortion rate introduced in the adaptive adversarial examples. The data is shown in \mbox{\Fig{fig:auc_distort}}, where every $<x$, $y>$ point denotes the average detection accuracy ($y$) for all the adaptive attacks whose distortions (MSE) is lower than or equal to a certain value ($x$). We find that overall the detection accuracy drops slightly as the distortion increases---an expected trend---although the trend is not strong, which is likely because the absolute distortion is too low (a desirable property) to demonstrate strong correlation with accuracy. We do verify that when the distortion is large enough to completely transform an image from one class to another, the detection accuracy would drop to 0, but at that point the input could not be considered an adversarial attack since the transformed image does not look like the original image.

\begin{figure}[t]
\begin{minipage}[t]{0.47\columnwidth}
  \centering
  \includegraphics[width=1.1\columnwidth]{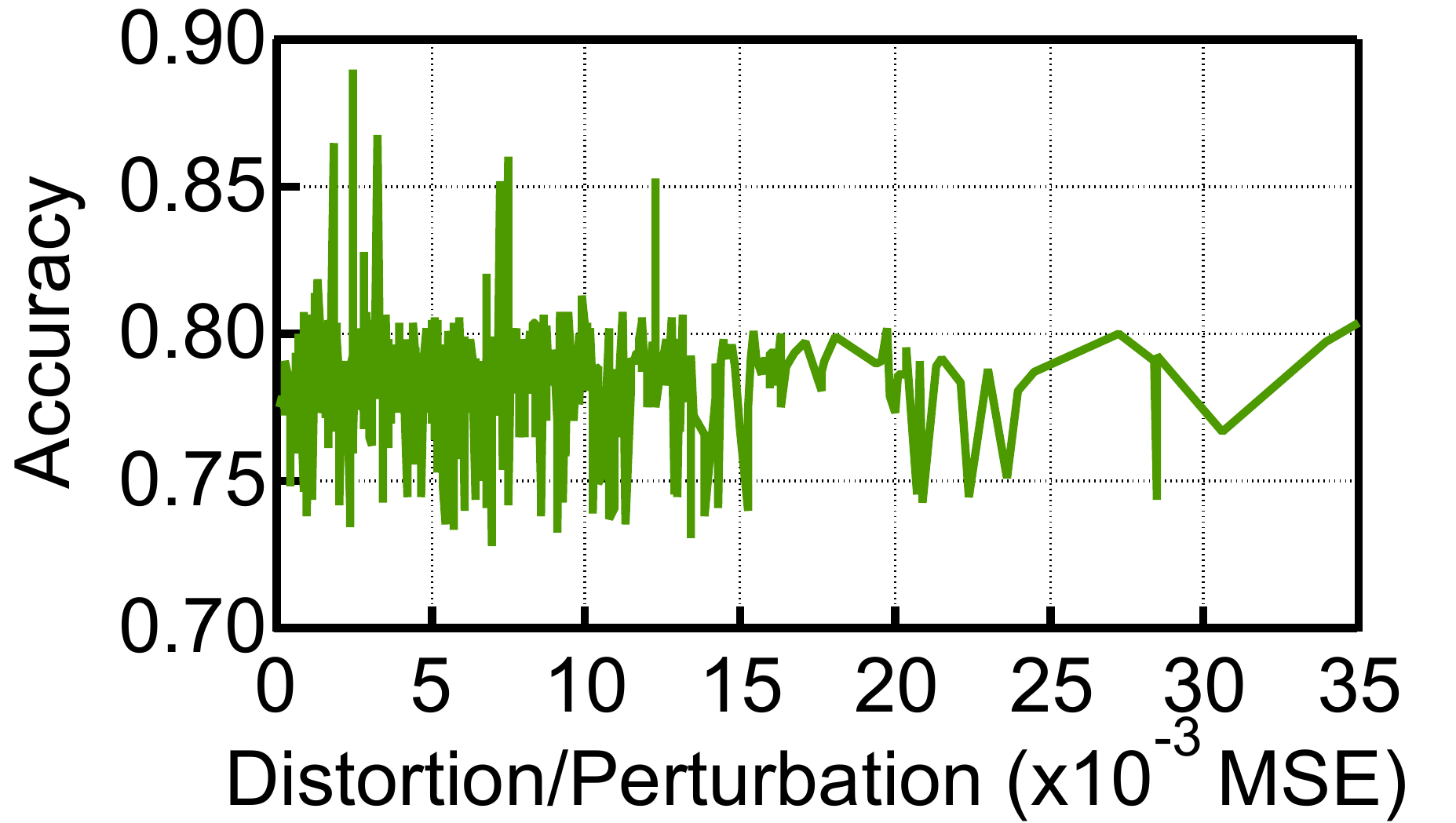}
  \caption{Detection accuracy of adaptive adversarial inputs under different distortions.}
  \label{fig:auc_distort}
\end{minipage}
\hspace{5pt}
\begin{minipage}[t]{0.47\columnwidth}
  \centering
  \includegraphics[width=1.1\columnwidth]{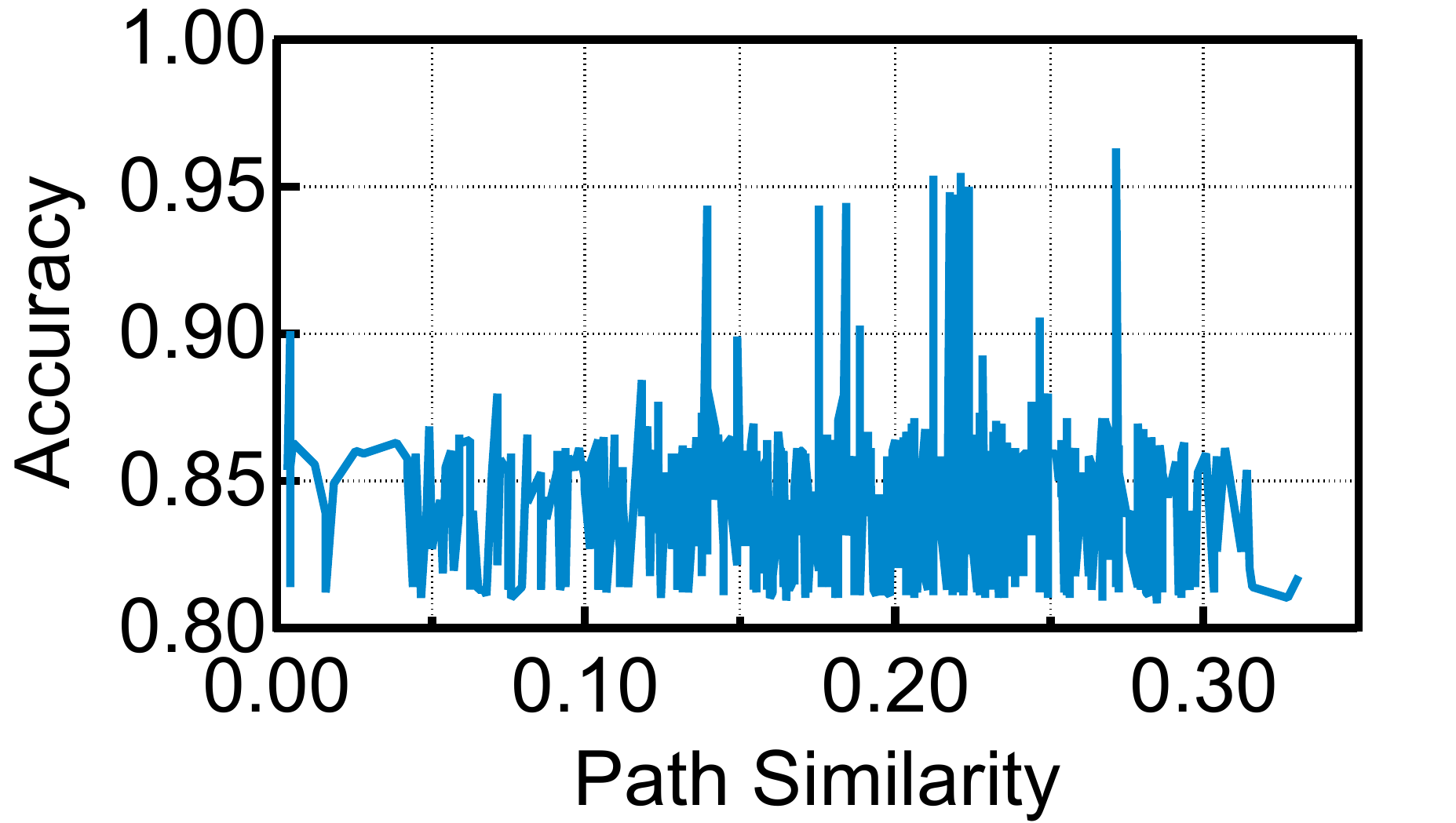}
  \caption{Detection accuracy of adaptive attacks under different path similarities.}
  \label{fig:auc_simi}
\end{minipage}
\end{figure}

%class similarity among adaptive adversarial examples and their targets (i.e., an adaptive adversarial example change original example belonging to $C_0$ to $C_i$, we try to find the relationship between detection accuracy and class path similarity $C_0$-$C_i$).
We also investigate how the detection accuracy is impacted by the path similarities between the original class and the target class. We show the results in \mbox{\Fig{fig:auc_simi}}, where every $<x$, $y>$ point denotes the average detection accuracy ($y$) for all the adaptive adversarial inputs whose path similarity between the original class and the target class is lower than or equal to a certain value ($x$). While the path similarity between the original class and the target class has a wide range (0.0 -- 0.34), the detection accuracy does not correlate strongly with the path similarity. This is a desirable property, as it suggests that \mbox{\proj} is not more vulnerable when the attacker simply targets a similar class when generating the attacks.
%The reason may be caused by the difference between class path similarity and example path similarity. Low class path similarity doesn't represent the adversarial example itself has a low example path similarity compared to the original example it tries to attack.

\paragraph{Discussion} The way we construct the adaptive attack is by approximating the hard path objective (i.e., forcing an adversarial sample to have the same activation path as a benign input) using a differentiable objective that constrains the individual activations. This relaxation let us formulate adversarial attack generation as an optimization problem that could be solved using effective optimization methods (e.g., PGD). If one were to force a hard constraint on the activation path, the objective function would not be differentiable.

In that case, a naive approach to generate adaptive attacks would be to exhaustively search all the possible perturbations. But without guidance such search would be prohibitively expensive (e.g., (${256^3}^{40,000}$ for an 8-bit color depth, 200$\times$200 resolution RGB image). We did try the exhaustive search method in a limited form, which generated results that add so much perturbation so that the resulted images do not look like the original images at all.

%\proj will not detect all adaptive attacks with 100\% accuracy --- just like any other defense mechanism. What we aimed to show is that 1) the particular defense mechanism used in Ptolemy makes it challenging to generate effective adaptive attacks, and 2) we have high detection accuracy even for adaptive attacks that could be successfully generated. Let us explain.

%As mentioned earlier, we have exercised our best due diligence to evaluate different strategies to approximate the object and what's reported is the best we find.

An interesting direction would be to investigate intelligent search heuristics (e.g., simulated annealing) to find perturbations that meets the hard path constraint while fooling \proj. We leave this to future work.

%\hl{We also tried to manually enforce a hard constraint on the activation path without using the optimization formulation, but the generated adversarial samples have a very low success rate in attacking the original models. This is because with a hard activation path constraint, an input generally becomes a completely different class (e.g., a dog image gets perturbed so much so it become a cat image and is indeed recognized as a cat).}

\subsection{Early-Termination and Late-Start}
\label{sec:eval:et}

\begin{figure}[t]
\vspace{-8pt}
\centering
\subfloat[\small{Accuracy.}]
{
  \includegraphics[trim=0 0 0 0, clip, height=0.9in]{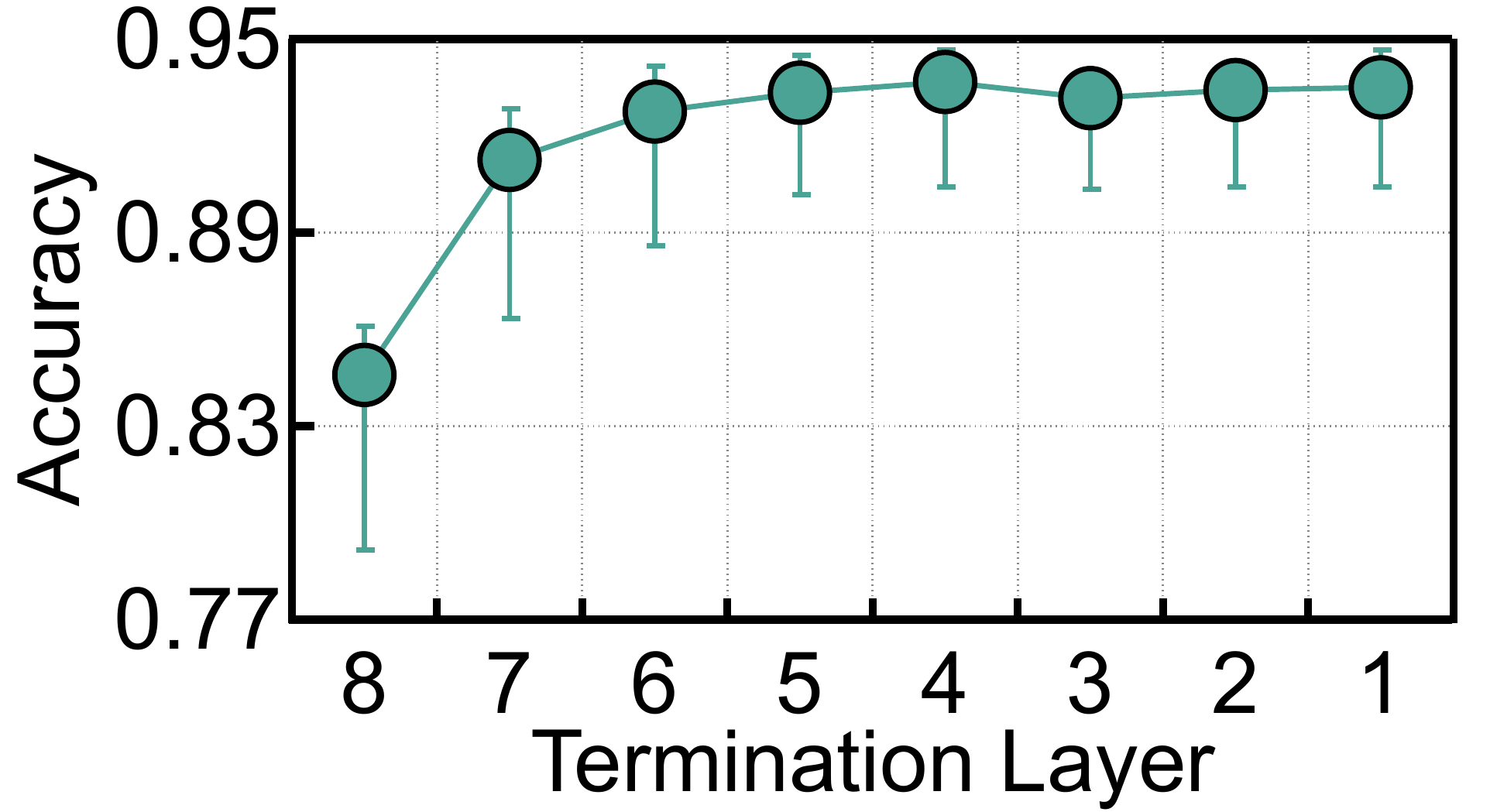}
  \label{fig:acc_layer_t2}
}
%\hspace{3pt}
\subfloat[\small{Latency and energy.}]
{
  \includegraphics[trim=0 0 0 0, clip, height=0.9in]{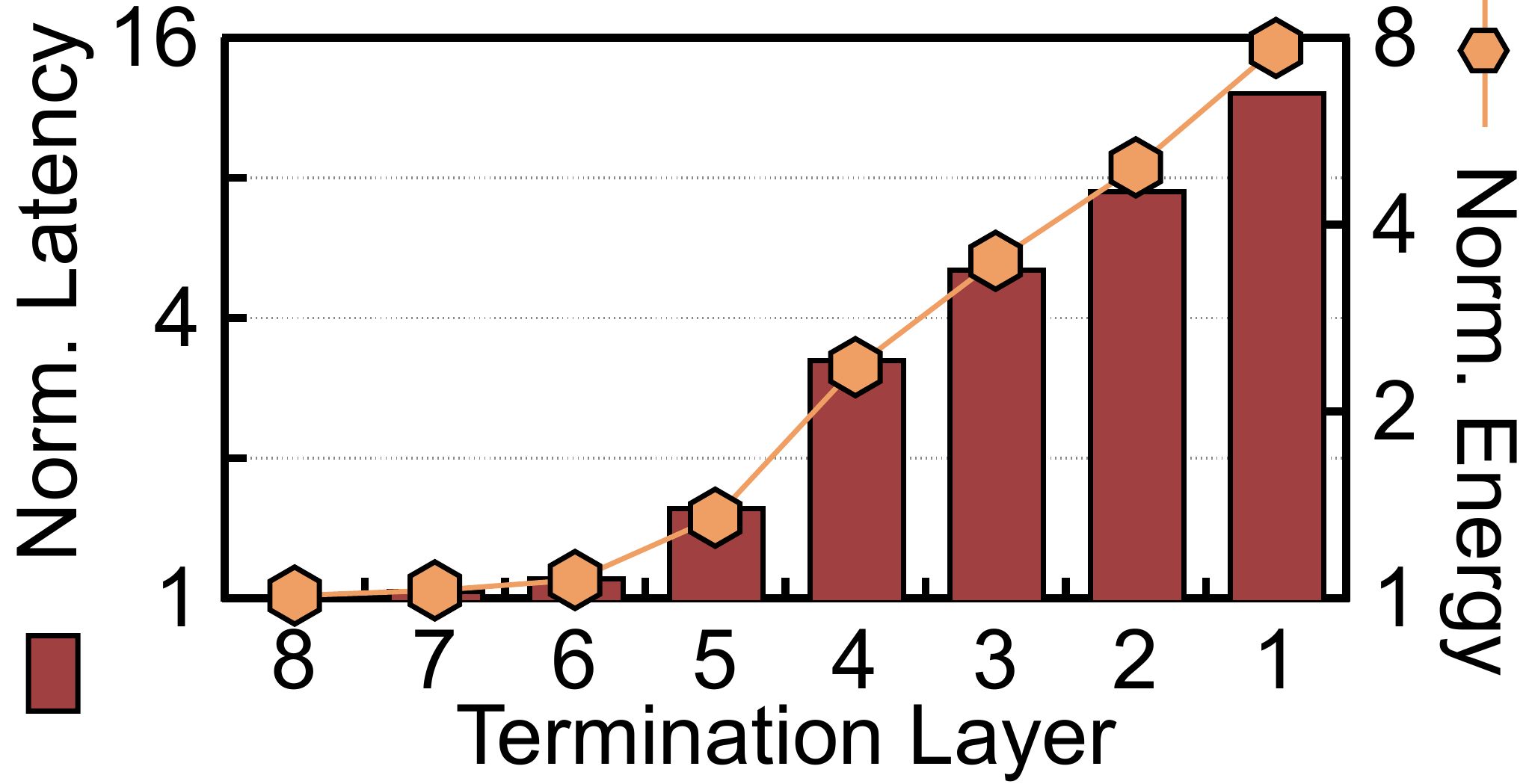}
  \label{fig:lat_layer_t2}
}
%\vspace{-5pt}
\caption{Accuracy, latency, and energy consumption under different termination layer in \sys{BwCu}.}
\label{fig:layer_diff_t2}
%\vspace{-5pt}
%\end{figure}
%
%\begin{figure}[t]
%\vspace{-5pt}
\centering
\subfloat[\small{Accuracy.}]
{
  \includegraphics[trim=0 0 0 0, clip, height=0.9in]{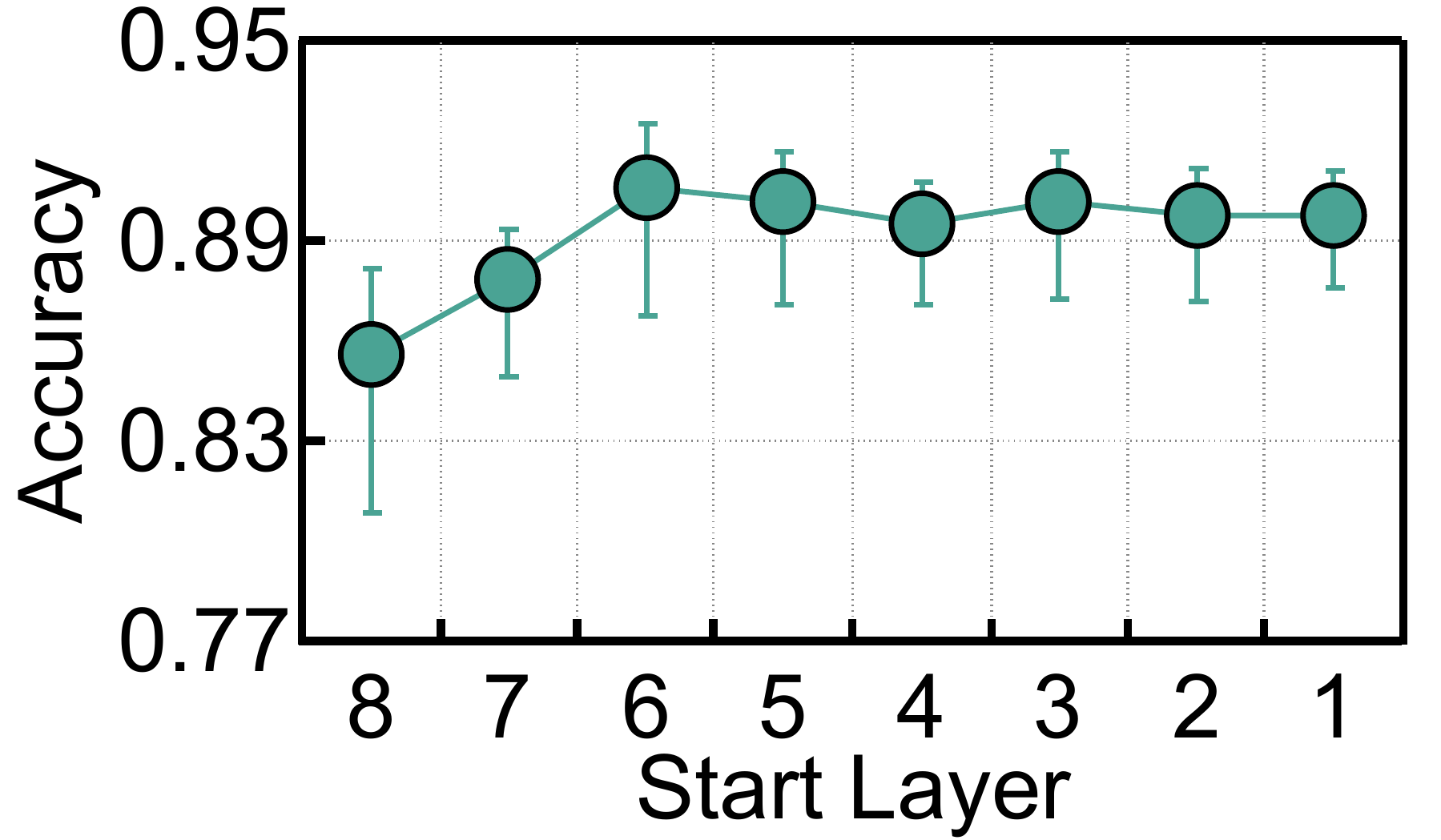}
  \label{fig:acc_layer_t8}
}
%\hfill
\subfloat[\small{Latency and energy.}]
{
  \includegraphics[trim=0 0 0 0, clip, height=0.9in]{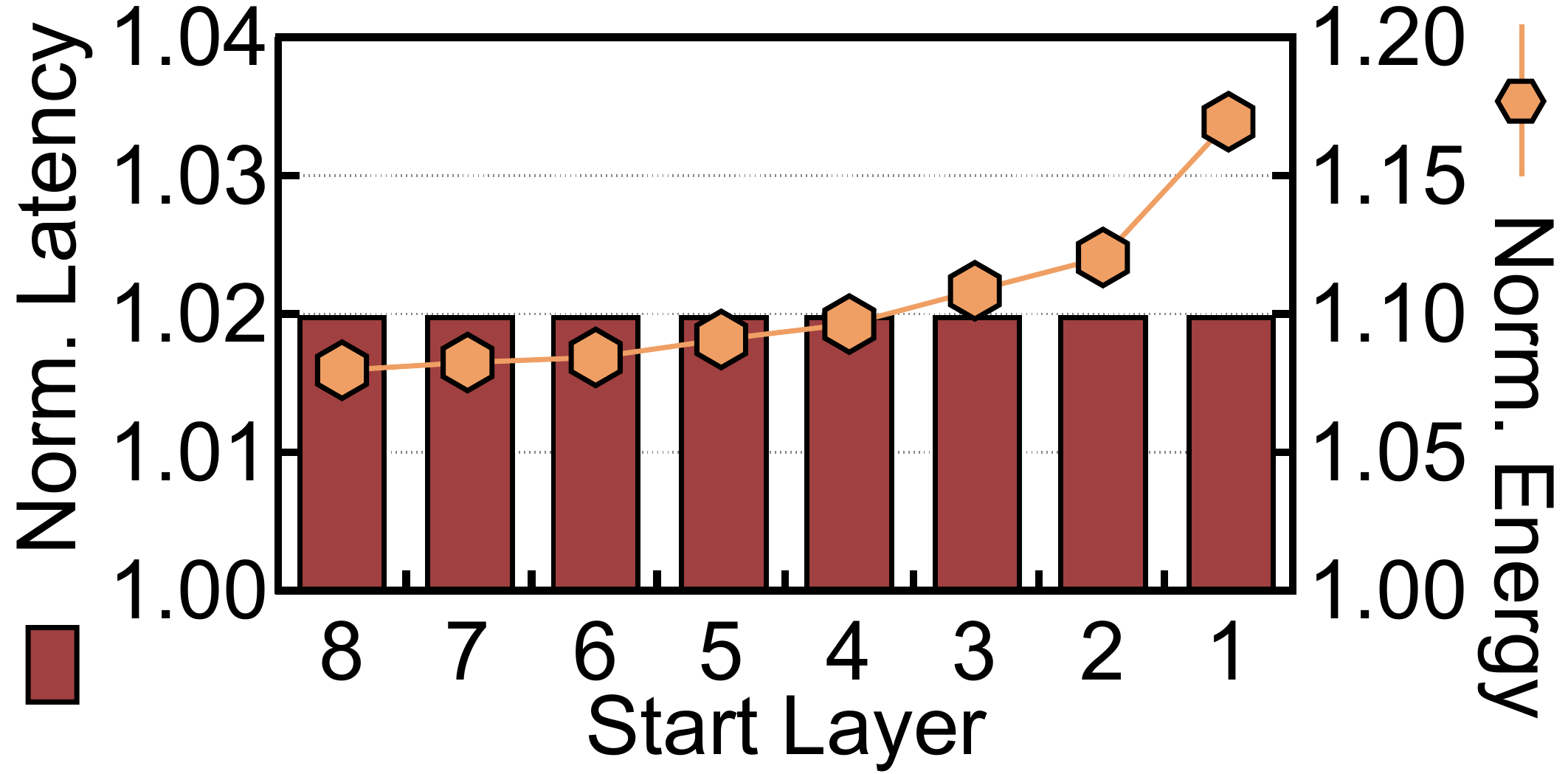}
  \label{fig:lat_layer_t8}
}
%\vspace{-5pt}
\caption{Accuracy, latency, and energy consumption under different start layer in \sys{FwAb}.}
\label{fig:layer_diff_t8}
%\vspace{-12pt}
\end{figure}

The \proj framework allows programmers to flexibly select which layers to extract important neurons from (\Sect{sec:sw:algo}). To trade accuracy for performance, programmers could start extracting important neurons later in forward extraction algorithms (as illustrated in~\Fig{fig:ex}), or terminate extraction earlier in backward extraction algorithms.

\paragraph{Early-Termination} We use \sys{BwCu} to showcase the trade-off that early-termination in backward extraction offers. For simplicity, we show only the results on AlexNet; ResNet18 has similar trends.~\Fig{fig:acc_layer_t2} shows how accuracy ($y$-axis) varies as the termination layer ($x$-axis) varies from 8 (the last layer) to 1 (the first layer). As AlexNet has 8 layers in total, terminating at layer 8 means extracting important neurons from only one layer. As extraction terminates later (further to the right on $x$-axis), more important neurons are captured and thus the accuracy increases. The accuracy increase eventually plateaus beyond layer 6, indicating marginal return of investment to extract more layers.

\Fig{fig:lat_layer_t2} shows how the latency and energy consumption varies with the termination layer. With virtually the same accuracy, extracting all the layers (i.e., terminating at layer 1) leads to 11.2$\times$ higher latency and 6.6$\times$ more energy compared to extracting only 3 layers (i.e., terminating after layer 6), which introduces only 1.1$\times$ and 1.1$\times$ latency and energy overhead over normal inference, respectively.

\paragraph{Late-Start} We use \sys{FwAb} as an example to demonstrate the trade-off that late-start provides to forward extraction-based methods.~\Fig{fig:acc_layer_t8} and~\Fig{fig:lat_layer_t8} show how the accuracy and latency/energy vary with the start layer, respectively.

Similar to early-termination, the accuracy increases as more layers are extracted, i.e., start earlier (further to the right). Interestingly, starting later does not help reduce the latency. This is because extraction latency is largely hidden behind the inference latency. However, starting later does reduce the energy consumption by 8.4\% because less work is done.

%Same with BWCU, the detection accuracy increases as start layer goes earlier but saturates at the third layer. The accuracy of using all layers even has marginal drop. \Fig{fig:lat_layer_t8} shows the performance variations with start layers. The latency overhead of \textit{Forward} remains the same as start layers goes higher since all the extraction latency was hided under the latency of next layer's inference, thus using late start policy will not help reducing latency overhead. However, tlehe late start policy contributes on 8.2\% energy saving starting at third layer compared with using all layers for detection. 

\subsection{Sensitivity and Scalability Studies}
\label{sec:eval:sen}

%The overhead brought by \proj is largely related to the hardware resources of the extractor. 
We show how \proj's performance varies with different hardware resource provisions in the path constructor. \no{We report power instead of energy results here to better reflect the impact of hardware resources.} We report only the results of \sys{BwCu} on AlexNet due to limited space.~\Fig{fig:merge_sens} shows how the latency and energy consumption (normalized to DNN inference) vary with the number of merge tree length (the number of partially sorted sequences that are merged simultaneously). As the merge tree length increases, the latency reduces (from 31.0$\times$ to 12.3$\times$), but the power consumption stays virtually the same. This is because a 16-length merge tree contributes to only 2\% of the total power.

%The latency improvements will almost saturate if the length of merge tree keep increasing because the DMA read and write to the merge buffer on-chip then become bottleneck. Thus, in \proj we pick length 16 as the merge tree length. 

\Fig{fig:sort_sens} shows how the latency and power consumption vary with the number of sort units. We find out latency decreases only marginally with more sort units, because sorting is memory-bound and thus increasing computing units has a marginal impact. The power consumption, however, increases significantly, because the sort unit contributes significantly (33.4\%) to the overall power in our design.

\begin{figure}[t]
\vspace{-10pt}
\centering
\subfloat[\small{Merge tree length.}]
{
  \includegraphics[trim=0 0 0 0, clip, width=0.45\columnwidth]{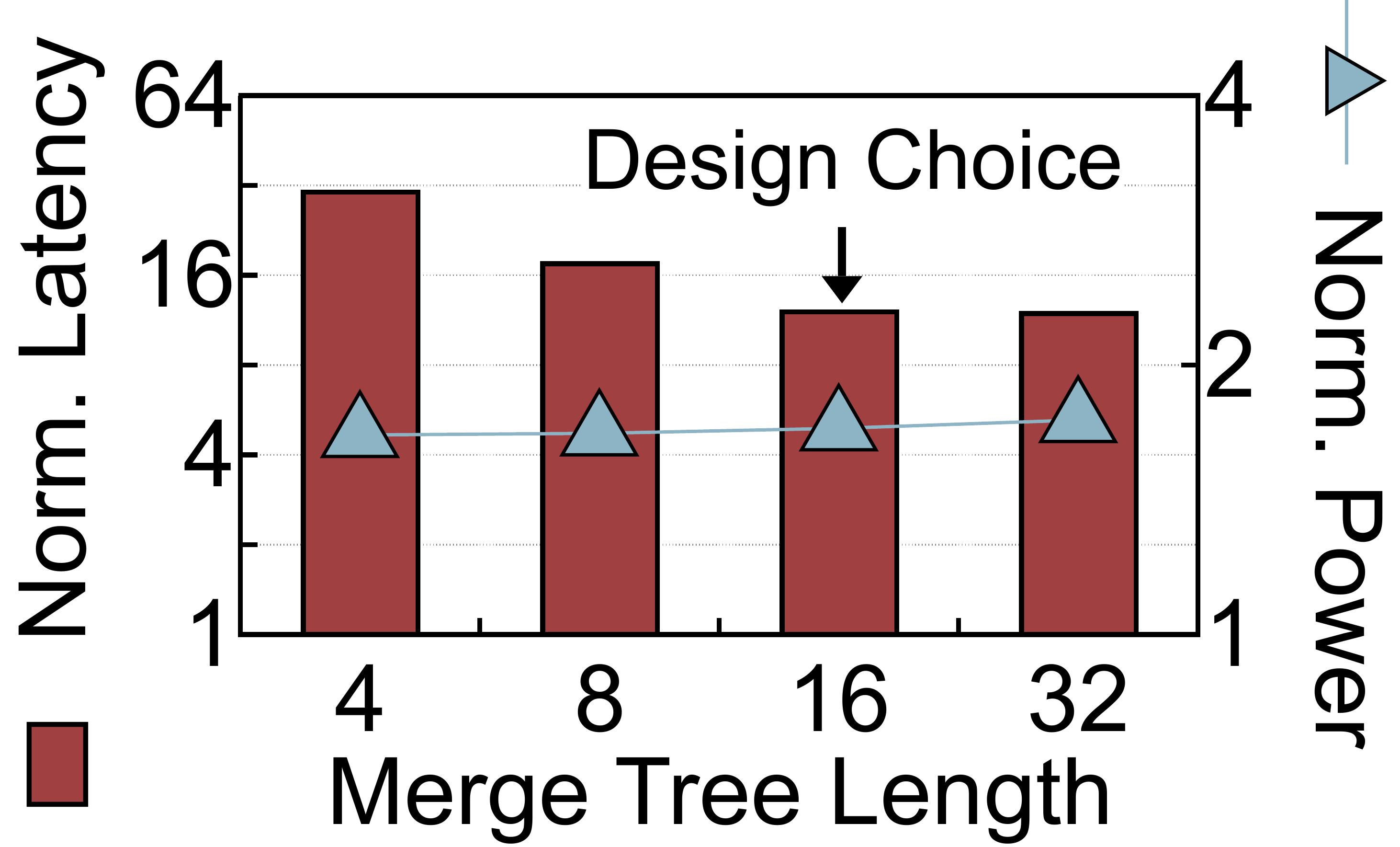}
  \label{fig:merge_sens}
}
\hfill
\subfloat[\small{Number of sort unit.}]
{
  \includegraphics[trim=0 0 0 0, clip, width=0.45\columnwidth]{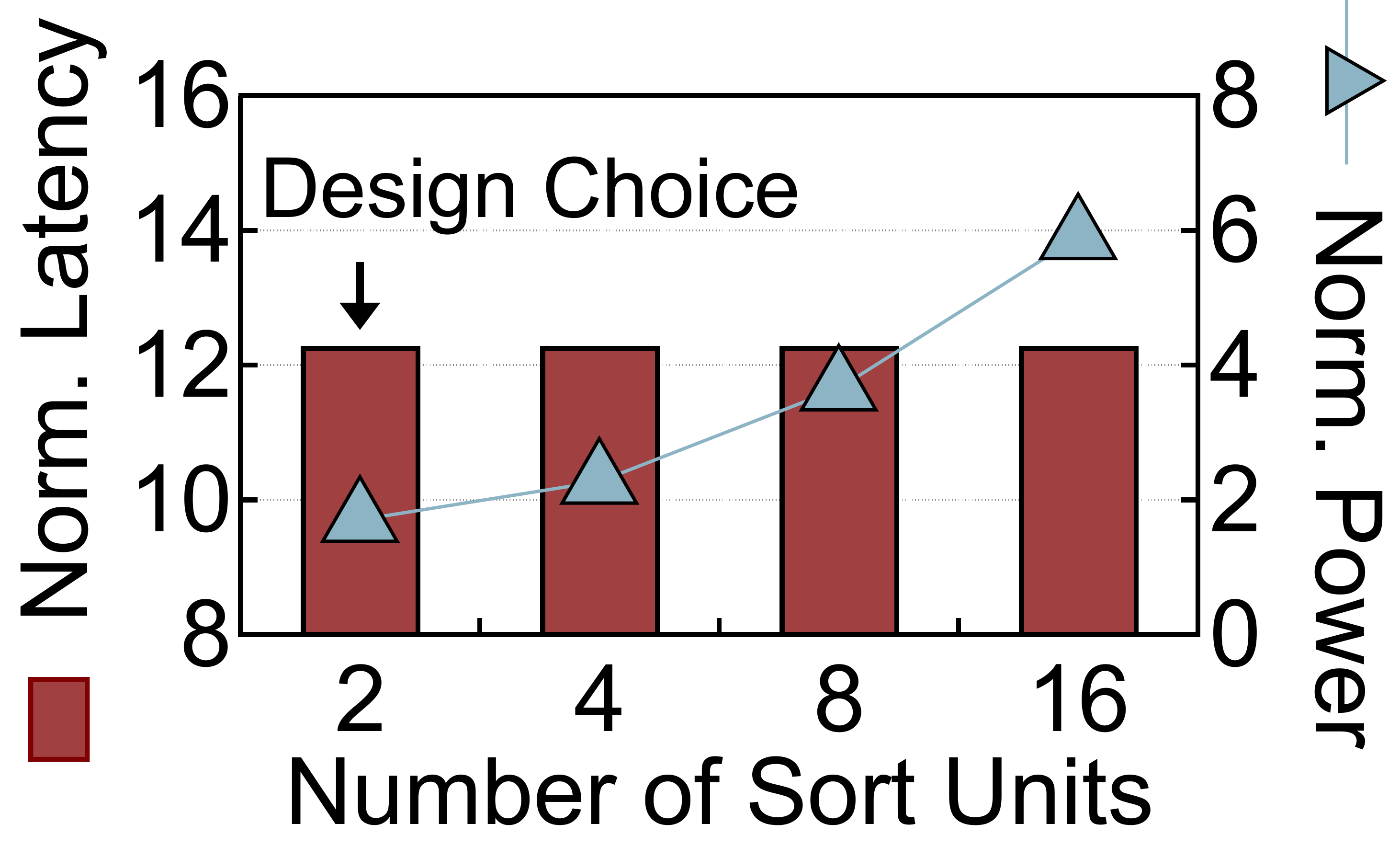}
  \label{fig:sort_sens}
}
\vspace{-5pt}
\caption{Performance vary with hardware resource.}
\label{fig:sens}
%\vspace{-12pt}
\end{figure}

While our original DNN accelerator uses 16-bit precision, we also evaluate \proj under a 8-bit design. The area overhead increases from 5.2\% to 5.5\%. For AlexNet, the 8-bit design has 2.1\% latency and 33.0\% energy overhead using \sys{FwAb}, comparable with 2.1\% and 16.0\% overhead of the original design. We also increase the MAC array size from 20$\times$20 to 32$\times$32. The area overhead increases from 5.2\% to 6.4\%. AlexNet has 4.4\% latency and 16.4\% energy overhead using \sys{FwAb}, on par with the original 2.1\% and 16.0\% value.

\subsection{Large Model Evaluation}
\label{sec:eval:large}
 
On VGG16~\cite{simonyan2014very} and Inception-V4~\cite{szegedy2017inception}, the average inter-class path similarity on ImageNet is only 41.5\% and 28.8\%, respectively, indicating that important neurons exist and class paths are unique in these models.
 
We also applied our detection scheme to DenseNet~\mbox{\cite{iandola2014densenet}}, and achieved 100\% detection accuracy with 0\% false positive rate (FPR), higher than the previously best accuracy at 96\% with 3.8\% FPR~\mbox{\cite{ma2019nic}}. We use the detection accuracy and false positive rate instead of AUC in order to directly compare with the referenced method.  We also evaluated ResNet50 on ImageNet using \mbox{\sys{BwCu}}. The accuracy is 0.900, which is more accurate than EP~\mbox{\cite{effectpath2019}} (0.898).

%To prove the effectiveness of \proj in realistic systems, we demonstrate our system on larger models including Resnet50~\cite{resnet}, Densenet~\cite{iandola2014densenet}, VGG16~\cite{simonyan2014very} and Inception-V4~\cite{szegedy2017inception}. We show the accuracy results in the table below.

%\begin{table}[t]
%\centering
%\caption{Accuracy results of \proj for large models.} 
%\renewcommand*{\arraystretch}{0.1}
%\renewcommand*{\tabcolsep}{3pt}
%\resizebox{\columnwidth}{!}
%{
%\begin{tabular}{c|c|c}
%\toprule[0.15em]
%\textbf{Network} & \textbf{BwCU} & \textbf{BwAB}  \\
%\midrule[0.05em]
%Resnet50   & \fixme{0.9} & \fixme{0.9}   \\ 
%Densenet  & \fixme{0.9} & \fixme{0.9}  \\ 
%VGG16   & \fixme{0.9}   & \fixme{0.9}   \\ 
%Inception-v4  & \fixme{0.9}    & \fixme{0.9}   \\ 
%\bottomrule[0.15em]
%\end{tabular}
%}
%\label{tbl:fpga}
%\end{table}

%To compare with existing works, for Resnet50, the \sys{BwCu} reaches \fixme{0.9} accuracy which is higher than the results in EP. For Densenet, \proj achieves 100\% detection rate with with 0\% false positive rate (FPR), higher than the previously best detection accuracy at 96\% with 3.8\% FPR from~\cite{ma2019nic}.

\section{Related Work and Discussion}
\label{sec:related}

Different mechanisms to counter adversarial attacks have been explored. One major class is to boost the DNN robustness at the training time through adversarial retraining~\cite{advtraining, fgsm, textadversarial, 2014towards}, which incorporates adversarial samples into the training data. However, adversarial retraining does not have the detection capability at inference time. It also requires accesses to the retraining data, which \proj does not. \proj can also be integrated with adversarial retraining.
Besides adversarial samples, we expect that \proj could also be used for detecting the execution errors of DNN accelerators caused by transient hardware errors~\cite{leng2020hpca, leng2015micro, margin2020tdmr}.

Detection mechanisms have also been extensively explored, ranging from using modular redundancies (e.g., input transformation~\cite{buckman2018thermometer, guo2017countering, thang2019image}, multiple models~\cite{rouhani2018deepfense}, and weights randomization~\cite{dhillon2018stochastic, xie2017mitigating}), to cascading a dedicated DNN to detect adversaries~\cite{ma2019nic, safetynet, gong2017adversarial, metzen2017detecting}. Wang et al.~\cite{wang2020dnnguard} proposes to spatially share the DNN accelerator resources between the original network and the detection network. \proj differs from them in two ways. First, we show that using \textit{path} as an explicit representation of the input, \proj can use a simple random forest classifier to detect adversarial inputs rather than complicated DNNs. Coupled with other performance optimizations, \proj provides very low (2\%) overhead to enable detection at inference-time while others introduce several folds higher overhead. Second, \proj provides an algorithm design framework that allows programmers to make trade-offs between detection efficiency and accuracy.
%not one single detection algorithm; rather it provides 

%\proj primarily focuses on classification tasks because they are by far the predominant target of today's adversarial attacks~\cite{akhtar2018threat}. However, the fundamental idea applies generally to other tasks such as object detection \cite{objectadv}, semantic segmentation \cite{objectadv,fischer2017adversarial,metzen2017universal}, and question answering \cite{qaadv,zhao2017generating}.

%Architecture community has also been looking at effectively detection techniques to achieve low overhead detection during deep learning inference. One example is utilizing hardware acceleration techniques such as parallel execution on existing detection algorithms ~\cite{wang2020dnnguard,rouhani2018deepfense}. Unlike above works, \proj co-design algorithms and hardware. Although with significant different design methodology, we try out best to compare \proj with \textbf{DNNGuard} ~\cite{wang2020dnnguard}. \proj results in less overhead on Resnet50 with Imagenet as the dataset (1.2 $\times$ compared with 1.5 $\times$).

\no{\proj detection algorithm falls in a class of detection schemes that are complete but introduce false positives/negatives. Another class of methods verifies the robustness properties of neural networks offline~\cite{anderson2019optimization, gehr2018ai2, katz2017reluplex, wang2018formal}. They are sound but incomplete. For instance, Anderson et al.~\cite{anderson2019optimization} propose the Charon framework that verifies (or falsifies) that all inputs in a given range are classified as the same class. \proj could be combined with these static techniques to improve runtime efficiency and detection accuracy.}

Carlini et al.~\mbox{\cite{carlini2019evaluating}} provides a checklist of best practices in evaluating defense mechanisms of adversarial attacks. This paper exercises the following red teaming:

\begin{itemize}
    \item Stated the threat model: attackers know everything (model, inputs, defense).
	\item Performed adaptive attacks (\Sect{sec:eval:adaptive}).
	\item Reported clean model accuracy (\Sect{sec:exp:setup}).
	\item Performed basic sanity checks (iterative attacks perform better than single-step attacks; increasing the perturbation budget strictly increases attack success rate; with ``high'' distortion, model accuracy reaches random guessing.).
	\item Analyzed success vs. distortion (perturbation) for our adaptive attack (\Sect{sec:eval:adaptive}).
	\item Showed that adaptive attacks are better (harder to be detected) than non-adaptive ones (\mbox{\Fig{fig:at}}).
	\item Showed attack hyper-parameters with the released code.
	\item Applied both non-adaptive attacks (covering all three types of input perturbation measures ($L_0$, $L_2$, and $L_\infty$)) and adaptive attacks (\Sect{sec:exp:setup}).
	\item For non-differentiable components (in adaptive attacks), applied differentiable techniques (\Sect{sec:eval:adaptive}).
	\item Verified that the attacks have converged under the selected hyper-parameters.
	%\item Compared against prior work and explained important differences.
\end{itemize}

\section{Conclusion}
\label{sec:conc}

Deep-learning driven applications are cultivating Software 2.0, an exciting software paradigm that is not robust to input perturbations. The robustness issue is further exacerbated by the lack of explainability in deep learning. Adversarial attacks exploit the robustness vulnerability, and represents one important instance of AI safety as AI techniques penetrate into mission-critical systems~\cite{zhao2019towards, zhu2017cognitive}.

\proj enables efficient and accurate adversarial detection at inference-time. The key is to exploit the program execution behaviors of DNN inference that are largely ignored before. We demonstrate a careful co-design of algorithmic framework, compiler optimizations, and hardware architecture. The concepts of important neuron and activation path complement existing explainable ML efforts~\cite{aireport, holzinger2017glass, biecek2018dalex, biecek2018dalex, sokol2018glass}, and could shed new light on interpreting DNNs.

%For non-classification tasks such as objective detection and segmentation tasks, the fundamental idea of extracting important neurons still applies. Since each object/pixel is classified individually, the overall activation path will aggregate the activation paths from all the objects/pixels, which we leave for future work. We hope that by disseminating the idea and releasing our code, the research community could explore the full potential of path-based adversary detection.

\section{Acknowledgement}

We thank the anonymous reviewers from ISCA 2020 and MICRO 2020 and the shepherd from MICRO for their valuable feedback and/or guidance. Jingwen Leng and Minyi Guo are the corresponding authors of the paper.

%%%%%%% -- PAPER CONTENT ENDS -- %%%%%%%%

%%%%%%%%% -- BIB STYLE AND FILE -- %%%%%%%%
\bibliographystyle{IEEEtranS}
\bibliography{refs}

% Generated by IEEEtranS.bst, version: 1.13 (2008/09/30)
\begin{thebibliography}{10}
\providecommand{\url}[1]{#1}
\csname url@samestyle\endcsname
\providecommand{\newblock}{\relax}
\providecommand{\bibinfo}[2]{#2}
\providecommand{\BIBentrySTDinterwordspacing}{\spaceskip=0pt\relax}
\providecommand{\BIBentryALTinterwordstretchfactor}{4}
\providecommand{\BIBentryALTinterwordspacing}{\spaceskip=\fontdimen2\font plus
\BIBentryALTinterwordstretchfactor\fontdimen3\font minus
  \fontdimen4\font\relax}
\providecommand{\BIBforeignlanguage}[2]{{%
\expandafter\ifx\csname l@#1\endcsname\relax
\typeout{** WARNING: IEEEtranS.bst: No hyphenation pattern has been}%
\typeout{** loaded for the language `#1'. Using the pattern for}%
\typeout{** the default language instead.}%
\else
\language=\csname l@#1\endcsname
\fi
#2}}
\providecommand{\BIBdecl}{\relax}
\BIBdecl

\bibitem{15nmcell}
\BIBentryALTinterwordspacing
``{15NM OPEN-CELL LIBRARY},'' http://www.si2.org/open-cell-library/. [Online].
  Available: \url{http://www.si2.org/open-cell-library/}
\BIBentrySTDinterwordspacing

\bibitem{xaviersoc}
\BIBentryALTinterwordspacing
``{NVIDIA Reveals Xavier SOC Details},'' https://bit.ly/2qq0TWp. [Online].
  Available:
  \url{https://www.forbes.com/sites/moorinsights/2018/08/24/nvidia-reveals-xavier-soc-details/amp/}
\BIBentrySTDinterwordspacing

\bibitem{aireport}
``{2016–2019 Progress Report: Advancing Artificial Intelligence R\&D},''
  \url{https://www.whitehouse.gov/wp-content/uploads/2019/11/AI-Research-and-Development-Progress-Report-2016-2019.pdf},
  2019.

\bibitem{akhtar2018threat}
N.~Akhtar and A.~Mian, ``Threat of adversarial attacks on deep learning in
  computer vision: A survey,'' \emph{IEEE Access}, vol.~6, pp.
  14\,410--14\,430, 2018.

\bibitem{allan1995software}
V.~H. Allan, R.~B. Jones, R.~M. Lee, and S.~J. Allan, ``Software pipelining,''
  \emph{ACM Computing Surveys (CSUR)}, vol.~27, no.~3, pp. 367--432, 1995.

\bibitem{athalye2018robustness}
A.~Athalye and N.~Carlini, ``On the robustness of the cvpr 2018 white-box
  adversarial example defenses,'' \emph{arXiv preprint arXiv:1804.03286}, 2018.

\bibitem{ball1996efficient}
T.~Ball and J.~R. Larus, ``Efficient path profiling,'' in \emph{Proceedings of
  the 29th annual ACM/IEEE international symposium on Microarchitecture}.\hskip
  1em plus 0.5em minus 0.4em\relax IEEE Computer Society, 1996, pp. 46--57.

\bibitem{biecek2018dalex}
P.~Biecek, ``Dalex: explainers for complex predictive models in r,'' \emph{The
  Journal of Machine Learning Research}, vol.~19, no.~1, pp. 3245--3249, 2018.

\bibitem{advtraining}
J.~Bradshaw, A.~G. d.~G. Matthews, and Z.~Ghahramani, ``Adversarial examples,
  uncertainty, and transfer testing robustness in gaussian process hybrid deep
  networks,'' \emph{arXiv preprint arXiv:1707.02476}, 2017.

\bibitem{buckman2018thermometer}
J.~Buckman, A.~Roy, C.~Raffel, and I.~Goodfellow, ``Thermometer encoding: One
  hot way to resist adversarial examples,'' 2018.

\bibitem{carlini2019evaluating}
N.~Carlini, A.~Athalye, N.~Papernot, W.~Brendel, J.~Rauber, D.~Tsipras,
  I.~Goodfellow, A.~Madry, and A.~Kurakin, ``On evaluating adversarial
  robustness,'' \emph{arXiv preprint arXiv:1902.06705}, 2019.

\bibitem{carlini2017adversarial}
N.~Carlini and D.~Wagner, ``Adversarial examples are not easily detected:
  Bypassing ten detection methods,'' in \emph{Proceedings of the 10th ACM
  Workshop on Artificial Intelligence and Security}.\hskip 1em plus 0.5em minus
  0.4em\relax ACM, 2017, pp. 3--14.

\bibitem{carlini2017towards}
N.~Carlini and D.~Wagner, ``Towards evaluating the robustness of neural
  networks,'' in \emph{2017 IEEE Symposium on Security and Privacy (SP)}.\hskip
  1em plus 0.5em minus 0.4em\relax IEEE, 2017, pp. 39--57.

\bibitem{cwl2}
N.~Carlini and D.~Wagner, ``Towards evaluating the robustness of neural
  networks,'' in \emph{2017 IEEE Symposium on Security and Privacy (SP)}.\hskip
  1em plus 0.5em minus 0.4em\relax IEEE, 2017, pp. 39--57.

\bibitem{chang1988trace}
P.~P. Chang and W.~Hwu, ``Trace selection for compiling large c application
  programs to microcode,'' in \emph{Proceedings of the 21st annual workshop on
  Microprogramming and microarchitecture}.\hskip 1em plus 0.5em minus
  0.4em\relax IEEE Computer Society Press, 1988, pp. 21--29.

\bibitem{chen2015energy}
R.~Chen, S.~Siriyal, and V.~Prasanna, ``Energy and memory efficient mapping of
  bitonic sorting on fpga,'' in \emph{Proceedings of the 2015 ACM/SIGDA
  International Symposium on Field-Programmable Gate Arrays}.\hskip 1em plus
  0.5em minus 0.4em\relax ACM, 2015, pp. 240--249.

\bibitem{imagenet}
J.~Deng, W.~Dong, R.~Socher, L.-J. Li, K.~Li, and L.~Fei-Fei, ``Imagenet: A
  large-scale hierarchical image database,'' in \emph{2009 IEEE conference on
  computer vision and pattern recognition}.\hskip 1em plus 0.5em minus
  0.4em\relax Ieee, 2009, pp. 248--255.

\bibitem{dhillon2018stochastic}
G.~S. Dhillon, K.~Azizzadenesheli, Z.~C. Lipton, J.~Bernstein, J.~Kossaifi,
  A.~Khanna, and A.~Anandkumar, ``Stochastic activation pruning for robust
  adversarial defense,'' \emph{arXiv preprint arXiv:1803.01442}, 2018.

\bibitem{donovan2000profile}
R.~J. Donovan, R.~R. Roediger, and W.~J. Schmidt, ``Profile driven optimization
  of frequently executed paths with inlining of code fragment (one or more
  lines of code from a child procedure to a parent procedure),'' Jun.~6 2000,
  uS Patent 6,072,951.

\bibitem{fisher1981trace}
J.~A. Fisher, ``Trace scheduling: A technique for global microcode
  compaction,'' \emph{IEEE transactions on computers}, no.~7, pp. 478--490,
  1981.

\bibitem{gong2017adversarial}
Z.~Gong, W.~Wang, and W.-S. Ku, ``Adversarial and clean data are not twins,''
  \emph{arXiv preprint arXiv:1704.04960}, 2017.

\bibitem{fgsm}
I.~J. Goodfellow, J.~Shlens, and C.~Szegedy, ``Explaining and harnessing
  adversarial examples,'' \emph{arXiv preprint arXiv:1412.6572}, 2014.

\bibitem{2014towards}
S.~Gu and L.~Rigazio, ``Towards deep neural network architectures robust to
  adversarial examples,'' \emph{arXiv preprint arXiv:1412.5068}, 2014.

\bibitem{guo2017countering}
C.~Guo, M.~Rana, M.~Cisse, and L.~Van Der~Maaten, ``Countering adversarial
  images using input transformations,'' \emph{arXiv preprint arXiv:1711.00117},
  2017.

\bibitem{he2017adversarial}
W.~He, J.~Wei, X.~Chen, N.~Carlini, and D.~Song, ``Adversarial example defense:
  Ensembles of weak defenses are not strong,'' in \emph{11th $\{$USENIX$\}$
  Workshop on Offensive Technologies ($\{$WOOT$\}$ 17)}, 2017.

\bibitem{holzinger2017glass}
A.~Holzinger, M.~Plass, K.~Holzinger, G.~C. Crisan, C.-M. Pintea, and
  V.~Palade, ``A glass-box interactive machine learning approach for solving
  np-hard problems with the human-in-the-loop,'' \emph{arXiv preprint
  arXiv:1708.01104}, 2017.

\bibitem{hoste2008cole}
K.~Hoste and L.~Eeckhout, ``Cole: compiler optimization level exploration,'' in
  \emph{Proceedings of the 6th annual IEEE/ACM international symposium on Code
  generation and optimization}.\hskip 1em plus 0.5em minus 0.4em\relax ACM,
  2008, pp. 165--174.

\bibitem{hu2020deepsniffer}
X.~Hu, L.~Liang, S.~Li, L.~Deng, P.~Zuo, Y.~Ji, X.~Xie, Y.~Ding, C.~Liu,
  T.~Sherwood \emph{et~al.}, ``Deepsniffer: A dnn model extraction framework
  based on learning architectural hints,'' in \emph{Proceedings of the
  Twenty-Fifth International Conference on Architectural Support for
  Programming Languages and Operating Systems}, 2020, pp. 385--399.

\bibitem{huang2005using}
J.~Huang and C.~X. Ling, ``Using auc and accuracy in evaluating learning
  algorithms,'' \emph{IEEE Transactions on knowledge and Data Engineering},
  vol.~17, no.~3, pp. 299--310, 2005.

\bibitem{iandola2014densenet}
F.~Iandola, M.~Moskewicz, S.~Karayev, R.~Girshick, T.~Darrell, and K.~Keutzer,
  ``Densenet: Implementing efficient convnet descriptor pyramids,'' \emph{arXiv
  preprint arXiv:1404.1869}, 2014.

\bibitem{jouppi2017datacenter}
N.~P. Jouppi, C.~Young, N.~Patil, D.~Patterson, G.~Agrawal, R.~Bajwa, S.~Bates,
  S.~Bhatia, N.~Boden, A.~Borchers \emph{et~al.}, ``In-datacenter performance
  analysis of a tensor processing unit,'' in \emph{2017 ACM/IEEE 44th Annual
  International Symposium on Computer Architecture (ISCA)}.\hskip 1em plus
  0.5em minus 0.4em\relax IEEE, 2017, pp. 1--12.

\bibitem{knuth2014art}
D.~E. Knuth, \emph{Art of computer programming, volume 3: Sorting and
  Searching}.\hskip 1em plus 0.5em minus 0.4em\relax Addison-Wesley
  Professional, 2014.

\bibitem{koch2011fpgasort}
D.~Koch and J.~Torresen, ``Fpgasort: A high performance sorting architecture
  exploiting run-time reconfiguration on fpgas for large problem sorting,'' in
  \emph{Proceedings of the 19th ACM/SIGDA international symposium on Field
  programmable gate arrays}.\hskip 1em plus 0.5em minus 0.4em\relax ACM, 2011,
  pp. 45--54.

\bibitem{cifar}
A.~Krizhevsky, G.~Hinton \emph{et~al.}, ``Learning multiple layers of features
  from tiny images,'' Citeseer, Tech. Rep., 2009.

\bibitem{alexnet}
A.~Krizhevsky, I.~Sutskever, and G.~E. Hinton, ``Imagenet classification with
  deep convolutional neural networks,'' in \emph{Advances in neural information
  processing systems}, 2012, pp. 1097--1105.

\bibitem{kurakin2016adversarial}
A.~Kurakin, I.~Goodfellow, and S.~Bengio, ``Adversarial examples in the
  physical world,'' \emph{arXiv preprint arXiv:1607.02533}, 2016.

\bibitem{mnist}
Y.~LeCun, ``The mnist database of handwritten digits,'' \emph{http://yann.
  lecun. com/exdb/mnist/}, 1998.

\bibitem{leng2020hpca}
J.~{Leng}, A.~{Buyuktosunoglu}, R.~{Bertran}, P.~{Bose}, Q.~{Chen}, M.~{Guo},
  and V.~{Janapa Reddi}, ``Asymmetric resilience: Exploiting task-level
  idempotency for transient error recovery in accelerator-based systems,'' in
  \emph{2020 IEEE International Symposium on High Performance Computer
  Architecture (HPCA)}, 2020, pp. 44--57.

\bibitem{leng2015micro}
J.~{Leng}, A.~{Buyuktosunoglu}, R.~{Bertran}, P.~{Bose}, and V.~J. {Reddi},
  ``Safe limits on voltage reduction efficiency in gpus: A direct measurement
  approach,'' in \emph{2015 48th Annual IEEE/ACM International Symposium on
  Microarchitecture (MICRO)}, 2015, pp. 294--307.

\bibitem{li2019adversarial}
J.~Li, F.~Schmidt, and Z.~Kolter, ``Adversarial camera stickers: A physical
  camera-based attack on deep learning systems,'' in \emph{International
  Conference on Machine Learning}, 2019, pp. 3896--3904.

\bibitem{liaw2002classification}
A.~Liaw, M.~Wiener \emph{et~al.}, ``Classification and regression by
  randomforest,'' \emph{R news}, vol.~2, no.~3, pp. 18--22, 2002.

\bibitem{safetynet}
J.~Lu, T.~Issaranon, and D.~Forsyth, ``Safetynet: Detecting and rejecting
  adversarial examples robustly,'' in \emph{Proceedings of the IEEE
  International Conference on Computer Vision}, 2017, pp. 446--454.

\bibitem{ma2019nic}
S.~Ma and Y.~Liu, ``Nic: Detecting adversarial samples with neural network
  invariant checking,'' in \emph{Proceedings of the 26th Network and
  Distributed System Security Symposium (NDSS 2019)}, 2019.

\bibitem{madry2017towards}
A.~Madry, A.~Makelov, L.~Schmidt, D.~Tsipras, and A.~Vladu, ``Towards deep
  learning models resistant to adversarial attacks,'' \emph{arXiv preprint
  arXiv:1706.06083}, 2017.

\bibitem{metzen2017detecting}
J.~H. Metzen, T.~Genewein, V.~Fischer, and B.~Bischoff, ``On detecting
  adversarial perturbations,'' \emph{arXiv preprint arXiv:1702.04267}, 2017.

\bibitem{textadversarial}
T.~Miyato, A.~M. Dai, and I.~Goodfellow, ``Adversarial training methods for
  semi-supervised text classification,'' \emph{arXiv preprint
  arXiv:1605.07725}, 2016.

\bibitem{deepfool}
S.-M. Moosavi-Dezfooli, A.~Fawzi, and P.~Frossard, ``Deepfool: a simple and
  accurate method to fool deep neural networks,'' in \emph{Proceedings of the
  IEEE conference on computer vision and pattern recognition}, 2016, pp.
  2574--2582.

\bibitem{mueller2012sorting}
R.~Mueller, J.~Teubner, and G.~Alonso, ``Sorting networks on fpgas,'' \emph{The
  VLDB Journal—The International Journal on Very Large Data Bases}, vol.~21,
  no.~1, pp. 1--23, 2012.

\bibitem{svhn}
Y.~Netzer, T.~Wang, A.~Coates, A.~Bissacco, B.~Wu, and A.~Y. Ng, ``Reading
  digits in natural images with unsupervised feature learning,'' 2011.

\bibitem{nguyen2015deep}
A.~Nguyen, J.~Yosinski, and J.~Clune, ``Deep neural networks are easily fooled:
  High confidence predictions for unrecognizable images,'' in \emph{Proceedings
  of the IEEE conference on computer vision and pattern recognition}, 2015, pp.
  427--436.

\bibitem{margin2020tdmr}
G.~{Papadimitriou}, A.~{Chatzidimitriou}, D.~{Gizopoulos}, V.~J. {Reddi},
  J.~{Leng}, B.~{Salami}, O.~S. {Unsal}, and A.~C. {Kestelman}, ``Exceeding
  conservative limits: A consolidated analysis on modern hardware margins,''
  \emph{IEEE Transactions on Device and Materials Reliability}, vol.~20, no.~2,
  pp. 341--350, 2020.

\bibitem{jsma}
N.~Papernot, P.~McDaniel, S.~Jha, M.~Fredrikson, Z.~B. Celik, and A.~Swami,
  ``The limitations of deep learning in adversarial settings,'' in \emph{2016
  IEEE European Symposium on Security and Privacy (EuroS\&P)}.\hskip 1em plus
  0.5em minus 0.4em\relax IEEE, 2016, pp. 372--387.

\bibitem{papernot2016distillation}
N.~Papernot, P.~McDaniel, X.~Wu, S.~Jha, and A.~Swami, ``Distillation as a
  defense to adversarial perturbations against deep neural networks,'' in
  \emph{2016 IEEE Symposium on Security and Privacy (SP)}.\hskip 1em plus 0.5em
  minus 0.4em\relax IEEE, 2016, pp. 582--597.

\bibitem{deepface}
O.~M. Parkhi, A.~Vedaldi, A.~Zisserman \emph{et~al.}, ``Deep face
  recognition.'' in \emph{bmvc}, vol.~1, no.~3, 2015, p.~6.

\bibitem{effectpath2019}
Y.~Qiu, J.~Leng, C.~Guo, Q.~Chen, C.~Li, M.~Guo, and Y.~Zhu, ``Adversarial
  defense through network profiling based path extraction,'' in
  \emph{Proceedings of the IEEE Conference on Computer Vision and Pattern
  Recognition}, 2019, pp. 4777--4786.

\bibitem{raghu2017svcca}
M.~Raghu, J.~Gilmer, J.~Yosinski, and J.~Sohl-Dickstein, ``Svcca: Singular
  vector canonical correlation analysis for deep learning dynamics and
  interpretability,'' in \emph{Advances in Neural Information Processing
  Systems}, 2017, pp. 6076--6085.

\bibitem{rouhani2018deepfense}
B.~D. Rouhani, M.~Samragh, M.~Javaheripi, T.~Javidi, and F.~Koushanfar,
  ``Deepfense: Online accelerated defense against adversarial deep learning,''
  in \emph{2018 IEEE/ACM International Conference on Computer-Aided Design
  (ICCAD)}.\hskip 1em plus 0.5em minus 0.4em\relax IEEE, 2018, pp. 1--8.

\bibitem{simonyan2014very}
K.~Simonyan and A.~Zisserman, ``Very deep convolutional networks for
  large-scale image recognition,'' \emph{arXiv preprint arXiv:1409.1556}, 2014.

\bibitem{smith2005virtual}
J.~Smith and R.~Nair, \emph{Virtual machines: versatile platforms for systems
  and processes}.\hskip 1em plus 0.5em minus 0.4em\relax Elsevier, 2005.

\bibitem{smith2000overcoming}
M.~D. Smith, ``Overcoming the challenges to feedback-directed optimization
  (keynote talk),'' in \emph{ACM SIGPLAN Notices}, vol.~35, no.~7.\hskip 1em
  plus 0.5em minus 0.4em\relax ACM, 2000, pp. 1--11.

\bibitem{sokol2018glass}
K.~Sokol and P.~A. Flach, ``Glass-box: Explaining ai decisions with
  counterfactual statements through conversation with a voice-enabled virtual
  assistant.'' in \emph{IJCAI}, 2018, pp. 5868--5870.

\bibitem{sorin2009fault}
D.~J. Sorin, ``Fault tolerant computer architecture,'' \emph{Synthesis Lectures
  on Computer Architecture}, vol.~4, no.~1, pp. 1--104, 2009.

\bibitem{deeppid}
Y.~Sun, D.~Liang, X.~Wang, and X.~Tang, ``Deepid3: Face recognition with very
  deep neural networks,'' \emph{arXiv preprint arXiv:1502.00873}, 2015.

\bibitem{resnet}
C.~Szegedy, S.~Ioffe, V.~Vanhoucke, and A.~A. Alemi, ``Inception-v4,
  inception-resnet and the impact of residual connections on learning,'' in
  \emph{Thirty-First AAAI Conference on Artificial Intelligence}, 2017.

\bibitem{szegedy2017inception}
C.~Szegedy, S.~Ioffe, V.~Vanhoucke, and A.~A. Alemi, ``Inception-v4,
  inception-resnet and the impact of residual connections on learning,'' in
  \emph{Thirty-first AAAI conference on artificial intelligence}, 2017.

\bibitem{intriguring}
C.~Szegedy, W.~Zaremba, I.~Sutskever, J.~Bruna, D.~Erhan, I.~Goodfellow, and
  R.~Fergus, ``Intriguing properties of neural networks,'' \emph{arXiv preprint
  arXiv:1312.6199}, 2013.

\bibitem{thang2019image}
D.~D. Thang and T.~Matsui, ``Image transformation can make neural networks more
  robust against adversarial examples,'' \emph{arXiv preprint
  arXiv:1901.03037}, 2019.

\bibitem{tramer2020adaptive}
F.~Tramer, N.~Carlini, W.~Brendel, and A.~Madry, ``On adaptive attacks to
  adversarial example defenses,'' \emph{arXiv preprint arXiv:2002.08347}, 2020.

\bibitem{adversarialtraining}
F.~Tram{\`e}r, A.~Kurakin, N.~Papernot, I.~Goodfellow, D.~Boneh, and
  P.~McDaniel, ``Ensemble adversarial training: Attacks and defenses,''
  \emph{arXiv preprint arXiv:1705.07204}, 2017.

\bibitem{triantafyllis2003compiler}
S.~Triantafyllis, M.~Vachharajani, N.~Vachharajani, and D.~I. August,
  ``Compiler optimization-space exploration,'' in \emph{Proceedings of the
  international symposium on Code generation and optimization:
  feedback-directed and runtime optimization}.\hskip 1em plus 0.5em minus
  0.4em\relax IEEE Computer Society, 2003, pp. 204--215.

\bibitem{wang2020dnnguard}
X.~Wang, R.~Hou, B.~Zhao, F.~Yuan, J.~Zhang, D.~Meng, and X.~Qian, ``Dnnguard:
  An elastic heterogeneous dnn accelerator architecture against adversarial
  attacks,'' in \emph{Proceedings of the Twenty-Fifth International Conference
  on Architectural Support for Programming Languages and Operating Systems},
  2020, pp. 19--34.

\bibitem{cdrp}
Y.~Wang, H.~Su, B.~Zhang, and X.~Hu, ``Interpret neural networks by identifying
  critical data routing paths,'' in \emph{The IEEE Conference on Computer
  Vision and Pattern Recognition (CVPR)}, June 2018.

\bibitem{xie2017mitigating}
C.~Xie, J.~Wang, Z.~Zhang, Z.~Ren, and A.~Yuille, ``Mitigating adversarial
  effects through randomization,'' \emph{arXiv preprint arXiv:1711.01991},
  2017.

\bibitem{zhao2019towards}
H.~Zhao, Y.~Zhang, P.~Meng, H.~Shi, L.~E. Li, T.~Lou, and J.~Zhao, ``Towards
  safety-aware computing system design in autonomous vehicles,'' \emph{arXiv
  preprint arXiv:1905.08453}, 2019.

\bibitem{improvingstability}
S.~Zheng, Y.~Song, T.~Leung, and I.~Goodfellow, ``Improving the robustness of
  deep neural networks via stability training,'' in \emph{Proceedings of the
  ieee conference on computer vision and pattern recognition}, 2016, pp.
  4480--4488.

\bibitem{zhu2017cognitive}
Y.~Zhu, V.~J. Reddi, R.~Adolf, S.~Rama, B.~Reagen, G.-Y. Wei, and D.~Brooks,
  ``Cognitive computing safety: The new horizon for reliability/the design and
  evolution of deep learning workloads,'' \emph{IEEE Micro}, vol.~37, no.~1,
  pp. 15--21, 2017.

\end{thebibliography}
%%%%%%%%%%%%%%%%%%%%%%%%%%%%%%%%%%%%

\end{document}